\documentclass[aps,twocolumn,showpacs,preprintnumbers,amsmath,floatfix,superscriptaddress,prd,longbibliography]{revtex4-1}
\usepackage{graphicx}% Include figure files
\usepackage{epsfig}
\usepackage{dcolumn}% Align table columns on decimal point
\usepackage{bbm}% bold math
\usepackage[colorlinks,linkcolor=red,anchorcolor=green,citecolor=blue,CJKbookmarks=True]{hyperref}
\usepackage{braket}
\usepackage{physics}
\usepackage{amssymb}

\begin{document}
\title{Decay properties of light $1^{-+}$ hybrids}
\author{\small Juzheng Liang}
\affiliation{\small Institute of High Energy Physics, Chinese Academy of Sciences, Beijing 100049, People's Republic of China}
\affiliation{\small School of Physics, University of Chinese Academy of Sciences, Beijing 100049, People's Republic of China}
\affiliation{\small School of Physics, University of Science and Technology of China, Hefei 230026, People's Republic of China}

\author{\small Siyang Chen}
\email{chensiyang@ihep.ac.cn}
\affiliation{\small Institute of High Energy Physics, Chinese Academy of Sciences, Beijing 100049, People's Republic of China}
\affiliation{\small School of Physics, University of Chinese Academy of Sciences, Beijing 100049, People's Republic of China}

\author{\small Ying Chen}
\email{cheny@ihep.ac.cn}
\affiliation{\small Institute of High Energy Physics, Chinese Academy of Sciences, Beijing 100049, People's Republic of China}
\affiliation{\small School of Physics, University of Chinese Academy of Sciences, Beijing 100049, People's Republic of China}
\affiliation{\small Center for High Energy Physics, Henan Academy of Sciences, Zhengzhou 450046, People's Republic of China}
\author{\small Chunjiang Shi}
\affiliation{\small Institute of High Energy Physics, Chinese Academy of Sciences, Beijing 100049, People's Republic of China}
\affiliation{\small School of Physics, University of Chinese Academy of Sciences, Beijing 100049, People's Republic of China}

\author{Wei Sun}
\affiliation{\small Institute of High Energy Physics, Chinese Academy of Sciences, Beijing 100049, People's Republic of China}

\def\modified#1{\red{#1}}
\begin{abstract}
We explore the decay properties of the isovector and isoscalar $1^{-+}$ light hybrids, $\pi_1$ and $\eta_1$, in $N_f=2$ lattice QCD at a pion mass $m_\pi \approx 417~\mathrm{MeV}$. The McNeile and Michael method is adopted to extract the effective couplings for individual decay modes, which are used to estimate the partial decay widths of $\pi_1(1600)$ and $\eta_1(1855)$ by assuming SU(3) symmetry. The partial decay widths of $\pi_1(1600)$ are predicted to be $(\Gamma_{b_1\pi}, \Gamma_{f_1(1285)\pi}, \Gamma_{\rho\pi}, \Gamma_{K^*\bar{K}}) = (325 \pm 75, \mathcal{O}(10), 52 \pm 7, 8.6 \pm 1.3)~\mathrm{MeV}$, and the total width is estimated to be $396 \pm 90~\mathrm{MeV}$, considering only statistical errors. If $\eta_1(1855)$ and the $4.4\sigma$ signal observed by BESIII (labeled as $\eta_1(2200)$) are taken as the two mass eigenstates of the isoscalar $1^{-+}$ light hybrids in SU(3), then the dominant decay channel(s) of $\eta_1(1855)$ ($\eta_1(2200)$) is $K_1(1270)\bar{K}$ ($K_1(1270)\bar{K}$ and $K_1(1400)\bar{K}$) through the $1^{+(-)}0^{-(+)}$ mode. The vector-vector decay modes are also significant for the two $\eta_1$ states. Using the mixing angle $\alpha \approx 22.7^\circ$ obtained from lattice QCD for the two $\eta_1$ states, the total widths are estimated to be $\Gamma_{\eta_1(1855)}=282(85)~\mathrm{MeV}$ and $\Gamma_{\eta_1(2200)}=455(143)~\mathrm{MeV}$. The former is compatible with the experimental width of $\eta_1(1855)$. Although many systematic uncertainties are not well controlled, these results are qualitatively informative for the experimental search for light hybrids.    
\end{abstract}
\maketitle
\section{Introduction}
QCD expects the existence of hybrid hadrons (hybrids), namely, bound states made up of both (constituent) quarks and gluons. The hybrid mesons with $J^{PC}=1^{-+}$ are most intriguing since this quantum number is prohibited for $q\bar{q}$ states in quark model. There have been several candidates for light $I^G J^{PC}=1^-1^{-+} $ hybrids, such as $\pi_1(1400)$, $\pi_1(1600)$ and $\pi_1(2015)$. The first evidence for a $I^G J^{PC}=1^-1^{-+}$ resonance dates back to 1988 when the GAMS/NA12 (IHEP-CERN) experiment observed $\pi_1(1400)$ in the $\eta\pi^0$ system~\cite{IHEP-Brussels-LosAlamos-AnnecyLAPP:1988iqi}. $\pi_1(1400)$ was also seen in $\eta\pi^-$ and $\eta\pi^0$ systems by later experiments, such as VES~\cite{VES:1992zkx,VES:1992zkx,VES:1993scg,VES:2001rwn,Amelin:2005ry}, E179 (KEK)~\cite{Aoyagi:1993kn}, E852~\cite{E852:1997gvf,E852:1999xev}, E862~\cite{E862:2006cfp} and Crystal Barrel~\cite{CrystalBarrel:1998cfz,CrystalBarrel:1999reg}. The OBELIX collaboration also observed $\pi_1(1400)$ in the $\rho\pi$ channel~\cite{OBELIX:2004oio}. Apart from $\pi_1(1400)$, many experiments also observed $\pi_1(1600)$ in $\eta'\pi$~\cite{VES:1993scg,Zaitsev:2000rc,Khokhlov:2000tk,E852:2004gpn,CLEO:2011upl,COMPASS:2014vkj}, $b_1\pi$~\cite{Zaitsev:2000rc,E852:2004rfa,Amelin:2005ry,Baker:2003jh}, $f_1(1285)\pi$~\cite{E852:2004gpn,Amelin:2005ry} and $\rho\pi$~\cite{E852:1998mbq,Zaitsev:2000rc,Chung:2002pu,COMPASS:2009xrl,COMPASS:2018uzl,COMPASS:2021ogp} systems. Theoretically, the Bose symmetry in the SU(3) limit prevents a hybrid from decaying into $\eta\pi$~\cite{Levinson:1964xx,Close:1987aw}, so it might be questionable for $\pi_1(1400)$ to be interpreted as a hybrid state. Moreover, a $\eta\pi-\eta'\pi$ coupled channel analysis of COMPASS data by JPAC indicates a single pole $(m,\Gamma)=(1564\pm 89, 492\pm 115)~\text{MeV}$~\cite{JPAC:2018zyd}, and a similar analysis of Crystal Barrel data leads also to a single pole around $(m,\Gamma)\sim (1623, 455)~\text{MeV}$~\cite{Kopf:2020yoa}. To date, $\pi_1(1600)$ is viewed as an established state by PDG with the parameters $(m,\Gamma)=(1661_{-11}^{+15}, 240\pm 50)~\text{MeV}$~\cite{Workman:2022ynf} (note the large discrepancy of this width with those from COMPASS and Crystal Barrel data). The 2024 version of the Review of Particle Physics (PDG 2024)~\cite{ParticleDataGroup:2024cfk} moves the previous $\pi_1(1400)$ entries into the $\pi_1(1600)$ section. As for the isoscalar counterpart of $\pi_1$, the BESIII collaboration reported recently the first observation of a $I^GJ^{PC}=0^+1^{-+}$ structure $\eta_1(1855)$ through the partial wave analysis of the $J/\psi\to \gamma \eta\eta'$ process~\cite{BESIII:2022riz,BESIII:2022iwi}. The resonance parameters of $\eta_1(1855)$ are determined to be $m_{\eta_1}=1855\pm 9_{-1}^{+6}$ MeV and $\Gamma_{\eta_1}=188\pm 18_{-8}^{+3}$ MeV. $\eta_1(1855)$ can be a candidate for an isoscalar $1^{-+}$ hybrid, and more experimental studies are under way.

Theoretically, light hybrid mesons are usually studied on the basis of the bag model~\cite{Barnes:1982tx,Chanowitz:1982qj}, potential models~\cite{Horn:1977rq,Ishida:1991mx}, QCD sum rules~\cite{Balitsky:1986hf,Latorre:1985tg,Govaerts:1986pp}, and the flux tube model~\cite{Isgur:1985vy,Close:1994hc}. In these models, a light hybrid is depicted either as a bound state of a pair of quark-antiquark ($q\bar{q}$) and a gluon, or a system that the constituent quark and antiquark are confined by an excited gluonic flux tube. The mass of the lightest $1^{-+}$ hybrid is predicted in a wide range from 1.3-2.5 GeV. On the other hand, many efforts from numerical lattice QCD calculations~\cite{Lacock:1996ny,MILC:1997usn,Mei:2002ip,Bernard:2003jd,Hedditch:2005zf,McNeile:2006bz,Dudek:2013yja,Woss:2020ayi,Chen:2022isv} have been devoted to predict the mass spectrum of light hybrids with the results that the mass of isovector $1^{-+}$ hybrid meson has a mass around 1.7-2.1 GeV, while the mass of the isoscalar is around 2.1-2.3 GeV~\cite{Dudek:2013yja}. These predictions are not far from the masses of $\pi_1(1600)$ and $\eta_1(1855)$ states.

The decay properties of hybrid mesons have been explored by various phenomenological models, among which the so-called triplet-P-zero (${}^3P_0$) model~\cite{Isgur:1985vy,Close:1994hc,Ackleh:1996yt,Barnes:1996ff} is the most commonly used one. In the ${}^3P_0$ model, a meson decays by producing a $q\bar{q}$ pair with vacuum quantum numbers ($J^{PC}=0^{++}$). It is found that the ${}^3P_0$ mechanism dominates most light-quark meson decays~\cite{Ackleh:1996yt}. Based on calculations using the ${}^3P_0$ model, a selection rule is proposed for hybrid decays suggesting that hybrids prefer to decay into an $L=0$ and an $L=1$ meson, while the decay modes involving two $L=0$ mesons are suppressed to the extent that the disconnected diagrams are not significant (OZI suppressed). Almost all models of hybrid mesons predict that they will not decay into identical pairs of mesons. These discussions lead to the often-quoted prediction for the decays of the $\pi_1$ hybrid:

\begin{eqnarray}\label{eq:width-ratio}
    &&\pi b_1 :\pi f_1:\pi \rho:\pi\eta:\pi\eta'\nonumber\\
    &=&170:60:5\sim 20:0\sim 10:0\sim 10.
\end{eqnarray}

However, $\pi_1(1600)$ is observed mainly in the $\rho\pi$ and $\eta'\pi$ systems, so this hierarchy pattern of $\pi_1$ decays needs to be validated by future experimental studies if $\pi_1(1600)$ is indeed a hybrid meson. Note that VES experiments give the estimate of the relative decay branching ratios $b_1\pi:f_1\pi:\rho\pi:\eta'\pi=(1.0\pm0.3):(1.1\pm0.3):< 0.3:1$~\cite{Amelin:2005ry}, and the E852 (BNL) results exhibit the ratio $f_1(1285)\pi:\eta'\pi=3.80\pm0.78$~\cite{E852:2001ikk,E852:2004gpn,Workman:2022ynf} for $\pi_1(1600)$. Right after the discovery of $\eta_1(1855)$, numerous theoretical studies on the properties of light hybrids have emerged in the literature.~\cite{Chen:2022qpd,Qiu:2022ktc,Shastry:2022mhk,Wang:2022sib,Swanson:2023zlm,Chen:2023ukh,Shastry:2023ths,Farina:2023oqk,Barsbay:2024vjt,Tan:2024grd,Giacosa:2024epf,Dong:2022cuw}. 

Hybrid decays can also be investigated through numerical lattice QCD studies. The state-of-the-art lattice QCD approach to study strong decays of hadrons is the L\"uscher method~\cite{Luscher:1986pf,Luscher:1990ux,Luscher:1991cf} and its generalization that takes coupled channel effects into account (see the review articles Ref.~\cite{Briceno:2017max,Mai:2022eur} and the references therein). To tackle the complicated coupled channel effects, the related study using the (generalized) L\"uscher method requires a substantial number of finite volume energy levels to be determined as precisely as possible. The calculation should be carried out on multiple lattice volumes and in different moving frames. This is numerically and computationally very challenging.

To date, only one lattice QCD study on the $\pi_1$ decay following this strategy has been carried out by the Hadron Spectrum Collaboration~\cite{Woss:2020ayi}. The calculation was performed in the limit of SU(3) flavor symmetry with $N_f=3$ dynamical strange quarks. The effective coupling of $\pi_1$ to different two-body decay modes was then obtained to predict the partial decay widths using physical kinematics. The sizable values are $\Gamma(b_1\pi,f_1(1285)\pi,\rho\pi,\eta'\pi)=(139\sim 529,0\sim24,0\sim 20,0\sim12)~\text{MeV}$, and they estimate the total width $\Gamma=139\sim 590~\text{MeV}$ of $\pi_1(1600)$. Despite the large variances, these results are in line with the phenomenological expectation and the total width is compatible with the PDG data~\cite{Workman:2022ynf}.

An alternative lattice QCD method for strong decays of hadrons is proposed by Michael and McNeile (M\&M)~\cite{McNeile:2002az,McNeile:2002fh}. The M\&M method calculates the tree-level transition amplitudes for two-body decays of a hadron, from which the effective couplings, and thereby the partial decay widths, can be estimated. This method has been applied to the studies of meson decays (and hadron-hadron mixings) with reasonable results~\cite{McNeile:2004rf,Michael:2005kw,McNeile:2006bz,McNeile:2006nv,Hart:2006ps,Michael:2006hf,Bali:2015gji,Zhang:2021xvl,Jiang:2022ffl,Shi:2023sdy}. The M\&M method is also applied to the study of the decay process $\Delta\to N\pi$~\cite{Alexandrou:2013ata,Alexandrou:2015hxa} and the results are consistent with those from the L\"{u}scher method~\cite{Andersen:2017una,Silvi:2021uya,Morningstar:2021ewk} and physical values~\cite{Pascalutsa:2005vq,Hemmert:1994ky}.

In Ref.~\cite{Bali:2015gji}, the M\&M method is applied to the decay process $\rho\to\pi\pi$ and obtains the effective coupling constants $g_{\rho\pi\pi}$ ranging from 5.2 to 8.4 (from different lattice volumes and different $\pi\pi$ kinetic configurations), which is compatible with the value $g_{\rho\pi\pi}\sim 6.0$ derived from the width of the $\rho$ meson~\cite{Workman:2022ynf} up to roughly a 40\% discrepancy.

In this work, we adopt the M\&M method to explore the decay properties of the isovector ($\pi_1$) and the isoscalar ($\eta_1$) $1^{-+}$ hybrids in the framework of $N_f=2$ lattice QCD. For $\pi_1$, we will compare the results from the M\&M method with those from the L\"uscher method as a consistency check. Then we will extend a similar study to the case of $\eta_1$ to provide the first lattice QCD prediction of $\eta_1$ decays. In $N_f=2$ QCD, the isoscalar $\eta_1$ is already a mass eigenstate, while in the $N_f=2+1$ QCD, there should be two mass eigenstates that are admixtures of $u\bar{u}+d\bar{d}$ and $s\bar{s}$ quark configurations (alternatively the flavor singlet and octet) through a mixing angle $\alpha$. So the connection of the $\eta_1$ results in this study with the physical $\eta_1$ states will also be discussed based on the value of $\alpha$ derived from a previous lattice QCD calculation~\cite{Dudek:2013yja}. 

Technically, the practical calculation of related correlation functions necessarily involves the annihilation diagrams of light quarks, which will be dealt with using the distillation method. This method provides a systematic scheme for the computation of the all-to-all quark propagators and the quark field smearing~\cite{Peardon:2009gh}.

This work is organized as follows. Sec.~\ref{sec:formalism} is devoted to a thorough introduction of the theoretical formalism for the extraction of the decay amplitudes of hybrids and the derivation of partial decay widths. The numerical procedures and results are presented in Sec.~\ref{sec:numerical}, which includes the basic information of the gauge ensemble and the construction of operators involved in this work. The lattice predictions of the decay properties of $\pi_1$ are presented in Sec.~\ref{sec:pi1-decay}. Section~\ref{sec:eta1-decay} is devoted to calculations of partial decay widths of $\eta_1(1855)$ and the possible $\eta_1(2200)$ based on SU(3) flavor symmetry. Sec.~\ref{sec:summary} is the summary of this work.

\section{Formalism}\label{sec:formalism}
\subsection{Transition matrix elements on lattice}
%\subsection{Decay amplitudes from correlation functions}
For a two-body decay process $h\to AB$ (without losing generality, $h$, $A$, and $B$ are assumed to be scalar particles for simplicity), the M\&M method starts with the Hamiltonian,
\begin{equation}\label{eq:H}
    H=\left(
    \begin{array}{cc} 
        E_h & x_{AB} \\
        x_{AB} & E_{AB}
    \end{array}
    \right),
\end{equation}
of the two-state system established by $|h\rangle\equiv(1,0)^T$ and $|AB\rangle\equiv(0,1)^T$, where $E_h$ and $E_{AB}$ are the energies of $|h\rangle$ and $|AB\rangle$ before the interaction, and $x\equiv\langle h|H|AB\rangle$ is the mixing energy or the transition amplitude from $|h\rangle$ to $|AB\rangle$. Thus, the first Fermi golden rule gives the decay width for $h\to AB$:
\begin{equation}\label{eq:Fermi}
    \Gamma(h\to AB)=2\pi |x_{AB}|^2 \rho_{AB},
\end{equation}
where $\rho_{AB}=dN_{AB}/dE$ is the state density of $AB$. The key point of the M\&M method is that the transition amplitude $x$ can be derived from the correlation function on the lattice:
\begin{eqnarray}
    C_{AB,h}(\vec{k},t)&=&\langle 0|\mathcal{O}_{AB}(\vec{k},t)\mathcal{O}_h^\dagger(0)|0\rangle\nonumber\\
    &=& \langle 0|\mathcal{O}_{AB}(\vec{k},0)e^{-Ht}\mathcal{O}_h^\dagger(0)|0\rangle,
\end{eqnarray}
where $\mathcal{O}_h$ is the interpolation field of $h$ and $\mathcal{O}_{AB}(\vec{k})$ is that for $AB$ with a relative momentum $\vec{k}$. The strategy is as follows. With the Hamiltonian in Eq.~(\ref{eq:H}), the exact expression of the time evolution operator $e^{-Ht}$ reads
\begin{multline}
    e^{-Ht}=e^{-\bar{E} t}\bigg[\cosh\left(\frac{\Delta}{2}t\right) I\\
    -\frac{1}{\Delta}\sinh\left(\frac{\Delta}{2}t\right)\sigma_3-\frac{2x_{AB}}{\Delta}\sinh\left(\frac{\Delta}{2}t\right)\sigma_1 \bigg],
\end{multline}
where $I,\sigma_i$ are the identity matrix and Pauli matrices, respectively, $\bar{E}=(E_h+E_{AB})/2$ is the average energy of $h$ and $AB$,
$\Delta=\sqrt{(E_h-E_{AB})^2+4x_{AB}^2}$ is the energy difference between two Hamiltanion eigenstates. First, we assume $\mathcal{O}_h$ couples exclusively to $|h\rangle$, while $\mathcal{O}_{AB}(\vec{k})$ couples exclusively to $|A(\vec{k})B(-\vec{k})\rangle\equiv|AB\rangle$, namely,
\begin{equation}\label{eq:overlap}
    \langle 0 |\mathcal{O}_a |b\rangle = Z_a^a\delta_{ab},
\end{equation}
with $a,b$ referring to either $h$ or $AB$, as is usually done in Refs.~\cite{McNeile:2002fh,McNeile:2002az,McNeile:2004rf,Michael:2005kw,McNeile:2006nv,Hart:2006ps,Michael:2006hf,Shi:2023sdy,Alexandrou:2013ata,Alexandrou:2015hxa}. Then it is easy to verify the relation:
\begin{equation}
    C_{AB,h}(\vec{k},t)\approx-\left(Z_h^h Z_{AB}^{AB}\right) \frac{2x_{AB}}{\Delta} \sinh\left(\frac{\Delta}{2}t\right) e^{-\bar{E}t}.
\end{equation}
When the relative momentum $\vec{k}$ of the $AB$ state is chosen appropriately such that $t\Delta$ is sufficiently small in a proper $t$ range, the decay amplitude $x_{AB}$ can be extracted from the ratio function
\begin{eqnarray}\label{eq:ratio1}
    R_{h,AB}(k,t)&=&\frac{C_{AB,h}}{\sqrt{C_{AB,AB}(k,t)C_{hh}(t)}}\nonumber\\
    &\approx& x_{AB}t\left(1+\frac{1}{24}(t\Delta)^2 \right),
\end{eqnarray}
where $C_{aa}(t)$ is the correlation function of $\mathcal{O}_a$ with $a$ referring to $h$ and $AB$. The partial decay width can then be predicted through Eq.~(\ref{eq:Fermi}) once the value of $x$ is determined. This is the major logic of the M\&M method that is applied in previous lattice calculations. Note that the expression in Eq.~(\ref{eq:ratio1}) will be slightly more complicated due to the polarization vectors if the spins of $\eta$, $A$, and $B$ are considered (see Eq.~(\ref{eq:rab}) below).

However, a small deviation from Eq.~(\ref{eq:overlap}) may induce corrections to Eq.~(\ref{eq:ratio1}) and thereby introduce systematic uncertainties to the transition matrix element $x$. To see this, we consider
\begin{eqnarray}
    \mathcal{O}_h^\dagger|0\rangle&=&Z_h^{h}(|h\rangle +\epsilon_1 |AB\rangle)\equiv Z_h^{h}
    \left(\begin{array}{c} 1\\\epsilon_1 \end{array}\right)
    \nonumber\\
    \mathcal{O}_{AB}^\dagger|0\rangle&=&Z_{AB}^{AB}(\epsilon_2|h\rangle+|AB\rangle)\equiv Z_{AB}^{AB}
    \left(\begin{array}{c} \epsilon_2\\1 \end{array}\right),
\end{eqnarray}
where $\epsilon_i\ll 1$ is assumed. In this case, we have
\begin{eqnarray}\label{eq:x-error}
    R_{h,AB}(k,t)&=& (\epsilon_1+\epsilon_2)\cosh\left(\frac{\Delta}{2}t\right)\nonumber\\
    &&-\frac{2x_{AB}+(\epsilon_1-\epsilon_2)\Delta}{\Delta} \sinh\left(\frac{\Delta}{2}t\right)\nonumber\\
    &\approx& (\epsilon_1+\epsilon_2) -\left(x_{AB}+\frac{\epsilon_1-\epsilon_2}{2}\Delta\right)t+\mathcal{O}((t\Delta)^2)\nonumber\\
    &\approx& a_0 + r_{AB} t +a_2 t^2 +\ldots.
\end{eqnarray}
In the practical data analysis, the above polynomial function with respect to $t$ is used to fit the numerical result of $R_{h,AB}(k,t)$, and the fit parameter $r_{AB}$ is an approximation of $x_{AB}$. The deviation $\delta x_{AB}=|r_{AB}-x|=|\frac{\epsilon_1-\epsilon_2}{2}|\Delta$ is taken as a systematic uncertainty that is estimated as $\delta x_{AB}\sim \sqrt{\overline{a^2_0 \Delta^2}}$.

It is important to note that the M\&M method introduced above is effective only when $ h, A $ and $ B $ are ground states in each channel~\cite{Michael:2005kw}. Contamination from excited states is anticipated to be suppressed in three primary ways. First, the distillation method we employ provides a smearing scheme for quark fields~\cite{Peardon:2009gh}, leading to operators constructed from smeared quark fields that significantly diminish couplings to excited states. Second, excited states are expected to contribute to both the numerator and the denominator of $ R_{AB}(t) $ in Eq.~(\ref{eq:ratio1}), and these contributions are anticipated to cancel each other out. Finally, the components of excited states in the correlation function are expected to be suppressed at large $ t $. Our fits are conducted within the time range where $ R_{AB}(t) $ exhibits the expected linear behavior, indicating that excited states are not significant in this range.

\subsection{Effective couples and partial decay widths}
After extracting the transition amplitude $ x_{AB} $ based on the aforementioned strategy, the transition rate on the lattice can be calculated utilizing Eq.~(\ref{eq:Fermi}) along with the lattice state density $ \rho_{AB} $. However, this transition rate cannot be considered a physical partial decay width, nor can it be employed in experimental studies, as lattice calculations are typically conducted at unphysical quark masses (and consequently unphysical hadron masses), leading to non-physical kinematics.

In quantum field theories, an effective interaction Lagrangian at the hadron level, $ \mathcal{L}_{h \to AB} \sim g_{AB} h AB $, is typically introduced for a specific decay process $ h \to AB $. The effective coupling $ g_{AB} $ is determined through experimental data, theoretical derivations, and symmetries. By defining the interaction Hamiltonian as $ H_I = -\int d^3 \vec{x} \mathcal{L}_{h \to AB}(x) $, one can demonstrate that the tree-level invariant amplitude $ \mathcal{M}_{AB} = \langle A(\vec{k}) B(-\vec{k}) | \mathcal{L}_{h \to AB} | h \rangle $ is related to $ x_{AB} $ by:
\begin{equation}\label{eq:match}
    x_{AB} =\frac{\mathcal{M}_{AB}}{(8L^3 m_h E_A(k)E_B(k))^{1/2}}
\end{equation}
Since $ \mathcal{M}_{AB} $ is typically derived using the relativistic state normalization $ \langle X(\vec{k}) | X(\vec{k}) \rangle = 2E_X(k) L^3 $ within a finite spatial volume of size $ L $, where $ X $ represents $ h, A, $ and $ B $, it follows that $ \mathcal{M}_{AB} $ is influenced by the effective coupling $ g_{AB} $. An alternative approach involves first extracting $ g_{AB} $ from $ x_{AB} $ and then using physical kinematics to predict the partial width. Here, we assume that $ g_{AB} $ is insensitive to light quark masses—and consequently to hadron masses—an assumption that is commonly made in the effective interaction analysis of hadron decays.

In this paper, the two-body strong decays of light $1^{-+}$ hybrids (denoted by $h$ here) are considered in the $N_f=2$ QCD formalism. The two-body final states can be an axial vector ($a_\mu$ for $J^{P(C)}=1^{+(+)}$ and $b_\mu$ for $1^{+(-)}$) and a pseudoscalar ($P$)), a vector ($V_\mu$) and a pseudoscalar ($P$) and two vectors ($V_\mu V_\nu$). The isospin symmetry and conservation of the charge conjugation ($\mathcal{C}$) impose strong constraints on the form of the effective interaction Lagrangian.

Let $C'(A)$ be the $\mathcal{C}$ transformation factor of $A$. For the decay process $h\to AB$ with $C'(A)C'(B)=-$, such as $\pi_1\to b_1\pi, \rho\pi$, the $\mathcal{C}$ conservation requires the effective Lagrangian responsible for the $\pi_1^0$ decay to be

\begin{eqnarray}
    \mathcal{L}_{\pi_1^0\to b_1 \pi}&=&m_{\pi_1}g_{\pi b_1}\pi_1^{0,\mu} \frac{1}{\sqrt{2}}\left(b_{1,\mu}^+\pi^- -b_{1,\mu}^-\pi^+\right)\nonumber\\
    \mathcal{L}_{\pi_1^0\to \rho\pi}&=&\frac{g_{\rho \pi}\epsilon^{\mu \nu \rho \sigma}}{\sqrt{2}m_{\pi_1}}  \left(\partial_{\mu} \pi_{1,\nu}^{0}\right)\left( \partial_{\rho} \rho_\sigma^{+} \pi^{-}-\partial_{\rho} \rho_{\sigma}^{-} \pi^{+}\right).\nonumber\\
\end{eqnarray}
where the constant factor $\frac{1}{\sqrt{2}}$ comes from the normalization of the isospin state $|II_3\rangle=|10\rangle$ of $AB$, namely, $\langle 10|10\rangle=1$.

Similarly, the effective Lagrangian for the decay processed $h\to AB$ with $C'(A)C'(B)=+$ reads 

\begin{eqnarray}\label{eq:Lagrangian-p}
    \mathcal{L}_{\pi_1^0\to f_1\pi }&=& m_{\pi_1}g_{f_1\pi} \pi_1^{0,\mu}f_{1,\mu}\pi^0\nonumber\\
    \mathcal{L}_{\pi_1^0\to a_1\eta}&=& m_{\pi_1}g_{a_1\eta} \pi_1^{0,\mu}a_{1,\mu}^0\eta\nonumber\\
    \mathcal{L}_{\pi_1^0 \to \pi\eta}&=&ig_{\pi\eta} \pi_1^{0,\mu}(\eta \overleftrightarrow{\partial}_{\mu} \pi^0)\nonumber\\
    \mathcal{L}_{\eta_1\to a_1\pi}&=& m_{\eta_1}g_{a_1\pi} \eta_1^{\mu}\frac{1}{\sqrt{3}}\left(a_{1,\mu}^+\pi^-+a_{1,\mu}^0\pi^0+a_{1,\mu}^-\pi^+\right)\nonumber\\
    \mathcal{L}_{\eta_1\to f_1\eta}&=& m_{\eta_1}g_{f_1\eta}\eta_1^{\mu}f_{1,\mu}\eta.
\end{eqnarray}
where $\overleftrightarrow{\partial}$ represents $\overleftarrow{\partial}-\overrightarrow{\partial}$.

The general expression of the effective Lagrangian for the decay mode $h\to VV'$ in the rest frame of $h$ reads %the Lorentz structure of $[hVV']$ in each term in Eq.~(\ref{eq:Lagrangian-p}) is
\begin{equation}\label{eq:hV1V2}
    \mathcal{L}_{h\to VV'}=h^\nu\left(
    g V^\mu\partial_\mu V'_\nu+g' V'^{\mu}\partial_\mu V_\nu+g_0 V_\mu\overleftrightarrow{\partial}_\nu V'^\mu
    \right),
\end{equation}
where three effective couplings $g,g',g_0$ are involved. $\eta_1$ can decay into (generalized) identical particle pairs $\rho\rho$ and $\omega\omega$. In this case one has $g=g'$ and $g_0=0$, and subsequently 
\begin{eqnarray}
    \mathcal{L}_{\eta_1\to\rho\rho}&=&\frac{g_{\rho\rho}}{\sqrt{3}}\eta_{1}^{\nu}(\rho^{\mu,+} \partial_{\mu} \rho_{\nu}^{-}+\rho^{\mu,0} \partial_{\mu} \rho_{\nu}^{0}+\rho^{\mu,-} \partial_{\mu} \rho_{\nu}^{+})\nonumber\\
    \mathcal{L}_{\eta_1\to\omega\omega}&=& g_{\omega\omega} \eta_{1}^{\nu}\omega^{\mu} \partial_{\mu} \omega_{\nu},
\end{eqnarray}
The relative $P$-wave ($L=1$) and the selection rule $L+S =2$ requires the total spin $S$ of the two vector meson is $S=1$ for two (generalized) identical vector mesons. One can see this from the desired structure of the decay amplitude for $\eta_1\to VV$ below.  

With the effective Lagrangian for each decay process $h \to AB$ and considering the polarization of (axial) vector mesons, the tree-level transition matrix element $M_{AB}^{(\lambda' \lambda'') \lambda}$ can be determined as follows:
\begin{equation}\label{eq:amplitude-sm}
    \begin{aligned}
        \mathcal{M}_{AP}^{\lambda'\lambda}=           & {g}_{AP} m_{h} \vec{\epsilon}_\lambda(\vec{0})\cdot \vec{{\epsilon}}^{~*}_{\lambda'}(\vec{k}),        \\
        \mathcal{M}_{PP}^\lambda=                     & 2{g}_{PP} \vec{{\epsilon}}_\lambda(\vec{0})\cdot \vec{k},                                             \\
        \mathcal{M}_{VP}^{\lambda'\lambda}=           & {g}_{VP} \vec{\epsilon}_\lambda(\vec{0})\cdot(\vec{\epsilon}_{\lambda'}^{~*}(\vec{k})\times \vec{k}), \\
        \mathcal{M}_{V V}^{\lambda''\lambda'\lambda}= &
        2g_{VV}\vec{\epsilon}_\lambda(\vec{0})\cdot\left( \vec{k}\times \left[\vec{\epsilon}^*_{\lambda'}(\vec{k})\times \vec{\epsilon}_{\lambda''}^*(-\vec{k})\right]\right)
    \end{aligned}
\end{equation}
where $\epsilon^\mu_\lambda(\vec{0})$,  $\epsilon_{\lambda'}(\vec{k})$ and $\epsilon_{\lambda''}(-\vec{k})$ are the polarization vectors
of $h$, $A$ (if (an axial) vector) and $B$ (if a (an axial) vector). Note that for a given kinetic configuration $A(\vec{k})B(-\vec{k})$, we use the normalized isospin wave function of the final state $|AB\rangle$ that has the same isospin quantum numbers as those of $h$. For example

\begin{eqnarray}\label{eq:isospin}
    |\rho\rho(I=0,I_3=0)\rangle&=&\frac{1}{\sqrt{3}}\left(|\rho^+\rho^-\rangle+|\rho^0\rho^0\rangle+|\rho^-\rho^+\rangle\right)\nonumber\\
    |b_1\pi(I=1,I_3=0)\rangle&=&\frac{1}{\sqrt{2}}\left(|b_1^+\pi\rangle-|b_1^-\pi^+\rangle\right).
\end{eqnarray}

%On the other hand,  the invariant amplitude $\mathcal{M}_{AB}$ is derived based on the relativistic state normalizations $\langle h|h\rangle=2m_h L^3$ and $\langle A(\vec{k})B(-\vec{k})|A(\vec{k})B(-\vec{k})\rangle=4L^6 E_A(k)E_B(k)$, so the connection between $\mathcal{M}$ and the transition amplitude $x$ in %Eq.~(\ref{eq:H}) reads
%\begin{equation}\label{eq:match}
%    x_{AB} =\frac{\mathcal{M}_{AB}}{(8L^3 m_h E_A(k)E_B(k))^{1/2}}.
%\end{equation}
%if all the particles involved are (pseudo-)scalar particles.

So, after $x_{AB}$ is obtained through Eq.~(\ref{eq:ratio1}), one can use Eq.~(\ref{eq:amplitude-sm}) and (\ref{eq:match}) to determine the effective coupling $g_{AB}$, from which the decay width is calculated as follows:
\begin{equation}\label{eq:decay-wid}
    \Gamma(h\to AB) = \frac{c}{8\pi} \frac{k_\mathrm{ex}}{m_h^2} \overline{|\mathcal{M}(h \to AB)|^2},
\end{equation}
where $c$ takes the value $c=1$ when $A$ and $B$ are different particles and $c=\frac{1}{2}$ when $A$ and $B$ are (generalized) identical particles (such as the $\rho\rho$ final state mode), and $k_\mathrm{ex}$ is the decay momentum,
\begin{eqnarray}\label{eq:decay-mom}
    k_\mathrm{ex}&=&\frac{1}{2m_h}\left(m_h^4+m_A^4+m_B^4\right.\nonumber\\
    &-& \left. 2m_h^2 m_A^2-2m_h^2 m_B^2-2m_A^2 m_B^2\right)^{1/2},
\end{eqnarray}
and $\overline{|\mathcal{M}(h \to AB)|^2}$ is the polarization-averaged transition amplitude at the tree level and is dictated by $g_{AB}$. The
explicit expressions are
\begin{equation}\label{eq:decay-amp}
    \begin{aligned}
        \overline{| \mathcal{M}(h\to AP)|^2} = & \frac{1}{3} {g}_{AP}^2 m_{h}^2 (3+\frac{{k}_\text{ex}^2}{m_A^2}), \\
        \overline{| \mathcal{M}(h\to PP)|^2} = & \frac{4}{3} {g}_{PP}^2 {k}_\text{ex}^2,                           \\
        \overline{| \mathcal{M}(h\to VP)|^2} = & \frac{2}{3} {g}_{VP}^2 {k}_\text{ex}^2,                           \\
        \overline{| \mathcal{M}(h\to VV)|^2} = & \frac{4}{3} g_{VV}^2 {k}_\text{ex}^2 \frac{m_{h}^2}{m_{V}^2}.
    \end{aligned}
\end{equation}

\section{Numerical details}\label{sec:numerical}
\subsection{Gauge ensemble}
The calculations in this work are performed on $N_f =2$ gauge ensembles generated using an anisotropic action with an aspect ratio $\xi \approx 5.0$. The lattice size is set to be $L^3 \times T=16^3 \times 128$ and the lattice spacing $a_s$ and pion mass are determined to be 0.1361 fm and 417 MeV, respectively \cite{Li:2024pfg}. The parameters of the gauge ensembles and perambulators are listed in Table~\ref{tab:config}. Since the two-body strong decays of a hybrid meson is governed by the gluon-$q\bar{q}$ transition, and there are quite a few isoscalar mesons involved in the decay processes, the quark annihilation diagrams need to be tackled. In doing so, we adopt the distillation method \cite{Peardon:2009gh} which facilitates a systematic treatment of the all-to-all quark propagators and smeared quark interpolation operators. On each timeslice of each configuration, we calculate $N_V=70$ eigenvectors of the gauge covariant Laplacian with the lowest eigenvalue $\{V_i(\vec{x},t),i=1,2,\ldots, N_V\}$ on the lattice, which span a Laplacian Heaviside subspace (LHS). The perambulators of light $u,d$ quarks, which encapsulate the all-to-all quark propagators, are calculated in the LHS. The $N_V$ eigenvectors also provide a LHS smearing scheme for the quark field, namely, $\psi^{(s)}(\vec{x},t)=\sum_i V_i(\vec{x},t)V_i^*(\vec{y},t)\psi(\vec{y},t)$, where $\psi^{(s)}$ is the LHS smeared quark field. Throughout this work, meson operators are built in terms of $u^{(s)}$ and $d^{(s)}$ fields and the superscripts are omitted for convenience in the following discussions.

\begin{table}[t]
    \renewcommand\arraystretch{1.5}
    \caption{Parameters of the gauge ensembles. $N_V$ is the number of the eigenvectors that span the Laplacian Heaviside subspace.~\cite{Peardon:2009gh}.}
    \label{tab:config}
    \begin{ruledtabular}
        \begin{tabular}{lcccccc}
            IE  & $N_s^3\times N_t$ & $a_t^{-1}$(GeV) & $\xi$  & $m_\pi$(MeV) & $N_V$ & $N_\mathrm{cfg}$ \\\hline
            L16M415 & $16^3 \times 128$ & $ 7.219$        & $5.0$ & $417$        & 70    & 400            \\
        \end{tabular}
    \end{ruledtabular}
\end{table}

\begin{table*}[t]
    \caption{Information for all particles involved. The upper indices of the notation of irreducible representations (irep) of $O_{h}$ denote the charge conjugate factor, while the lower indices $u(g)$ denote the parity. The notation of operators follows Ref. \cite{Dudek:2007wv}. The masses of $\pi_{1}$ and $\eta_1$ are taken from Ref. \cite{JPAC:2018zyd} and Ref. \cite{BESIII:2022riz} respectively. Other experimental masses are taken from the PDG \cite{Workman:2022ynf}.}
    \begin{ruledtabular}
        \label{tab:operators}
        \begin{tabular}{lccccc}
            $I^GJ^{PC}$   & Particle   & Irep         & Operator                        &  $m^{\mathrm{lat}}$ (GeV) (this work)      & $m^{\mathrm{exp}}$(GeV)~\cite{Workman:2022ynf}\\\hline
            $1^+0^{-(+)}$    & $\pi$      & $A_{1u}^{+}$ & $\gamma_5$                 & 0.4176(13)       & 0.135     \\
            $0^-0^{-+}$      & $\eta(\gamma_5)$     & $A^{+}_{1u}$ & $\gamma_5$            & 0.731(34)      & 0.958    \\
            $1^+1^{--}$   & $\rho$     & $T^{-}_{1u}$ & $\gamma_i$                & 0.8461(42)       & 0.775      \\
            $1^-1^{-+}$   & $\pi_1$    & $T_{1u}^{+}$ & $\mathbf{\rho}\times\mathbf{B}$                & 1.980(21)      & 1.661            \\
            $0^+1^{-+}$   & $\eta_1$   & $T_{1u}^{+}$ & $\mathbf{\rho}\times\mathbf{B}$                & 2.253(54)      & 1.855            \\
            $1^-1^{+(+)}$ & $a_1$      & $T_{1g}^{+}$ & $\gamma_5 \gamma_i$        & 1.300(12)      & 1.230        \\
            $0^+1^{++}$   & $f_1$      & $T^{+}_{1g}$ & $\gamma_5\gamma_i$         & 1.516(14)      & 1.282        \\
            $1^+1^{+(-)}$ & $b_1$      & $T_{1g}^{-}$ & $\gamma_5\gamma_i\gamma_4$ & 1.340(19)      & 1.230      \\
        \end{tabular}
    \end{ruledtabular}
\end{table*}

\begin{figure}[t]
    \centering
    \includegraphics[width=0.8\linewidth]{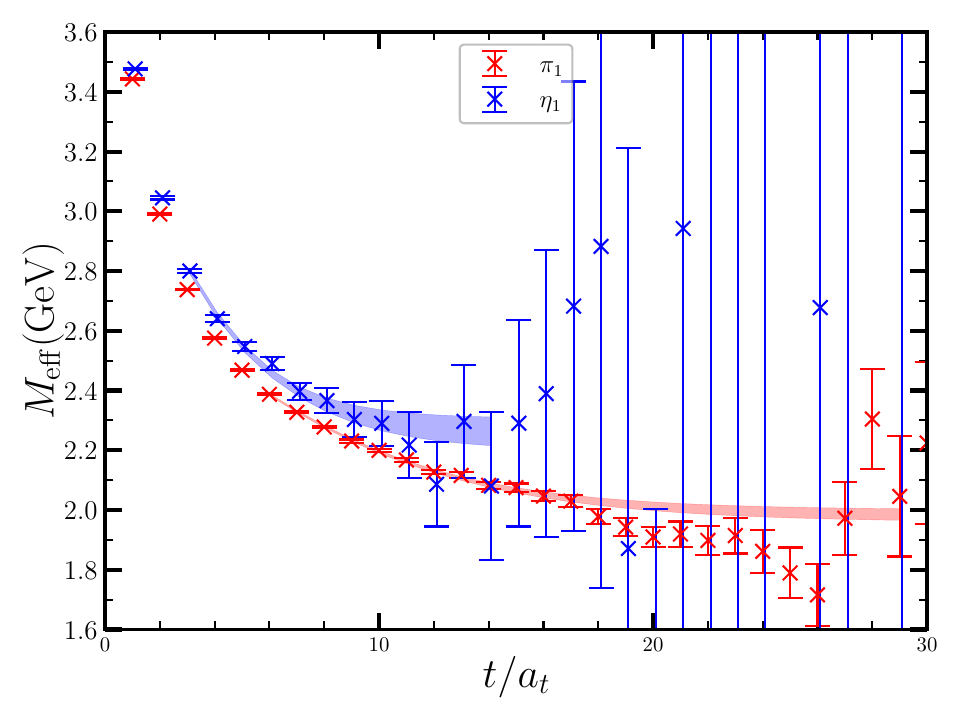}
    \includegraphics[width=0.8\linewidth]{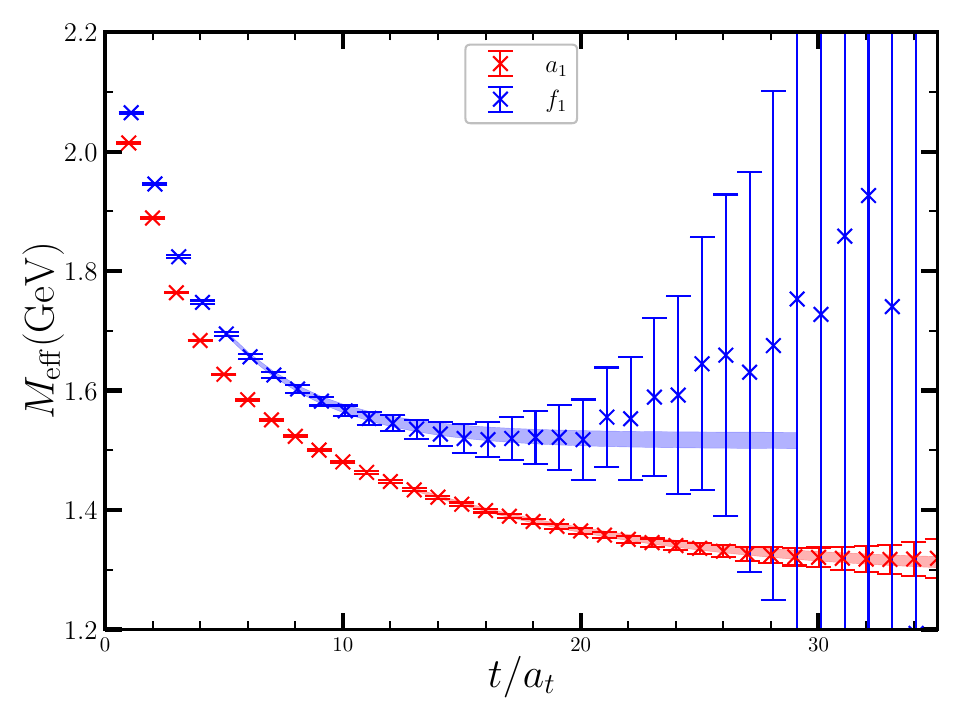}
    \includegraphics[width=0.8\linewidth]{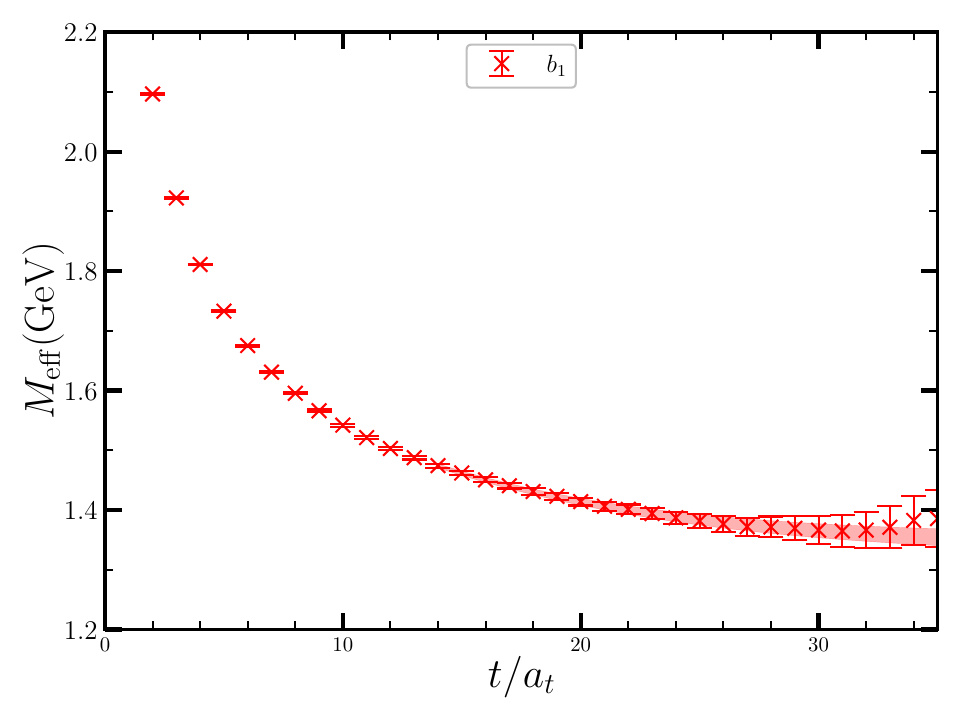}
    \caption{Effective masses $M_\mathrm{eff}(t)$ of some mesons. $M_\mathrm{eff}(t)$ is defined through $M_\mathrm{eff}(t) = a_t^{-1} \ln \frac{C_{XX}(t)}{C_{XX}(t + a_t)}$, where $C_{XX}(t)$ is the correlation function of the particle $X = \pi_1, \eta_1, a_1, f_1, b_1$. The colored bands illustrate the fit results and the fit ranges through two-state function forms. The large errors of the signals of isoscalars ($\eta_1$ and $f_1$) are due to the inclusion of disconnected diagrams in the calculation of corresponding correlation functions $C_{XX}(t)$.
}
    \label{fig:effmass}
\end{figure}

\subsection{Meson Operators and light hadron spectrum}
For mesons with the conventional quantum numbers $I^G J^{PC}$, the lattice operators are quark bilinears $\bar{\psi_1}\Gamma\psi_2$, with the quantum numbers reflected by the quark flavors $\psi_i$ and the gamma matrix $\Gamma$. For the light hybrids $\pi_1$ and $\eta_1$ of $1^-1^{-+}$ and $0^+1^{-+}$ quantum numbers, respectively, we use the quark bilinear operators $\bar{\psi}_1\vec{\gamma}\times \vec{B} \psi_2$ (denoted by $\rho\times\vec{B}$ in Table~\ref{tab:operators}) with $\vec{B}$ being the chromomagnetic field strength and defined through the lattice covariant derivatives, namely, $\vec{B}\sim \vec{D}\times\vec{D}$~\cite{Dudek:2007wv}. The hybrid operators have the quantum numbers $\Lambda^{P(C)}=T_1^{-(+)}$ on the lattice. The specific operators for the particles involved are listed in Table~\ref{tab:operators}. 

Particles with non-zero momentum can be projected to different irreducible representation of the little group. For the $f_1$ and $a_1$ meson in flight with an on-axis momentum orientation, the operators in the $A_1$ representation (with longitudinal polarization) of the little group $\mathrm{C}_{4v}$ are not taken into account to avoid the mixing from a $0^{-+}$ state, and the $b_1$ in the $E$ representation (with transverse polarization) of $\mathrm{C}_{4v}$ is also excluded to avoid the mixing from the $1^{--}$ states. 

From the correlation functions, we obtain the masses of mesons involved in this study, as shown in Table~\ref{tab:operators}. The mass of the isoscalar pseudoscalar $\eta$, $m_\eta=731(34)~\mathrm{MeV}$, is consistent with previous lattice results with a similar lattice setup~\cite{Shi:2024fyv,Jiang:2022gnd,Jiang:2022ffl} (note that $\eta(N_f=2)$ is different from $\eta$ and $\eta'$ in the physical $N_f=2+1$ case). The masses of the light vector and axial vector mesons
\begin{eqnarray}
    m_\rho &=& 0.846(4)~\mathrm{GeV}\nonumber\\
    m_{a_1}&=&1.300(12)~\mathrm{GeV}\nonumber\\
    m_{f_1}&=&1.516(14)~\mathrm{GeV}\nonumber\\
    m_{b_1}&=&1.340(19)~\mathrm{GeV},
\end{eqnarray}
are also consistent with previous lattice results~\cite{Dudek:2013yja} but a little higher than the physical masses possibly owing to the higher pion mass in this study compared to the physical one. On our lattice, the $\rho$ meson is stable since it decays into $P$-wave $\pi\pi$ states whose minimum energy is higher than $m_\rho$. The $f_1 \to 4\pi$ and $a_1 \to 3\pi$ decays are not open either, so $f_1$ and $a_1$ can also be considered stable. The $b_1$ lies a little higher than the $\omega\pi$ threshold ($m_\omega \approx m_\rho$) and therefore is unstable on our lattice. Its resonance properties may introduce some systematic uncertainties when taken as a stable particle. We tentatively ignore this uncertainty in the present study. The hybrid meson masses are
\begin{eqnarray}
    m_{\pi_1}&=&1.977(36)~\mathrm{GeV},\nonumber\\
    m_{\eta_1}&=&2.275(48)~\mathrm{GeV},
\end{eqnarray}
which are also compatible with previous lattice results with similar lattice setups~\cite{Dudek:2013yja,Chen:2022isv}. Figure~\ref{fig:effmass} shows the effective masses defined through $M_\mathrm{eff}(t)=a_t^{-1} \ln \frac{C_{XX}(t)}{C_{XX}(t+a_t)}$ with $C_{XX}(t)$ being the correlation function of the particle $X= \pi_1, \eta_1, a_1, f_1, b_1$, where the colored bands illustrate the fit results through two-state function forms. It is seen that the mass splittings of $\pi_1-\eta_1$ and $a_1-f_1$ are large and signal the importance of the inclusion of the disconnected diagrams in the calculation of $C_{\eta_1\eta_1}(t)$ and $C_{f_1f_1}(t)$. The meson masses involved in this study are collected in Table~\ref{tab:operators} and are compared with the physical mass values~\cite{Workman:2022ynf}.

When considering the two-body decays of hybrids $h\to AB$, the two-particle operator for $AB$ is required. 
We use the partial-wave method to construct the interpolating meson-meson (labeled as $A$ and $B$) operators for the specific quantum numbers $J^P$~\cite{Feng:2010es, Wallace:2015pxa, Prelovsek:2016iyo} if all the corresponding irreps of the little group are selected.
In general, let $\mathcal{O}_{X}^{M_{X}}(\vec{k})$ be the operator for the particle $X=A$ or $B$ with spin $S_X$ and spin projection $M_{X}$ in the $z$-direction, then, for the total angular momentum $J$ and the $z$-axis projection $M$, the relative orbital angular momentum $L$, and the total spin $S$, the explicit construction of the $AB$ operator is expressed as
\begin{widetext}
    \begin{equation}\label{eq:two-meson}
        \begin{aligned}
            \mathcal{O}_{AB;JLSP}^{M}(\hat{k}) = & \sum\limits_{M_L, M_S, M_{A}, M_{B}} \langle L, M_L; S,M_S| JM \rangle
            \langle S_AM_{A}; S_BM_{B}| S, M_S \rangle                                                                                                                                   \\
                                                 & \times \sum\limits_{R\in O_h} Y^*_{LM_L}(R\circ\vec{k}) \mathcal{O}_A^{M_{A}}(R\circ \vec{k}) \mathcal{O}_B^{M_{B}}(-R\circ \vec{k})%/\textcolor{red}{\sqrt{\sum\limits_{R \in O_h}|Y^*_{LM_L}(R\circ\vec{k})|^2}}
                                                 ,
        \end{aligned}
    \end{equation}
\end{widetext}
where $\hat{k}=(n_1,n_2,n_3)$ is the momentum mode of $\vec{k}=\frac{2\pi}{La_s}\hat{k}$ with $n_1\le n_2\le n_3$ by convention, $R\circ\vec{k}$ is the spatial momentum rotated from $\vec{k}$ by $R\in O_h$ with $O_h$ being the lattice symmetry group, $|S,M_S\rangle$ is the total spin state of the two particles involved, $|LM_L\rangle$ is the relative orbital angular momentum state, $|JM\rangle$ is the total angular momentum state, and $ Y^*_{LM_L}(R\circ\vec{k})$ is the spherical harmonic function of the direction of $R\circ \vec{k}$. The precise expressions of $\mathcal{O}_{AB;JLSP}^{M=0}(\hat{k})$ for specific $AB$ and specific momentum modes $\hat{k}$ are given in the Appdendix, where 
one can see that, each term of $\mathcal{O}_{AB;JLSP}^{M=0}(\hat{k})$ has a definite operator combination
\begin{equation}\label{eq:oper3}
    [\mathcal{O}_A^{(i)}(\vec{k})\mathcal{O}_B^{(j)}(-\vec{k})]\equiv \mathcal{O}_{AB}^3(\vec{k})
\end{equation}
for a specifically rotated momentum $\vec{k}$ of the $\hat{k}$ mode, with the superscript of $\mathcal{O}_{A(B)}$ being void for $A(B)$ to be a pseudoscalar or taking the values $i=1,2,3$ for $A(B)$ to be a (an axial) vector. 

\begin{figure}[t]
    \centering
    \includegraphics[width=\linewidth]{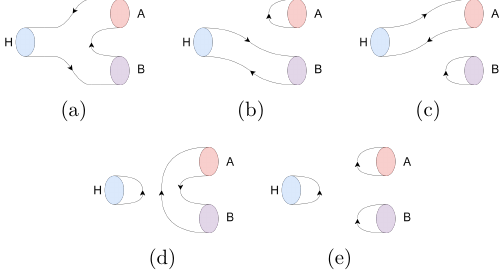}
    % \vspace{-0.10in}
    \caption{Schematic diagrams for $C_{AB,h}^{33}(\vec{k},t)$ after the Wick's contraction. The lines with arrows represent the quark propagators, and the colored ellipses stand for the operator structures (listed in Table~\ref{tab:operators}) of individual mesons. The type (a) diagram are universal for all the correlation functions $C_{AB,h}^{33}(\vec{k},t)$. Diagrams of type (b), (c) and (d) are additional ones if $h$, $A$, $B$ are isoscalars (flavor singlets), respectively. The type (e) operator also contributes to $C_{AB,h}^{33}(\vec{k},t)$ if $h$, $A$ and $B$ are all isoscalars (flavor singlets).}

    \label{fig:schamatic-diagrams}
\end{figure}

\subsection{Ratio function $R_{AB}(\vec{k},t)$}
In practice, we calculate the two point functions
\begin{eqnarray}
    C^{33}_{AB,h}(\vec{k},t)&=&\langle 0|\mathcal{O}_{AB}^{3}(\vec{k},t)\mathcal{O}^{3,\dagger}_h(\vec{0},0)|0\rangle\nonumber\\
    C^{(33)}_{hh}(\vec{k},t)&=&\langle 0|\mathcal{O}_{h}^{3}(\vec{0},t)\mathcal{O}_h^{3,\dagger}(\vec{0},0)|0\rangle\nonumber\\
    C^{(ii)}_{AA}(\vec{k},t)&=&\langle 0|\mathcal{O}_{A}^{(i)}(\vec{k},t)\mathcal{O}_A^{(i),\dagger}(\vec{k},0)|0\rangle\nonumber\\
    C^{(ii)}_{BB}(\vec{k},t)&=&\langle 0|\mathcal{O}_{B}^{(i)}(\vec{k},t)\mathcal{O}_B^{(i),\dagger}(\vec{k},0)|0\rangle, 
\end{eqnarray}
from which we define the ratio function
\begin{equation}
    R_{AB}(\vec{k},t)=\frac{C_{AB,h}^{33}(\vec{k},t)}{\sqrt{C_{hh}^{(33)}(\vec{0},t)C_{AA}^{(ii)}(\vec{k},t)C_{BB}^{(jj)}(-\vec{k},t)}}. 
\end{equation}

With polarization involved, we have:
\begin{eqnarray}
\langle 0|\mathcal{O}_A^{i}(\vec{k})|V,\vec{k},\lambda\rangle&\equiv& Z_A(\vec{k})\epsilon^i_\lambda(\vec{k})
\end{eqnarray}
if $A$ is a (an axial) vector. Therefore, when $R_{AB}$ is parameterized as:
\begin{equation}\label{eq:r-para}
    R_{AB}(\vec{k},t)\approx a_0 + r_{AB} t + a_2 t^2
\end{equation}
the transition matrix element can be calculated as:
\begin{equation}\label{eq:rab}
    r_{AB}=\sum\limits_{\lambda,\lambda'} \frac{\mathcal{M}_{AB}^{(\lambda'\lambda'')\lambda} [\epsilon^{(i)}_{(\lambda')}(\vec{k})\epsilon^{(j)}_{(\lambda'')}(\vec{k})]\epsilon^{3*}_\lambda(\vec{0})} {\sqrt{(8L^3 m_h E_A(k)E_B(k))\mathcal{P}_A^{(ii)}(\vec{k})\mathcal{P}_B^{(jj)}(-\vec{k})}},
\end{equation}
where the indices $i,j$ are those in Eq.~(\ref{eq:oper3}), the polarization vector $\epsilon^{(i)}_{(\lambda)}(\vec{k})$ is replaced by one for a pseudoscalar $A$ or $B$, $\mathcal{P}_A^{(ii)}(\vec{k})$ takes a value of unity for a pseudoscalar $A$ and $1+k^ik^i/m_A^2$ for a (an axial) vector $A$. 
%%%%%%%%%%%%%%%%%%%%%%%%%%%%%%%%%%%%%%%%%%%%%%%%%%%%%%%%%%%%%%%%%%%%%%%%%%%%%%%%%%%%%%%%%%%%%%%%%%%%%%%%%%%%%%%%%%%
\begin{figure}[t]
%    \centering
    \includegraphics[width=\linewidth]{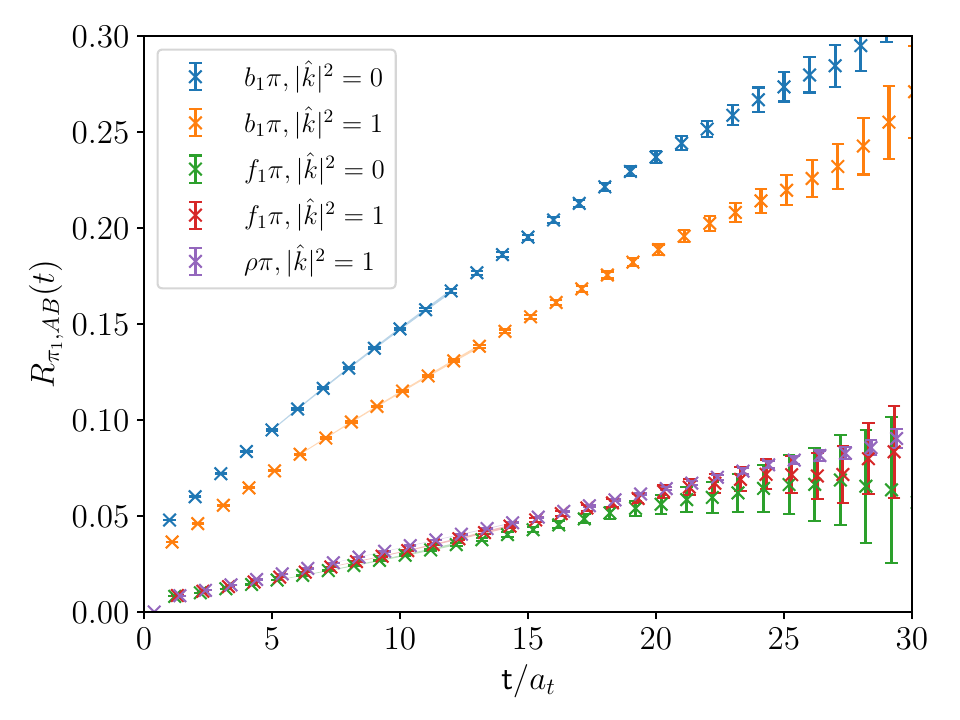}\\
    \caption{The ratio function $R_{AB}(t)$ for $\pi_1 \to AB$ with momentum mode $\hat{k}$. The shaded bands illustrate our fit for $R_{AB}(t)$ using the polynomial function $R_{AB}(t) = a_0 + r_{AB} t + a_2 t^2$.}
    \label{fig:fit-pi1}
\end{figure}
%%%%%%%%%%%%%%%%%%%%%%%%%%%%%%%%%%%%%%%%%%%%%%%%%%%%%%%%%%%%%%%%%%%%%%%%%%%%%%%%%%%%%%%%%%%%%%%%%%%%%%%%%%%%%%%%%%%
\begin{table*}[t]
%    \centering
    \caption{Fit results of the ratio function $R_{AB}(\hat{k}, t)$. The parameters are those involved in the polynomial function form $R_{AB}(t) = a_0 + r_{AB} t + a_2 t^2$. The energy difference $\Delta = \sqrt{(E_h - E_{AB})^2 + 4x^2}$ is also shown for each decay channel. The `contraction' column shows the quark diagrams (illustrated in Fig.~\ref{fig:schamatic-diagrams}) that contribute to the correlation function $C_{AB,h}^{33}(\vec{k}, t)$. The fit ranges $[t_\mathrm{min}, t_\mathrm{max}]$ and the values of $\chi^2/\mathrm{d.o.f}$ are also given for the final fit results.}
    \begin{ruledtabular}
        \begin{tabular}{ccccccccc}
    %            \hline
    Decay modes                  & mode $\hat{k}$ & contractions & $a_t\Delta(\times 10^{-3})$ & $a_0(\times 10^{-3})$ & $a_t r_{AB}(\times 10^{-3})$ & $a_t^2 a_2(\times 10^{-4})$ & fit range & $\chi^2 / \mathrm{d.o.f}$ \\ \hline
    $\pi_1\rightarrow b_1\pi$    & (0,0,0)        & (a)          & $30.5\pm 4.3$               & 36.91(39)             & 12.15(14)                    & -1.06(14)                   & [5,13]    & 0.87                      \\
    $\pi_1\rightarrow b_1\pi$    & (0,0,1)        & (a)          & $-25.5\pm 4.3$              & 28.23(52)             & 9.44(22)                     & -0.73(12)                   & [6,14]    & 0.95                      \\
    $\pi_1\rightarrow f_1\pi$    & (0,0,0)        & (a,b)        & $12.1\pm 9.3$               & 5.82(30)              & 1.95(11)                     & 0.41(12)                    & [5,13]    & 0.38                      \\
    $\pi_1\rightarrow f_1\pi$    & (0,0,1)        & (a,b)        & $-42.6\pm 8.9$              & 7.25(85)              & 1.92(23)                     & 0.56(15)                    & [7,15]    & 0.47                      \\
    $\pi_1\rightarrow \rho\pi$   & (0,0,1)        & (a)          & $35.2\pm 3.3$               & 5.17(26)              & 2.895(71)                    & 0.030(52)                   & [7,15]    & 1.3                       \\
    \hline
    $\eta_1\rightarrow a_1\pi$   & (0,0,0)        & (a,d)        & $51\pm 28$                  & 11.28(83)             & 2.68(29)                     & 1.06(26)                    & [5,13]    & 1.09                      \\
    $\eta_1\rightarrow a_1\pi$   & (0,0,1)        & (a,d)        & $-14\pm 28$                 & 11.64(95)             & 3.30(33)                     & 0.63(29)                    & [5,13]    & 0.89                      \\
    $\eta_1\rightarrow f_1\eta$  & (0,0,0)        & (a,b,c,d,e)  & $-20\pm 29$                 & 7.53(86)              & 4.06(33)                     & -0.98(32)                   & [4,12]    & 0.43                      \\
    $\eta_1\rightarrow f_1\eta$  & (0,0,1)        & (a,b,c,d,e)  & $-14\pm 28$                 & 1.6(1.8)              & 4.81(52)                     & -1.35(39)                   & [6,14]    & 0.38                      \\
    $\eta_1\rightarrow \rho\rho$ & (0,0,1)        & (a,d)        & $3\pm 28$                   & 8.85(78)              & 3.50(27)                     & -0.39(23)                   & [5,13]    & 0.75                      \\
    $\eta_1\rightarrow \rho\rho$ & (0,1,1)        & (a,d)        & $-38\pm 28$                 & 10.06(91)             & 3.77(32)                     & -0.41(28)                   & [5,13]    & 0.46                      \\
\end{tabular}
    \end{ruledtabular}
    \label{tab:fit}

\end{table*}
\begin{figure*}[t]
\includegraphics[width=0.45\linewidth]{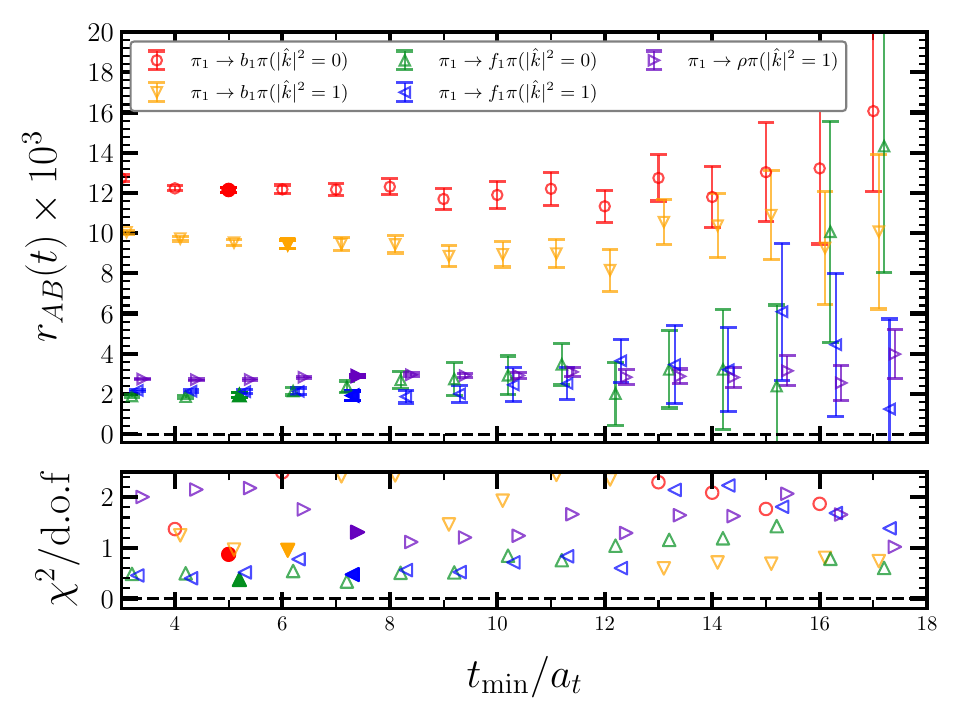}
\includegraphics[width=0.45\linewidth]{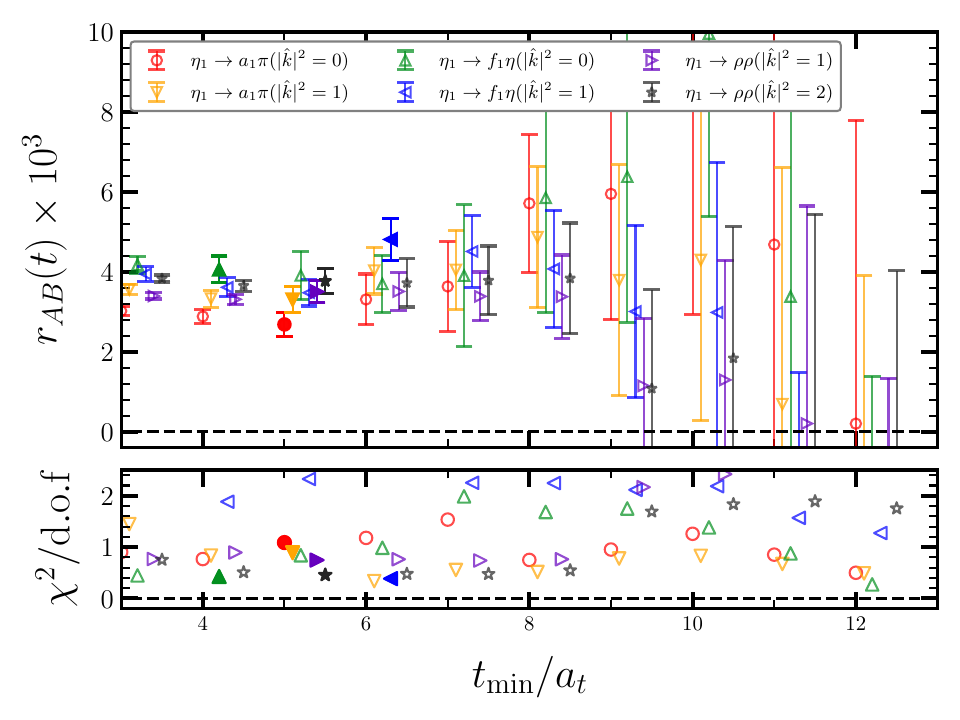}
\caption{Check of the fit stability of $R^{33}_{AB}(t)$ using Eq.~(\ref{eq:r-para}). The left two panels are for $\pi_1$ decays, and the right ones are for $\eta_1$ decays. In the upper two panels, the data points are the fitted results of $r_{AB}$ in time intervals $[t_{\min}/a_t, t_{\min}/a_t + 8]$}, and the horizontal axis represents the different values of $t_{\min}$. The lower two panels show the values of $\chi^2/\mathrm{d.o.f}$ for each fit. The solid points represent the final values of $r_{AB}$ adopted to extract the corresponding effective couplings.
\label{fig:fit-stability}
\end{figure*}
It is easy to see that each term in the two particle helicity operator $\mathcal{O}_{AB;JLSP}^{M=0}(\hat{k})$ gives the same $r_{AB}$.
So we average over all the terms in $\mathcal{O}_{AB;JLSP}^{M=0}(\hat{k})$ to increase the statistics. Note that the flavor structure of $\mathcal{O}_{AB}$ are properly normalized according to the flavor wave function similar to Eq.~(\ref{eq:isospin}) in the calculation of $C_{AB,h}^{33}(\vec{k},t)$. 

According to the Wick's contraction, there are five types of quark diagrams involved in the calculation of $C_{AB,h}^{33}(\vec{k},t)$, as shown in Fig.~\ref{fig:schamatic-diagrams}. In each diagram, the filled lines with arrows represent the quark propagators (actually quark perambulators in the formalism of the distillation method), and the colored ellipses stand for the operator structures (listed in Table~\ref{tab:operators}) of individual mesons. The type (a) diagram is universal for all the correlation functions $C_{AB,h}^{33}(\vec{k},t)$. Diagrams of type (b), (c), and (d) are additional ones if $h$, $A$, or $B$ are isoscalars (flavor singlets), respectively. The type (e) diagram also contributes to $C_{AB,h}^{33}(\vec{k},t)$ if $h$, $A$, and $B$ are all isoscalars (flavor singlets). The quark diagrams involved in an individual $C_{AB,h}^{33}(\vec{k},t)$ are shown in Table~\ref{tab:fit}.

For $\pi_1$ decays, we consider the decay modes $AB=b_1\pi, f_1\pi, \rho\pi$. The final states $b_1\pi$ and $f_1\pi$ are in relative $S$-wave, so we calculate $R_{AB}(\vec{k},t)$ at the relative momentum modes $\hat{k}=(0,0,0)$ and $(0,0,1)$ for a self-consistent check of the derived effective coupling $g_{AB}$. The final state $\rho\pi$ is in the relative $P$-wave, and we calculate $R_{AB}(\vec{k},t)$ at $\hat{k}=(0,0,1)$, which has $E_A(\hat{k})+E_B(\hat{k})$ very close to $m_{\pi_1}$ on our lattice. The ratio functions $R_{AB}(\vec{k},t)$ for these $AB$ modes are plotted in Fig.~\ref{fig:fit-pi1} as data points. The polynomial fits using Eq.~(\ref{eq:r-para}) are also illustrated by the color bands. It can be seen that the functional form describes the data well for all the $AB$ modes considered.

The fit stability of $R_{AB}(\vec{k},t)$ is also checked by varying the fit $t$ window. In doing so, we fix the length of the fit window to be 10 and conduct the fit in the time range $t\in [t_\text{min},t_\text{min}+10]$ by varying $t_\text{min}$ from 5 to 25. The the values of $r_{AB}$ and $\chi^2/\text{d.o.f}$ values of the fits are illustrated in Fig.~\ref{fig:fit-stability}, where the left panels are the results for $\pi_1$ decays and the right ones are for the $\eta_1$ decays. Obviously, the central values of $r_1$ for all the decay modes are stable when $t_\text{min}>4$, while the errors increase with the increasing of $t_\text{min}$. The $\chi^2/\text{d.o.f}$ values are acceptable for all the fits and manifest the feasibility of the function form in Eq.~(\ref{eq:r-para}). We take the fitted values of $r_1$ in the time ranges $[t_\text{min},t_\text{max}]$ that have relatively small $\chi^2/\mathrm{d.o.f}$ (listed in Table~\ref{tab:fit}) as our final results. 
 The fitted results of the parameters $a_0$, $r_{AB}$, and $a_2$ for all the $AB$ modes are collected in Table~\ref{tab:fit} along with the corresponding fit windows $[t_\mathrm{min},t_\mathrm{max}]$ and the $\chi^2/\mathrm{d.o.f}$.  

\begin{figure}[t]
    \centering
    \includegraphics[width=\linewidth]{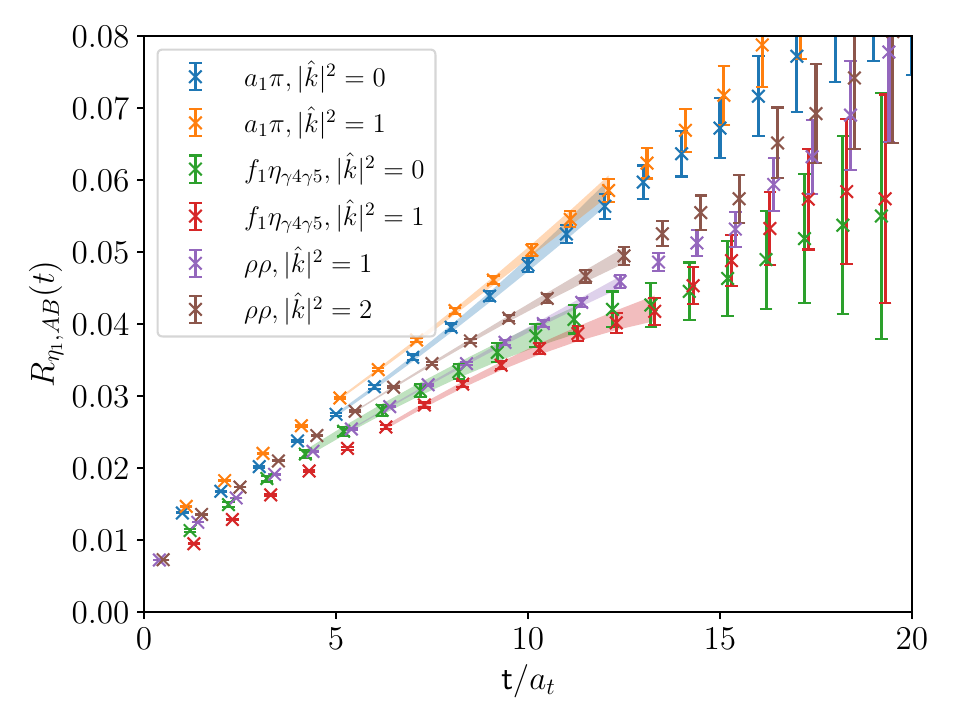}
    \caption{The ratio function $R_{AB}(t)$ for $\eta_1 \to AB$ with momentum mode $\hat{k}$. The shaded bands illustrate our fit for $R_{AB}(t)$ using the polynomial function $R_{AB}(t) = a_0 + r_{AB} t + a_2 t^2$.}
    \label{fig:fit-eta1}
\end{figure}

For $\eta_1$ decays, we consider the modes $AB=a_1\pi, f_1\eta, \rho\rho$. Similar to that of $\pi_1$ decays, we calculate $R_{AB}(\vec{k},t)$ at relative momentum modes $\hat{k}=(0,0,0),(0,0,1)$ for the $S$-wave $a_1\pi$ and $f_1\eta$ decays, and $\hat{k}=(0,0,1),(0,1,1)$ for the $P$-wave $\rho\rho$ decay. Figure~\ref{fig:fit-eta1} shows the ratio functions $R_{AB}(\vec{k},t)$ for these $AB$ modes, where the lattice results are indicated by data points and the polynomial fits using Eq.~(\ref{eq:r-para}) are illustrated by colored bands. The statistical errors in this case are larger than those for $\pi_1$ decay modes, since multiple disconnected quark diagrams contribute when the isoscalar $\eta_1$, $\eta$, and $f_1$ mesons are involved. The fitted results of the parameters $a_0$, $r_{AB}$, and $a_2$ for all the $AB$ modes are also listed in Table~\ref{tab:fit} along with the corresponding fit windows $[t_\mathrm{min},t_\mathrm{max}]$ and the $\chi^2/\mathrm{d.o.f}$.

%%%%%%%%%%%%%%%%%%%%%%%%%%%%%%%%%%%%%%%%%%%%%%%%%%%%%%%%%%%%%%%%%%%%%%%%%%%%%%%%%%%%%%%%%%%%%%%%%%%%%%%%%%%%%%%%%%%%%%%%%%%%%%%
%%%%%%%%%%%%%%%%%%%%%%%%%%%%%%%%%%%%%%%%%%%%%%%%%%%%%%%%%%%%%%%%%%%%%%%%%%%%%%%%%%%%%%%%%%%%%%%%%%%%%%%%%%%%%%%%%%%%%%%%%%%%%%%
After $r_{AB}$ is determined from the slope of $R_{AB}(\vec{k},t)$ using Eq.~(\ref{eq:r-para}), one can derive the effective coupling $g_{AB}$ from $r_{AB}$ by combining Eq.~(\ref{eq:decay-amp}) and (\ref{eq:rab}). As expressed in Eq.~(\ref{eq:x-error}) in Sec.~\ref{sec:formalism}, when the transition matrix element $x_{AB}^{\lambda'\lambda}$ is determined from $r_{AB}$, the systematic uncertainty due to the deviation from the assumption $\langle 0|\mathcal{O}_a|b\rangle=Z_a^b \delta_{ab}$ must be considered and can be estimated by $\delta x \sim |a_0\Delta|$. The values of $a_t \Delta$ for all the channels are also shown in Table~\ref{tab:fit}. With these values of $\Delta$ and the fitted values of $a_0$, this kind of systematic uncertainty is estimated and added in quadrature to the total error of each $g_{AB}$. The final results of $g_{AB}$ are listed in Table~\ref{tab:gAB}.

It is seen that for $\pi_1\to b_1\pi$ and $\eta_1\to\rho\rho$, the effective coupling $g_{AB}$ derived at different $\hat{k}$ are consistent with each other. For other decay modes, the values of $g_{AB}$ deviate from each other at the two $\hat{k}$'s, signaling the systematic uncertainties of the M\&M method to some extent. This kind of uncertainty was also observed in Ref.~\cite{Bali:2015gji} in the derivation of $g_{\rho\pi\pi}$ for the $\rho\to\pi\pi$ decay using the M\&M method, where the value of $g_{\rho\pi\pi}$, obtained at different relative momenta and different moving frames of the $\pi\pi$ system on the lattices used, varied from 5.2 to 8.4, manifesting roughly a 40\% discrepancy from $g_{\rho\pi\pi}\approx 6.0$ determined from the $\rho$ width. Anyway, as a ballpark estimate of the effective couplings for the two-body decays of $\pi_1$ and $\eta_1$, we average the values of $g_{AB}$ at different $\hat{k}$ (if available) and take the largest discrepancy as the systematic uncertainty of the M\&M method, namely,
\begin{eqnarray}
\bar{g}_{AB}&=&\frac{1}{2}\left(g_{AB}(p=0)+g_{AB}(p=1)\right),\nonumber\\
\delta\bar{g}_{AB}&=& \frac{1}{2} \left(\mathrm{max}(g_{AB}+\delta g_{AB})-\mathrm{min}(g_{AB}-\delta g_{AB})\right).\nonumber
\end{eqnarray}
The values of $\bar{g}_{AB}$ are also shown in Table~\ref{tab:gAB}.

\begin{table}[t]
    \caption{The effective couplings and partial decay widths of $\pi_1$ and $\eta_1$. The average of $g_{AB}$ over different $\hat{k}$ (if available) gives the effective coupling $\bar{g}_{AB}$, whose uncertainty is estimated through $\delta\bar{g}_{AB} = [(\mathrm{max}(g_{AB} + \delta g_{AB}) - \mathrm{min}(g_{AB} - \delta g_{AB}))]/2$.}
    \label{tab:gAB}
    \begin{ruledtabular}
        \begin{tabular}{llc}
            mode                                      & $g_{AB}$  &  $\bar{g}_{AB}$\\ \hline
            $\pi_1\rightarrow b_1\pi(\hat{k}^2=0)$    & 4.84(46)   &  4.72(54) \\
            $\pi_1\rightarrow b_1\pi(\hat{k}^2=1)$    & 4.69(38)  &  \\
            &&\\
            $\pi_1\rightarrow f_1\pi(\hat{k}^2=0)$    & 0.81(6)   &  0.96(28)\\
            $\pi_1\rightarrow f_1\pi(\hat{k}^2=1)$    & 1.08(22)  &\\
            &&\\
            $\pi_1\rightarrow \rho\pi(\hat{k}^2=1)$   & 4.54(31)   &  4.54(31)\\
            &&\\
            \hline
            $\eta_1\rightarrow a_1\pi(\hat{k}^2=0)$   & 1.02(28)  &  1.30(55)\\
            $\eta_1\rightarrow a_1\pi(\hat{k}^2=1)$   & 1.59(25)  &\\
            &&\\
            $\eta_1\rightarrow f_1\eta(\hat{k}^2=0)$  & 2.20(26)  &  2.28(36)\\
            $\eta_1\rightarrow f_1\eta(\hat{k}^2=1)$  & 2.37(28)  \\
            &&\\
            $\eta_1\rightarrow \rho\rho(\hat{k}^2=1)$ & 2.79(32)  & 2.90(51)\\
            $\eta_1\rightarrow \rho\rho(\hat{k}^2=2)$ & 3.01(48)  &\\
        \end{tabular}
    \end{ruledtabular}
\end{table}

\section{Results of $\pi_1$ decays}\label{sec:pi1-decay}
Now we are ready to discuss the partial decay widths of the decay processes $\pi_1\to b_1\pi, f_1\pi, \rho\pi$ using the derived effective couplings $\bar{g}_{AB}$. Here we assume the quark mass dependence on $\bar{g}_{AB}$ is negligible, as is usually done in phenomenological studies and also in Ref.~\cite{Woss:2020ayi}. In the 2024 version of the Review of Particle Physics, the pole parameter of $\pi_1(1600)$ is given to be $m_{\pi_1}-i\Gamma/2=(1480-1680)-i(150-300)~\mathrm{MeV}$, which is in a fairly large range. So we use the PDG 2022 value $m_{\pi_1}=1661_{-11}^{+15}~\mathrm{MeV}$ of $\pi_1(1600)$~\cite{Workman:2022ynf} to estimate the partial widths of $\pi_1(1600)$ along with experimental mass values of $b_1$, $f_1(1285)$, $\rho$, and $\pi$. Eq.~(\ref{eq:decay-wid}), (\ref{eq:decay-mom}), and (\ref{eq:decay-amp}) give the partial decay widths
\begin{eqnarray}
    \Gamma_{b_1\pi}&=&325(75)~\mathrm{MeV}~~(
    \bar{g}_{b_1\pi}=4.72(54))\nonumber\\
    \Gamma_{\rho\pi}&=&52(7)~\mathrm{MeV}~~~~~(\bar{g}_{\rho\pi}=4.54(31)).
\end{eqnarray}

Experimentally, there are two $0^+1^{++}$ states, $f_1(1285)$ and $f_1(1420)$, which are admixtures of the light quark component $|f_1^{(l)}\rangle=|(u\bar{u}-d\bar{d})/\sqrt{2}\rangle$ and the strange quark component $|f_1^{(s)}\rangle=|s\bar{s}\rangle$ through a mixing angle $\alpha_A$, namely,
\begin{equation}
    \begin{pmatrix}
        |f_1(1285)\rangle \\
        |f_1(1420)\rangle
    \end{pmatrix}
    =
    \begin{pmatrix}
        \cos\alpha_A & -\sin\alpha_A \\
        \sin\alpha_A & \cos\alpha_A
    \end{pmatrix}
    \begin{pmatrix}
        |f_1^{(l)}\rangle \\
        |f_1^{(s)}\rangle
    \end{pmatrix}.
\end{equation}
A previous lattice QCD calculation gives $\alpha_A\approx 30^\circ$ at $m_\pi=391~\mathrm{MeV}$~\cite{Dudek:2013yja}, while the PDG recommends $|\sin\alpha_A|\approx \sin (90^\circ-23^\circ-54.7^\circ)=\sin 12.3^\circ$~\cite{Workman:2022ynf}. Both values of $\alpha_A$ indicate that the lower state $f_1(1285)$ is dominated by the $|f_1^{(l)}\rangle$ component. So with $\bar{g}_{f_1\pi}=0.98$, we estimate the partial decay width 
\begin{eqnarray}
    \Gamma_{f_1(1285)\pi}&\sim& \mathcal{O}(10)~\mathrm{MeV},\nonumber\\
    \Gamma_{f_1(1420)\pi}&\sim& \mathcal{O}(1)~\mathrm{MeV}.
\end{eqnarray}

Regarding the large coupling $\bar{g}_{\rho\pi}=4.54(31)$, it is expected that $\pi_1(1600)$ has a sizeable decay fraction to $K\bar{K}^*$. So we can use $\bar{g}_{\rho\pi}=4.54(31)$ derived in the $N_f=2$ QCD to estimate the partial decay width of $\pi_1(1600)\to K\bar{K}^*$. In the two-body decays of a meson, the additional constituent quarks in the final states are generated by gluonic excitations. In the $N_f=2$ QCD, gluons couple equally to $u\bar{u}$ and $d\bar{d}$, while in the $N_f=3$ QCD, gluons couple approximately equally to $u\bar{u}$, $d\bar{d}$, and $s\bar{s}$ if the quark mass effect is ignored. The SU(3) flavor symmetry implies $\bar{g}_{\rho\pi}=\sqrt{2} \bar{g}_{K\bar{K}^*}$. Thus we obtain the partial decay width
\begin{equation}
    \Gamma_{K\bar{K}^*}\approx 8.6(1.3)~\mathrm{MeV}
\end{equation}
using the physical masses of $K$ and $K^*$.

\begin{table}[t]
    \caption{
   The partial decay widths are calculated using $\bar{g}_{AB}$ and the experimental values~\cite{Workman:2022ynf} of the mesons involved. The previous lattice QCD results through the L\"uscher method (labelled by LM)~\cite{Woss:2020ayi} are also shown for comparison.}
    \begin{ruledtabular}
        \begin{tabular}{lc|c}
            $\Gamma_i$                           & $\Gamma_{AB}$(MeV) & $\Gamma_{AB}$(MeV)\cite{Woss:2020ayi} \\ \hline
            $\Gamma(\pi_1\rightarrow b_1\pi)$       & $325(75)$         & 139-529                               \\
            $\Gamma(\pi_1\rightarrow f_1(1285)\pi)$ & $\mathcal{O}(10)$  & 0-24                                  \\                    
            $\Gamma(\pi_1\rightarrow f_1(1420)\pi)$ & $\mathcal{O}(1)$   & 0-2                                  \\                    
            $\Gamma(\pi_1\rightarrow \rho\pi)$      & $52(7)   $         & 0-20                                  \\
            $\Gamma(\pi_1\rightarrow K\bar{K}^*)$   & $8.6(1.3)$           & 0-2                                  \\  \hline
            $\sum\limits_i \Gamma_i$                      &   $\sim 396(90)$          & 139-590
        \end{tabular}
    \end{ruledtabular}
    \label{tab:pi1width-shi}
\end{table}

The decays of $\pi_1(1600)$ have been investigated by $N_f=3$ lattice QCD using the L\"{u}scher method~\cite{Woss:2020ayi}, where the flavor $SU(3)$ symmetry is exact with the pion mass being set to $m_\pi\approx 700~\mathrm{MeV}$. By assuming the couplings derived at this pion mass are insensitive to light quark masses (and also the hadron masses involved), the partial decay widths of $\pi_1$ are predicted using the physical kinematics, as also shown in Table~\ref{tab:pi1width-shi}. These partial widths vary in a large range but are consistent with our results from the M\&M method. There is a slight difference in the partial decay widths of the $\rho\pi$ decay mode in that we obtain a relatively larger value $\Gamma_{\rho\pi}=52(7)~\mathrm{MeV}$. The consistency of our result with the previous lattice study using the L\"{u}scher method also indicates the feasibility of the M\&M method in studying the strong decays of hybrid mesons.

The major pattern for the $\pi_1(1600)$ two-body decay from lattice QCD calculations is that $b_1\pi$ is the largest and even dominant decay process. This is more or less in line with the expectation from the phenomenological studies based on the flux tube models which expect the decay modes composed by a $P$-wave meson (axial vector meson) and a $S$-wave meson is preferable~\cite{Close:1994hc} and the ratios of the partial decay widths are expected to be
\begin{eqnarray}
    &&\pi b_1 :\pi f_1:\pi \rho:\pi\eta:\pi\eta'\nonumber\\
    &=&170:60:5\sim 20:0\sim 10:0\sim 10.
\end{eqnarray}
Although the very large partial width of the $b_1\pi$ decay is consistent with the expectation of the phenomenological studies, the lattice results of the $f_1\pi$ is much smaller than that expected by the phenomenological result. This should be understood in the future.

\begin{figure}[t]
    \centering
    \includegraphics[width=\linewidth]{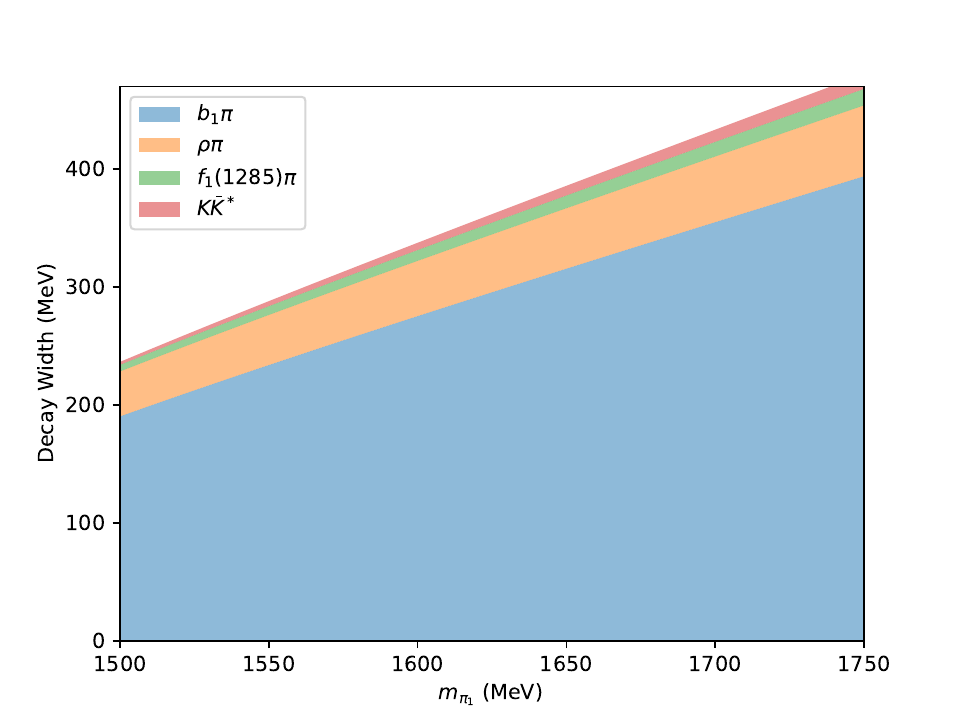}\\
    \caption{The partial decay widths of $\pi_1$ versus the $m_{\pi_1}$. The partial decay widths of $b_1\pi$ (blue),  $\rho\pi$(orange), $f_1\pi$(green), and $K\bar{K}^*$(red) are calculated using the same coupling constants obtained in this study with $m_{\pi_1}$ varying from 1.5 GeV to 1.75 GeV. The height of each colored region shows the partial decay width of each decay mode, and the top line also illustrates the total width for the $m_{\pi_1}$.}
    \label{fig:widthvsmass}
\end{figure}

The large value of $\Gamma_{\rho\pi}$ we obtain also comply with the fact $\pi_1(1600)$ is observed in the $\rho\pi$ system by different experiments. Considering the experimental value of $\pi_1$ mass varies in a large range from 1564 MeV to roughly 1700 MeV, which result in very different phase space factors of two-body decays, especially for the $P$-wave final states $\rho\pi$ and $\eta\pi$. So we also calculated the partial decays widths using the same coupling constants and and a varying $\pi_1$ mass from 1.5 GeV to 1.75 GeV. The results are illustrated in Fig.~\ref{fig:widthvsmass}.

Since $\pi_1(1600)$ is likely below the $K_1 \bar{K}$ threshold and the decay mode $K^*\bar{K}$ (in $P$-wave) is suppressed by the centrifugal barrier, the total width of $\pi_1(1600)$ can be estimated by adding up the partial decays of $b_1\pi$, $f_1\pi$, and $\rho\pi$, which gives
\begin{equation}
    \Gamma(\pi_1(1600))= 396(90) ~\mathrm{MeV}.
\end{equation}
Note that this total width does not consider the $\eta(\eta')\pi$ decays. This width is larger than the PDG value $\Gamma(\pi_1(1600))=240\pm 50~\mathrm{MeV}$~\cite{Workman:2022ynf}, but compatible with the COMPASS result $\Gamma(\pi_1(1600))=580_{-230}^{+100}~\mathrm{MeV}$~\cite{COMPASS:2009xrl}, the B852 result $\Gamma(\pi_1(1600))=403\pm 80\pm 115~\mathrm{MeV}$~\cite{E852:2004gpn} and $340\pm40\pm 50~\mathrm{MeV}$~\cite{E852:2001ikk}. It is important to note that the PDG value incorporates the smaller value $\Gamma(\pi_1(1600))=185\pm 25\pm 28~\mathrm{MeV}$ from E852 experiments~\cite{E852:2004rfa}.

\section{Results of $\eta_1$ decays}\label{sec:eta1-decay}

Let us switch to the two-body strong decays of the isoscalar $1^{-+}$ hybrid $\eta_1$. By following a similar procedure as in the case of $\pi_1$ decay, we calculate the related ratio functions $R_{AB}(\vec{k},t)$ for $\eta_1$ decaying into two-body modes $AB$. A slight complication arises due to the involvement of more isoscalar particles, leading to the appearance of quark annihilation diagrams in several instances. Specifically, the correlation functions $C^{33}_{AB,\eta_1}$ for $AB = a_1\pi, \rho\rho$ include the diagrams in panels (a) and (d) of Fig.~\ref{fig:schamatic-diagrams}, while $C^{33}_{AB,\eta_1}$ for $f_1\eta$ includes diagrams in panels (a), (b), (c), (d), and (e). The contribution from annihilation diagrams makes the $C^{33}_{AB,\eta_1}(\vec{k},t)$ more noisy in the large $t$ region. The corresponding ratio functions $R^{(33)}_{AB}(k,t)$ with different momentum modes $\hat{k}$ are shown in Fig.~\ref{fig:fit-eta1}. Fortunately, approximate linear behaviors appear in the time region for $t_\mathrm{max}/a_t < 18$. We then fit the ratio functions with the polynomial function form in Eq.~(\ref{eq:r-para}) to obtain the parameters $a_0$, $r_{AB}$, and $a_2$, which are collected in Table~\ref{tab:fit}. Subsequently, we extract the effective couplings $g_{AB}$ from $r_{AB}$ for the $AB$ modes $a_1\pi$, $f_1\eta$, and $\rho\rho$, as shown in Table~\ref{tab:gAB}.

The situation becomes more complicated when predicting the partial decay widths of $\eta_1$ using the effective coupling derived here. In the $N_f=2$ QCD, there is only one isoscalar $\eta_1$, but this state cannot be connected with the possible hybrid meson $\eta_1(1855)$ observed by BESIII~\cite{BESIII:2022riz}. In the physical $N_f=2+1$ case, there should be two isoscalar $1^{-+}$ states, $\eta_1^{(l)}$ and $\eta_1^{(s)}$, on the flavor basis, where $\eta_1^{(l)}$ and $\eta_1^{(s)}$ have the flavor wave functions
\begin{equation}
    |\eta_1^{(l)}\rangle = \frac{1}{\sqrt{2}} (|u\bar{u}\rangle + |d\bar{d}\rangle), \quad |\eta_1^{(s)}\rangle = |s\bar{s}\rangle.
\end{equation}

In order to estimate the two-body decay widths of $\eta_1^{(L,H)}$ using the effective couplings obtained in the $N_f=2$ QCD, we consider the expectations from SU(3) flavor symmetry. First, we introduce the matrix form of the flavor nonet $X$,
\begin{equation}
    X = \begin{pmatrix}
        \frac{\pi^0_X + \eta_{X}^{(l)}}{\sqrt{2}} & \pi_X^+                                  & K_X^+           \\
        \pi_X^-                                   & \frac{-\pi^0_X + \eta_{X}^{(l)}}{\sqrt{2}} & K_X^0           \\
        K_X^-                                     & \bar{K}_X^0                              & \eta_{X}^{(s)}
    \end{pmatrix},
\end{equation}
where $X$ stands for the hybrid nonet (denoted by $H$) to which $\pi_1$ and $\eta_1^{(l,s)}$ belong, as well as the nonets (denoted by $A$ and $B$) to which $A$ and $B$ belong. Here, $(\pi_X^+, \pi_X^0, \pi_X^-)$ form the $I=1$ multiplet, $(K_X^+, K_X^0)$ and $(\bar{K}_X^0, K_X^-)$ are the two $I=1/2$ doublets, and $\eta_X^{(l)}$ and $\eta_X^{(s)}$ are the two isoscalars with quark configurations $\frac{1}{\sqrt{2}} \left(|u\bar{u}\rangle + |d\bar{d}\rangle\right)$ and $|s\bar{s}\rangle$, respectively.

Let $C'(X)$ be the charge conjugation transformation ($\mathcal{C}$) factor of the nonet $X$, which is defined by $\mathcal{C}|X\rangle = C'(X)|\bar{X}\rangle$. This factor takes the value $C'(X) = C'(\pi_X^0)$, where $C'(\pi_X^0)$ is the $\mathcal{C}$-parity of $\pi_X^0$. For the decay modes $AB$ with $C'(A)C'(B) = -$, flavor symmetry and $\mathcal{C}$-conservation require the effective interaction Lagrangian to take the form (with Lorentz indices and possible derivative operators omitted here):
\begin{eqnarray}\label{eq:flavor_eff-minus}
    \mathcal{L}^{(-)}_{HAB} = \frac{g^{(-)}}{2} \Tr(H[A,B]),
\end{eqnarray}
where $g^{(-)}$ is the unique effective coupling constant. This type of interaction is OZI-favored (no quark annihilation diagrams contribute) and applies to the decay modes:
\begin{eqnarray}
    \pi_1 &\to& b_1 \pi, \nonumber\\
    \pi_1 &\to& K_1 \bar{K}, \nonumber\\
    \pi_1 &\to& \rho \pi, \nonumber\\
    \pi_1 &\to& K^* \bar{K}, \nonumber\\
    \eta_1^{(l/s)} &\to& K_1 \bar{K}, \nonumber\\
    \eta_1^{(l/s)} &\to& K^* \bar{K}.
\end{eqnarray}

Flavor symmetry implies that we can use the effective couplings $\bar{g}_{b_1\pi}$ and $\bar{g}_{\rho\pi}$ to estimate the partial decay widths of $\eta_1^{(l/s)} \to K_1 \bar{K}$ and $\eta_1^{(l/s)} \to K^* \bar{K}$ (see below). Since $\pi_1(1600)$ lies below the $K_1 \bar{K}$ threshold, we do not consider the $\pi_1(1600) \to K_1 \bar{K}$ decay in Sec.~\ref{sec:pi1-decay}.

For the decay modes $AB$ with $C'(A)C'(B)=+$, the effective Lagrangian takes the form:
\begin{eqnarray}\label{eq:flavor_eff-plus}
    \mathcal{L}^{(+)}_{HAB} &=& \frac{g}{2} \Tr(H \{A, B\}) - g_H \Tr H \Tr(AB) \nonumber \\
    && - g_A \Tr A \Tr(BH) - g_B \Tr B \Tr(HA) \nonumber \\
    && + g_3 \Tr H \Tr A \Tr B,
\end{eqnarray}
where five effective couplings $g$, $g_H$, $g_A$, $g_B$, and $g_3$ are involved. Since the trace `$\mathrm{Tr}$' is taken in the flavor space, each `$\mathrm{Tr}$' operation implies a constituent quark loop and contributes a minus sign, which results in the relative signs of the five terms in the Lagrangian. The quark loops are flavor singlets and are necessarily connected by gluons, so different terms in the effective Lagrangian above manifest different dynamics that are described by the individual effective couplings and are responsible for the $H \to AB$ decays. Specifically, the effective coupling $g$ describes the decay dynamics of the fully connected quark diagrams, while $g_X$ with $X=H,A,B$ accounts for the annihilation effect of the quarks in the initial hybrid state $X$. The coupling $g_3$ describes the fully annihilation effects when the three particles $H$, $A$, and $B$ are all isospin singlets. In other words, the five terms have a qualitative one-to-one correspondence to the schematic quark diagrams (a), (d), (b), (c), and (e) in Fig.~\ref{fig:schamatic-diagrams}, respectively.

If we introduce the following notations:
\begin{eqnarray}\label{eq:two-meson-comb}
    (\pi_A\pi_B)_1^- &=& \frac{1}{\sqrt{2}} \left(\pi_A^+ \pi_B^- - \pi_A^- \pi_B^+\right) \nonumber\\
    (\pi_A\pi_B)_0^+ &=& \frac{1}{\sqrt{3}} \left(\pi_A^+ \pi_B^- + \pi_A^0 \pi_B^0 + \pi_A^- \pi_B^+\right) \nonumber\\
    (K_A K_B)_1^- &=& \frac{1}{2} \left(K_A^+ K_B^- - K_A^0 \bar{K}_B^0 - K_A^- K_B^+ + \bar{K}_A^0 K_B^0\right) \nonumber\\
    (K_A K_B)_0^- &=& \frac{1}{2} \left(K_A^+ K_B^- + K_A^0 \bar{K}_B^0 - K_A^- K_B^+ - \bar{K}_A^0 K_B^0\right) \nonumber\\
    (K_A K_B)_1^+ &=& \frac{1}{2} \left(K_A^+ K_B^- - K_A^0 \bar{K}_B^0 + K_A^- K_B^+ - \bar{K}_A^0 K_B^0\right) \nonumber\\
    (K_A K_B)_0^+ &=& \frac{1}{2} \left(K_A^+ K_B^- + K_A^0 \bar{K}_B^0 + K_A^- K_B^+ + \bar{K}_A^0 K_B^0\right) \nonumber\\
\end{eqnarray}
where the subscripts represent $I=0,1$ and the superscripts denote the sign of $C'(A)C'(B)$, we then obtain the explicit expressions for the effective Lagrangian governing the decays $\pi_1^0 \to AB$.

\begin{eqnarray}\label{eq:pi1AB}
        \mathcal{L}_{\pi_1^0\to AB}= \pi_1^0 \bigg[ &&\frac{1}{\sqrt{2}}g(K_A K_B)_1^+
        +\frac{1}{\sqrt{2}}g^{(-)}(K_A K_B)_1^-                                                                            \nonumber  \\
                                                       & +&  g^{(-)} (\pi_A\pi_B)_1^-                                       \nonumber \\
                                                       & +&  (\frac{1}{\sqrt{2}}(g-2g_A)\eta_A^{(l)}-g_A\eta_A^{(s)})\pi_B^0  \nonumber \\
                                                       & +&  \pi_A^0(\frac{1}{\sqrt{2}}(g-2g_B)\eta_B^{(l)}-g_B\eta_B^{(s)}) \bigg], 
\end{eqnarray}
from which one can infer the relations of the effective couplings for $\pi_1\to \rho\pi, K^*\bar{K}$ and $\pi_1\to b_1\pi, (K_1\bar{K})_1^-$,
\begin{eqnarray}\label{eq:lAB}
    g_{b_1\pi}:g_{(K_1 \bar{K})_1^-}&=&\sqrt{2}:1\nonumber\\
    g_{\rho\pi}:g_{K^*\bar{K}} &=&\sqrt{2}:1,
\end{eqnarray}
the latter of which has been used in Sec.~\ref{sec:pi1-decay} to estimate the partial decay width $\Gamma_{K^*\bar{K}}$ of $\pi_1(1600)$.

The effective Lagrangian for the $\eta_1^{(l/s)}$ decays reads
\begin{eqnarray}\label{eq:sAB}
    \mathcal{L}_{\eta_1^{(l)}\to AB}= \eta_1^{(l)}  \bigg[ && \frac{1}{\sqrt{2}}g^{(-)}(K_A K_B)_0^-\nonumber\\
                                                        &+& \frac{1}{\sqrt{2}}(g-4g_H) (K_A K_B)_0^+\nonumber\\
                                                       & +&  \sqrt{\frac{3}{2}} (g-2g_H) (\pi_A\pi_B)_0^+  \nonumber\\
                                                       & +&  \frac{1}{\sqrt{2}}(g-2g_H-2g_A\nonumber\\
                                                       &&-2g_B+4g_3)\eta_A^{(l)} \eta_B^{(l)} \nonumber\\
                                                       & +&  (-g_B+2g_3)\eta_A^{(l)} \eta_B^{(s)}\nonumber\\
                                            & +&(-g_A+2g_3)\eta_A^{(s)} \eta_B^{(l)}\nonumber\\
                                                       & +&  \sqrt{2}(-g_H+g_3)\eta_A^{(s)} \eta_B^{(s)} \bigg]
\end{eqnarray}
\begin{eqnarray}
    \mathcal{L}_{\eta_1^{(s)}\to AB}= \eta_1^{(s)}  \bigg[ &  -&g^{(-)}(K_A K_B)_0^-\nonumber\\
                                                       &+& (g-2g_H) (K_A K_B)_0^+\nonumber\\
                                                      & - & \sqrt{3}g_H (\pi_A\pi_B)_0^+ \nonumber\\
                                                      & + & (-g_H+2g_3)\eta_A^{(l)} \eta_B^{(l)} \nonumber\\
                                                      & + & \sqrt{2}(-g_A+g_3)\eta_A^{(l)} \eta_B^{(s)} \nonumber\\
                                                      &+&\sqrt{2}(-g_B+g_3)\eta_A^{(s)} \eta_B^{(l)}\nonumber\\
                                                      & + &  (g-g_H-g_A-g_B+g_3)\eta_A^{(s)} \eta_B^{(s)} \bigg].\nonumber\\
\end{eqnarray}

The interaction terms involving the effective coupling $g^{(-)}$ do not include contributions from quark annihilation effects. Therefore, $g^{(-)}$ in the physical $N_f=2+1$ case can be approximated by the $N_f=2$ value obtained in this study. This approximation is justified as the $N_f$ dependence is actually embodied in the strong coupling constant $\alpha_s$ due to the vacuum polarization of quarks. This argument may also apply to the effective coupling $g$, which describes the dynamics of the fully connected diagrams of valence quarks. However, when the M\&M method is adopted to extract the specific effective coupling $g_{AB}$ for an individual decay process $H \to AB$, the physical observable is the correlation function $C_{AB,H}$, which includes contributions from all the quark diagrams after Wick contraction. Hence, the effective couplings $g$, $g_H$, $g_A$, $g_B$, and $g_3$ are entangled together according to the combinations in the Lagrangian above and collectively contribute to the total $g_{AB}$.

To estimate the contribution of each diagram in Fig.~\ref{fig:schamatic-diagrams} to $g_{AB}$, we tentatively calculate each diagram of $C_{AB,H}$ and extract $g$, $g_H$, $g_A$, $g_B$, and $g_3$ following a similar procedure for the extraction of $g_{AB}$ for each decay process $H \to AB$ in $N_f=2$ QCD. The results are shown in Table~\ref{tab:individual}. For the axial vector-pseudoscalar decay modes ($AP$), the values of $g$ from different decay modes at different momentum modes are close to each other, as required by SU(2) flavor symmetry, and the values of $g_H$ and $g_A$ are much smaller than $g$. Notably, when $\eta$ is involved, the values of $g_A$ and $g_H$ have larger central values but also much larger uncertainties. This may be attributed to the exclusion of the disconnected diagrams, which are important for $\eta$. The large value of $g_P$ signals the significant role played by the $\mathrm{U}_A(1)$ anomaly when gluons couple to the isoscalar $\eta$ in $N_f=2$ QCD. The $g_H$ for the $H \to VV$ decay modes is also much smaller than $g$ and can be understood by the OZI suppression. The coupling constant $g_3$, which accounts for the fully annihilation diagrams and only appears in the decays $\eta_1^{(l/s)} \to f_1 \eta$, is observed to be negligible.

\begin{table}[t]
    \caption{The coupling constant of different channels. The contribution of each schematic diagram is presented separately. In Fig. \ref{fig:schamatic-diagrams} $g$ refers to the son of two connected diagrams (labeled as (a)). $g_X$ refers to the diagrams in which particle $X$ is disconnected from other particles. \label{tab:individual}}
    \begin{ruledtabular}
        \begin{tabular}{ccc}
            \multicolumn{3}{c}{ $H\rightarrow AP$ }                    \\\hline
%            $g^{(-)}$ & $\pi_1\rightarrow b_1\pi(p=0)$    & -4.406(75) \\
%                      & $\pi_1\rightarrow b_1\pi(p=1)$    & -4.59(10) \\\hline
                      & $\pi_1\rightarrow f_1\pi(p=0)$    & -0.985(30) \\
                      & $\pi_1\rightarrow f_1\pi(p=1)$    & -1.638(67) \\
            $g$       & $\eta_1\rightarrow a_1\pi(p=0)$   & -0.870(82) \\
                      & $\eta_1\rightarrow a_1\pi(p=1)$   & -1.44(12) \\
                      & $\eta_1\rightarrow f_1\eta(p=0)$  & -1.44(31) \\
                      & $\eta_1\rightarrow f_1\eta(p=1)$  & -2.32(64)  \\\hline
                      & $\eta_1\rightarrow a_1\pi(p=0)$   & 0.009(39)  \\
            $g_H$     & $\eta_1\rightarrow a_1\pi(p=1)$   & 0.008(58)  \\
                      & $\eta_1\rightarrow f_1\eta(p=0)$  & 0.32(23)  \\
                      & $\eta_1\rightarrow f_1\eta(p=1)$  & 0.33(40)  \\\hline
                      & $\pi_1\rightarrow f_1\pi(p=0)$    & 0.075(29)  \\
            $g_A$     & $\pi_1\rightarrow f_1\pi(p=1)$    & -0.001(70)  \\
                      & $\eta_1\rightarrow f_1\eta(p=0)$  & 0.11(10)  \\
                      & $\eta_1\rightarrow f_1\eta(p=1)$  & -0.11(12) \\\hline
            $g_P$     & $\eta_1\rightarrow f_1\eta(p=0)$  & 0.35(30)   \\
                      & $\eta_1\rightarrow f_1\eta(p=1)$  & 0.60(24)  \\\hline
%            \multicolumn{3}{c}{ $H\rightarrow VP$ }                    \\\hline
%            $g^{(-)}$ & $\pi_1\rightarrow \rho\pi(p=1)$   & 4.378(54) \\\hline
            \multicolumn{3}{c}{ $H\rightarrow VV$ }                    \\\hline
            $g$       & $\eta_1\rightarrow \rho\rho(p=1)$ & -2.03(15)  \\
                      & $\eta_1\rightarrow \rho\rho(p=2)$ & -2.36(23)  \\\hline
            $g_H$     & $\eta_1\rightarrow \rho\rho(p=1)$ & 0.110(57)  \\
                      & $\eta_1\rightarrow \rho\rho(p=2)$ & 0.10(14)  \\
        \end{tabular}
    \end{ruledtabular}
\end{table}
%%%%%%%%%%%%%%%%%%%%%%%%%%%%%%%%%%%%%%%%%%%%%%%%%%%%%%%%%%%%%%%%%%%%%%%%%%%%%%%%%%%%%%%%%%%%%%%%%%%%%%

Based on the observations mentioned above and according to the expressions of the Lagrangian in Eqs.~(\ref{eq:pi1AB}), (\ref{eq:lAB}), and (\ref{eq:sAB}), we have the following approximate effective couplings for $\eta_1^{(l)}$ decays:
\begin{eqnarray}\label{eq:eta-light}
    g_{\eta_1^{(l)}(K_1\bar{K})_0^-} &=& \frac{1}{\sqrt{2}} g^{(-)} \approx \frac{1}{\sqrt{2}} \bar{g}_{b_1\pi} \nonumber \\
    g_{\eta_1^{(l)}a_1\pi} &=& \sqrt{\frac{3}{2}}(g - 2g_H) \approx \bar{g}_{a_1\pi} \nonumber \\
    g_{\eta_1^{(l)}(K_1\bar{K})_0^+} &=& \frac{1}{\sqrt{2}} (g - 4g_H) \approx \frac{1}{\sqrt{3}} \bar{g}_{a_1\pi} \nonumber \\
    g_{\eta_1^{(l)}\rho\rho} &=& \sqrt{\frac{3}{2}}(g - 2g_H) \approx \bar{g}_{\rho\rho} \nonumber \\
    g_{\eta_1^{(l)}(K^*\bar{K}^*)_0^+} &=& \frac{1}{\sqrt{2}} (g - 4g_H) \approx \frac{1}{\sqrt{3}} \bar{g}_{\rho\rho} \nonumber \\
    g_{\eta_1^{(l)}\omega\omega} &=& \frac{1}{2}(g - 2g_H - \ldots) \approx \frac{1}{\sqrt{3}} \bar{g}_{\rho\rho} \nonumber \\
    g_{\eta_1^{(l)}K^*\bar{K}} &=& \frac{1}{\sqrt{2}} g^{(-)} \approx \frac{1}{\sqrt{2}} \bar{g}_{\rho\pi},
\end{eqnarray}
where $g_H \ll 1$ is assumed to be zero(see Table~\ref{tab:individual}) and $g_A$, $g_B$, $g_3$ in $\eta_1^{(l)} \to \omega\omega$ are also negligible because both $A$ and $B$ are the vector meson $\omega$. The couplings $\bar{g}_{AB}$ take the values in Table~\ref{tab:gAB}. Similarly, the effective couplings for $\eta_1^{(s)}$ decays are approximated as:
\begin{eqnarray}\label{eq:eta-heavy}
    g_{\eta_1^{(s)}(K_1\bar{K})_0^-} &=& -g^{(-)} \approx -\bar{g}_{b_1\pi} \nonumber \\
    g_{\eta_1^{(s)}a_1\pi} &=& -\sqrt{3}g_H \approx 0 \nonumber \\
    g_{\eta_1^{(s)}(K_1\bar{K})_0^+} &=& (g - 2g_H) \approx \sqrt{\frac{2}{3}} \bar{g}_{a_1\pi} \nonumber \\
    g_{\eta_1^{(s)}\rho\rho} &=& -\sqrt{3}g_H \approx 0 \nonumber \\
    g_{\eta_1^{(s)}(K^*\bar{K}^*)_0^+} &=& (g - 2g_H) \approx \sqrt{\frac{2}{3}} \bar{g}_{\rho\rho} \nonumber \\
    g_{\eta_1^{(s)}\phi\phi} &=& (g - g_H - \ldots) \approx \sqrt{\frac{2}{3}} \bar{g}_{\rho\rho} \nonumber \\
    g_{\eta_1^{(s)}K^*\bar{K}} &=& -g^{(-)} \approx -\bar{g}_{\rho\pi}.
\end{eqnarray}
The decay modes involving $\eta(\eta')$ will be discussed elsewhere.

Experimentally, there are two $K_1$ states, namely, $K_1(1270)$ and $K_1(1400)$, which are nearly equal mixtures of the $1^{+(+)}$ state $K_{1A}$ and the $1^{+(-)}$ state $K_{1B}$. This mixing can be expressed as:
\begin{equation}\label{eq:K-mixing}
    \begin{pmatrix}
        K_1(1270) \\
        K_1(1400)
    \end{pmatrix}
    =
    \begin{pmatrix}
        \cos\theta_K & \sin\theta_K \\
        -\sin\theta_K & \cos\theta_K
    \end{pmatrix}
    \begin{pmatrix}
        K_{1B} \\
        K_{1A}
    \end{pmatrix},
\end{equation}
where $\theta_K$ is the mixing angle. Phenomenological analyses indicate that $|\theta_K|$ is around either $35^\circ$ or $55^\circ$~\cite{Suzuki:1993yc,Burakovsky:1997dd,Cheng:2003bn}. For simplicity, we take the approximate value $\theta_K \approx 45^\circ$. The $K_{1A}$ component of $K_1(1270)/K_1(1400)$ is responsible for the $(K_1\bar{K})_0^+$ decay mode, while $K_{1B}$ enters the $(K_1\bar{K})_0^-$ mode.

Meson observed in experiments are only mass eigenstates. For $\eta_1$ states, there should be two mass eigenstates, namely, $\eta_1^{(L)}$ and $\eta_1^{(H)}$, which are admixtures of $\eta_1^{(l)}$ and $\eta_1^{(s)}$ through a mixing angle $\alpha$ 
\begin{equation}
    \begin{pmatrix}
        \eta_1^{(L)} \\
        \eta_1^{(H)}
    \end{pmatrix}
    =
    \begin{pmatrix}
        \cos\alpha & -\sin\alpha \\
       \sin\alpha & \cos\alpha
   \end{pmatrix}
    \begin{pmatrix}
        \eta_1^{(l)} \\
       \eta_1^{(s)}
    \end{pmatrix},
\end{equation}
or the admixtures of the singlet $\eta_1^{(1)}\sim (|u\bar{u}\rangle+|d\bar{d}\rangle+|s\bar{s}\rangle)/\sqrt{3}$ and $\eta_1^{(8)}\sim (|u\bar{u}\rangle+|d\bar{d}\rangle-2|s\bar{s}\rangle)/\sqrt{6}$ through the mixing angle $\theta$
\begin{equation}
    \begin{pmatrix}
        \eta_1^{(L)} \\
        \eta_1^{(H)}
    \end{pmatrix}
    =
    \begin{pmatrix}
        \cos\theta & -\sin\theta \\
       \sin\theta & \cos\theta
   \end{pmatrix}
    \begin{pmatrix}
        \eta_1^{(8)} \\
       \eta_1^{(1)}
    \end{pmatrix}.
\end{equation}
One can easily show that $\theta$ is related to $\alpha$ by $\theta=\alpha-54.7^\circ$. 

BESIII observed for the first time a $I^GJ^{PC} = 0^+1^{-+}$ structure, $\eta_1(1855)$, through partial wave analysis of the $J/\psi \to \gamma \eta \eta'$ process~\cite{BESIII:2022riz,BESIII:2022iwi}. The resonance parameters of $\eta_1(1855)$ are determined to be $m_{\eta_1} = 1855 \pm 9_{-1}^{+6}$ MeV and $\Gamma_{\eta_1} = 188 \pm 18_{-8}^{+3}$ MeV. $\eta_1(1855)$ can be a candidate for an isoscalar $1^{-+}$ hybrid. However, the existence of another $\eta_1$ state is crucial for unraveling the nature of $\eta_1(1855)$. In fact, BESIII also reported a weak ($4.4\sigma$) signal of a $1^{-+}$ component around 2.2 GeV~\cite{BESIII:2022iwi}, which needs to be confirmed in future experiments. On the other hand, a previous lattice QCD calculation~\cite{Dudek:2013yja} predicted the mixing angle to be $\alpha = 22.7^\circ$, and the masses of $\eta_1^{(L)}$ and $\eta_1^{(H)}$ to be around 2.17 GeV and 2.35 GeV at $m_\pi \approx 391~\mathrm{MeV}$. The mass of $\eta_1^{(L)}$ is consistent with our result $m_{\eta_1} \approx 2275(48)~\mathrm{MeV}$. Therefore, we tentatively assign $\eta_1(1855)$ as the lighter state $\eta_1^{(L)}$ and the structure around 2.2 GeV (labeled as $\eta_1(2200)$) as a candidate for the higher state $\eta_1^{(H)}$. We then explore the decay properties of $\eta_1(1855)$ and $\eta_1(2200)$ based on the discussion above and the effective couplings obtained in this work.

%%%%%%%%%%%%%%%%%%%%%%%%%%%%%%%%%%%%%%%%%%%%%%%%%%%%%%%%%%%%%%%%%%%%%%%%%%%%%%%%%%%%%%%%%%%%%%%%%%%%%%%%%%%%%%
\subsection{$\eta_1(1855)$ decays}

If $\eta_1(1855)$ is the lighter state $\eta_1^{(L)}$, its wave function reads
\begin{equation}
    |\eta_1(1855)\rangle=\cos\alpha |\eta_1^{(l)}\rangle-\sin\alpha |\eta_1^{(s)}\rangle.
\end{equation}
We treat $\alpha$ as a free parameter and use the physical masses of the mesons involved to discuss the decay properties of $\eta_1(1855)$.

First, we consider the decay process $\eta_1(1855)\to (K_1\bar{K})_0^-$. According to Eqs.~(\ref{eq:eta-light}), (\ref{eq:eta-heavy}) and (\ref{eq:K-mixing}), the effective coupling is expressed as 
\begin{eqnarray}
    g_{\eta_1^{(L)}(K_1\bar{K})_0^-}&=& \left(\frac{1}{\sqrt{2}}\cos\alpha+\sin\alpha\right) \bar{g}_{b_1\pi}\cos\theta_K\nonumber\\
    &\approx& \bar{g}_{b_1\pi} \cos(\alpha-54.7^\circ) \sqrt{\frac{3}{2}} \sqrt{\frac{1}{2}},
\end{eqnarray}
for $\eta_1(1855)\to (K_1(1270)\bar{K})_0^-$, where $\cos\theta_K\approx \cos45^\circ=1/\sqrt{2}$ is used. This coupling also indicates that
the $(K_1\bar{K})_0^-$ decay of $\eta_1(1855)$ takes place only through its octet component because of $\cos\theta=\cos(\alpha-54.7^\circ)$. The coupling for $\eta_1(1855)\to (K_1(1400)\bar{K})_0^-$ can be derived similarly with $\cos\theta_K$ being replaced by $\sin\theta_K$, although this decay does not take place since $\eta_1(1855)$ is below the $K_1(1400)\bar{K}$ threshold. Then with the value $g_{b_1\pi}=4.68(51)$ and the expressions Eq.~(\ref{eq:decay-wid}) and (\ref{eq:decay-amp}), we estimate the partial decay width to be
\begin{equation}
    \Gamma_{(K_1(1270)\bar{K})_0^-}\approx (189(45)~\mathrm{MeV}) \cos^2(\alpha-54.7^\circ).
\end{equation}

The decay $\eta_1(1855)\to a_1\pi$ takes place mainly from the $\eta_1^{(l)}$ component of $\eta_1(1855)$. Thus using the value of the coupling constant $\bar{g}_{a_1\pi}=1.42(53)$, we estimate
\begin{equation}
    \Gamma_{a_1\pi}=(36(30)~\mathrm{MeV})\cos^2\alpha.
\end{equation}

On the other hand, $\eta_1(1855)$ also decays into $K_1(1270)\bar{K}$ through the $(K_1\bar{K})_0^+$ mode with the $K_{1A}$ component playing the role. The effective coupling is 
\begin{eqnarray}
    g_{\eta_1^{(L)}(K_1\bar{K})_0^+}&\approx& \left(\frac{1}{\sqrt{3}}\cos\alpha-\sqrt{\frac{2}{3}}\sin\alpha\right) \bar{g}_{a_1\pi}\sin\theta_K\nonumber\\
    &\approx& \bar{g}_{a_1\pi} \cos(\alpha+54.7^\circ)\sqrt{\frac{1}{2}},
\end{eqnarray}
which results in the partial decay width
\begin{equation}
    \Gamma_{(K_1(1270)\bar{K})^+_0}\approx (10(8)~\mathrm{MeV}) \cos^2(\alpha+54.7^\circ),
\end{equation}
Obviously, this partial width is much smaller than $\Gamma_{(K_1(1270)\bar{K})_0^-}$ when $0<\alpha<35.3^\circ$.

Now we consider the $\eta_1(1855)\to VV$ decays. We calculate the effective coupling $\eta_1\to\rho\rho$ in the $N_f=2$ QCD and obtain $\bar{g}_{\rho\rho}=2.93(64)$. This value can be applied to the physical $N_f=2+1$ case when the quark annihilation effect is neglected. Obviously, the decays $\eta_(1855)\to \rho\rho, \omega\omega$ take place through the $\eta_1^{(l)}$ component of $\eta_1(1855)$, and therefore we estimate 
\begin{eqnarray}
    \Gamma_{\rho\rho}&=&(49(18)~\mathrm{MeV})\cos^2\alpha\nonumber\\
    \Gamma_{\omega\omega}&\approx& \frac{1}{3} \Gamma_{\rho\rho},
\end{eqnarray}
where the phase space factor $1/2$ has been considered for the two (generalized) identical particles in the final state. As indicated by the effective Lagrangian in Eq.~(\ref{eq:lAB}) and (\ref{eq:sAB}), $\eta_1(1855)$ also decays into $K^*\bar{K}^*$. Similar to the derivation of $g_{\eta_1^{(L)}(K_1\bar{K})_0^+}$, we have the estimation
\begin{equation}
        g_{\eta_1^{(L)}(K^*\bar{K}^*)_0^+}\approx \bar{g}_{\rho\rho} \cos(\alpha+54.7^\circ),
\end{equation}
which gives a very small partial decay width
\begin{equation}
    \Gamma_{K^*\bar{K}^*}=(5(3)~\mathrm{MeV})\cos^2(\alpha+54.7^\circ).
\end{equation}
owing to the phase space suppression. 

$\eta_1(1855)$ cannot decay into $\rho\pi$ but can decay into $K^*\bar{K}$ through its flavor octet component with the effective coupling 
\begin{equation}
    g_{\eta_1^{(L)}K^*\bar{K}}=\bar{g}_{\rho\pi}\cos(\alpha-54.7^\circ)\sqrt{\frac{3}{2}}.
\end{equation}
With the value $\bar{g}_{\rho\pi}=4.54(31)$ we have 
\begin{equation}
    \Gamma_{K^*\bar{K}}=(52(7)~\mathrm{MeV})\cos^2(\alpha-54.7^\circ).
\end{equation}

The decay $\eta_1(1855)\to f_1(1285)\eta$ is also kinetically permitted. However, we are unable to get very solid results of the effective couplings for the decays involving the isoscalar pseudoscalar meson $\eta$ in the $N_f=2$ QCD. So we can only give a rougher estimate of the partial decay width $f_1(1285)\eta$. As we addressed in Sec.~\ref{sec:pi1-decay}, experiments~\cite{Workman:2022ynf} and a previous lattice QCD calculation~\cite{Dudek:2013yja} indicate that $f_1(1285)$ has mainly a $(u\bar{u}+d\bar{d})/\sqrt{2}$ component. On the other hand, $\eta$ is mainly an octet pseudoscalar, so we use the effective coupling $\bar{g}_{f_1\pi}=0.98(26)$ to approximate the effective coupling for $\eta_1(1855)\to f_1(1285)\eta$. Then according to the Lagrangian in Eqs.~(\ref{eq:pi1AB}), (\ref{eq:lAB}), and (\ref{eq:sAB}), we 
estimate
\begin{equation}
    g_{\eta_1^{(L)}f_1(1285)\eta}\approx \bar{g}_{f_1\pi} \cos\alpha \cos\alpha_P,
\end{equation}
where $\alpha_P\approx 54.7^\circ+\theta_P$ with $\theta_P$ being the singlet-octet mixing angle of the pseudoscalar meson $\theta_P\approx -11.3^\circ$ (quadratic mass relation) or $-24.5^\circ$ (linear mass relation)~\cite{Workman:2022ynf}. Then the partial decay width reads
\begin{equation}
    \Gamma_{f_1(1285)\eta}=(5(2)~\mathrm{MeV})\cos^2\alpha\cos^2\alpha_P.
\end{equation}

%%%%%%%%%%%%%%%%%%%%%%%%%%%%%%%%%%%%%%%%%%%%%%%%%%%%%%%%%%%%%%%%%%%%%%%%%%%%%%%%%%
\begin{table*}[t]
    \caption{The partial decay widths of $\eta_1(1855)$ and $\eta_1(2200)$. The explicit expressions of partial widths in terms of the mixing angle $\alpha$ are shown in the second column. The third column column are the values of partial widths at $\alpha$ is set as $22.7^\circ$. The sum of these values gives an estimate of total width of $\eta_1(1855)$ and $\eta_1(2200)$ with the error being just a simple sum over the errors of the partial widths. \label{tab:partial-width-eta1}}
    \begin{ruledtabular}
        \begin{tabular}{llr}
            mode                                     & $\Gamma_i(\alpha)$ (MeV)             &  $\Gamma_i(\alpha\approx22.7^\circ)$ (MeV) \\ 
            \hline
            $\eta_1(1855)\to K_1(1270)\bar{K}$        & $189(45)\times\cos^2(\alpha-54.7^\circ)+10(8)\times\cos^2(\alpha+54.7^\circ) $ &   136(32)\\
            $\eta_1(1855)\to a_1\pi$                  & $36(30)\times\cos^2\alpha$                  &   31(26) \\
            $\eta_1(1855)\to \rho\rho$                & $49(18)\times \cos^2\alpha$                 &   42(15)\\
            $\eta_1(1855)\to \omega\omega$            & $15(7)\times \cos^2 \alpha$                 &   13(5) \\
            $\eta_1(1855)\to K^*\bar{K}$              & $52(7)\times \cos^2(\alpha-54.7^\circ)$     &   37(5)  \\
            $\eta_1(1855)\to \eta \eta'$              &                                             &   $\sim 20$\\
            $\eta_1(1855)\to f_1(1285)+\eta$          & $5(2)\times \cos^2\alpha\cos^2\alpha_P$     &   $\mathcal{O}(1)$  \\    
            $\eta_1(1855)\to K^*\bar{K}^*$            & $5(3)\times \cos^2(54.7^\circ+\alpha)$      &   $\sim 0$\\\hline
                                                      &                                             &   $\sum_i \Gamma_i\approx 282(85)$\\
            \hline
            $\eta_1(2200)\to K_1(1270)\bar{K}$        & $450(100)\times\sin^2(\alpha-54.7^\circ)+23(19)\times\sin^2(\alpha+54.7^\circ) $ &   149(47)\\
            $\eta_1(2200)\to K_1(1400)\bar{K}$        & $350(80)\times\sin^2(\alpha-54.7^\circ)+18(15)\times\sin^2(\alpha+54.7^\circ) $ &   115(37)\\
            $\eta_1(2200)\to a_1\pi$                  & $57(48)\times\sin^2\alpha$                  &   8(7) \\
            $\eta_1(2200)\to \rho\rho$                & $176(62)\times\sin^2\alpha$                 &   26(9)\\
            $\eta_1(2200)\to \omega\omega$                & $56(20)\times\sin^2\alpha$                 &   8(3)\\
            $\eta_1(2200)\to K^*\bar{K}^*$            & $76(27)\times\sin^2(54.7^\circ+\alpha)$     &   72(25)\\          
            $\eta_1(2200)\to \phi\phi$                & $10(4)\times \cos^2 \alpha$                 &   8(3) \\
            $\eta_1(2200)\to K^*\bar{K}$              & $100(14)\times\sin^2(\alpha-54.7^\circ)$      &   28(4)  \\
            $\eta_1(2200)\to \eta \eta'$              &                                             &   $\sim 26$\\
            $\eta_1(2200)\to f_1(1285)+\eta$          & $23(13)\times(0.43\sin\alpha+0.36\cos\alpha)^2$     &   6(3)  \\    
            $\eta_1(2200)\to f_1(1420)+\eta$          & $17(10)\times(0.25\sin\alpha-0.61\cos\alpha)^2$     &  7(4)  \\  \hline
                                                      &                                             &   $\sum_i \Gamma_i\approx 455(143)$\\
        \end{tabular}
    \end{ruledtabular}
\end{table*}

At last, we discuss the partial width for $\eta_1(1855)\to \eta\eta'$. We do not get a reliable result of the effective coupling for the $\pi_1\to\pi\eta$ decay, and there is only one isoscalar pseudoscalar meson $\eta$ in the $N_f=2$ QCD. Given that $\eta_1(1855)$ is the lighter state $\eta_1^{(L)}$, a previous lattice QCD study predicts the partial decay width $\Gamma(J/\psi\to\gamma \eta_1(1855)=(2.0\pm0.7)~\mathrm{eV}$~\cite{Chen:2022isv}, which gives an estimate of the branching fraction of $\eta_1(1855)\to\eta\eta'$ to be $(13\pm 5)\%$ using the measured branching fraction $\mathrm{Br}(J/\psi\to\gamma \eta_1(1855)\to\gamma\eta\eta')=(2.70\pm 0.41_{-0.35}^{+0.16})\times 10^{-6}$ by BESIII~\cite{BESIII:2022riz}. If this is true, the partial width of $\eta(1855)\to\eta\eta'$ can be estimated to be $\Gamma_{\eta\eta'}\approx 20~\mathrm{MeV}$. 

All the $\alpha$-dependent partial widths are collected in Table~\ref{tab:partial-width-eta1}. We assume that the two-body decay widths derived above saturate approximately the total decay width of $\eta_1(1855)$. After summing up them, we can obtain the total width $\Gamma(\alpha)$ in terms of the mixing angle $\alpha$. Figure~\ref{fig:widthvsalpha} shows the $\alpha$ dependence of the total decay width $\Gamma(\alpha)$ in the interval $\alpha\in [0,45^\circ]$. It is interesting to see that $\Gamma(\alpha)$ varies in a very narrow range
\begin{equation}
    \Gamma(\alpha)\in [209(76),306(79)]~\mathrm{MeV}.
\end{equation}
If we use the lattice QCD result $\alpha\approx 22.7(1.0)^\circ$~\cite{Dudek:2013yja}, the total width of $\eta_1(1855)$ is estimated to be 
\begin{equation}
\Gamma(\eta_1(1855))=\sum\limits_i \Gamma_i \approx 282(85)~\mathrm{MeV}    
\end{equation}, 
whose central value is larger than the physical value $\Gamma(\eta_1(1855))=188\pm 18_{-8}^{+3}~\mathrm{MeV}$ by roughly 50\%. These results indicate that the hybrid assignment of $\eta_1(1855)$ is compatible with our study.

Note that according to the isospin symmetry, the partial decay width of $\eta_1\to \omega\omega$
is $\Gamma_{\omega\omega}\approx \Gamma_{\rho\rho}/3\approx 20-30~\mathrm{MeV}$. Considering the result above is obtained in the flavor SU(3) symmetric limit, and the effective couplings have large systematic uncertainties from the present calculation, this prediction is a ballpark theoretical result.

Obviously, the $K_1(1270)\bar{K}$, $a_1\pi$, $\rho\rho$, and $K^*\bar{K}$ are dominant decay modes. Although the large partial decay width is understandable for the $S$-wave $K_1(1270)\bar{K}$ and $a_1\pi$ decay,
the large $\Gamma_{\rho\rho}$ is totally unexpected, since the $\rho\rho$ decay is usually thought to be highly suppressed in the phenomenological flux tube picture~\cite{Isgur:1984bm,Close:1994hc,Page:1996rj,Page:1998gz} where the decay mode of two identical particles is prohibited for a $1^{-+}$ hybrid meson. A similar
decay pattern is observed by the lattice QCD study on the two-body decays of the charmoniumlike $1^{-+}$ hybrid $\eta_{c1}$~\cite{Shi:2023sdy}. This can be checked for experiments to search for $\eta_1(1855)$ in the $\rho\rho$ and $\omega\omega$ systems.

\subsection{$\eta_1(2200)$ decays}
Theoretically, there must exist the other mass eigenstate of $I^GJ^{PC}=0^+1^{-+}$. A previous lattice QCD study indicates that the state with a larger $s\bar{s}$ component has a higher mass~\cite{Dudek:2013yja}. Experimentally, BESIII observes a $4.4\sigma$ signal at 2.2 GeV of the same quantum numbers as that of $\eta_1(1855)$~\cite{BESIII:2022iwi}.
So we take this structure as the higher state $\eta_1^{(H)}$, labelled as $\eta_1(2200)$, which has the wave function
\begin{equation}
    |\eta_1(2200)\rangle=\sin\alpha |\eta_1^{(l)}\rangle+\cos\alpha |\eta_1^{(s)}\rangle.
\end{equation}

Different from the $\eta_1(1855)$ case, $\eta_1(2200)$ can decay into both $K_1(1270)\bar{K}$ and $K_1(1400)\bar{K}$ states, since it lies above both thresholds. Similar to the $(K_1\bar{K})_0^-$ decay modes of $\eta_1^{(L)}$, the effective couplings for these two processes have the same form
\begin{equation}
    g_{\eta_1^{(H)}\to (K_1\bar{K})_0^-}\approx \bar{g}_{b_1\pi} \sin(\alpha-54.7^\circ) \sqrt{\frac{3}{2}} \sqrt{\frac{1}{2}},
\end{equation}
and the corresponding decay widths are
\begin{eqnarray}
    \Gamma_{(K_1(1270)\bar{K})_0^{-}}&\approx& (450(100)~\mathrm{MeV}) \sin^2(\alpha-54.7^\circ)\nonumber\\
    \Gamma_{(K_1(1400)\bar{K})_0^-}&\approx& (350(80)~\mathrm{MeV}) \sin^2(\alpha-54.7^\circ).
\end{eqnarray}

The effective coupling for $(K_1\bar{K})_0^+$ decay mode of $\eta_1^{(H)}$ reads
\begin{equation}
    g_{\eta_1(1855)\to (K_1\bar{K})_0^+}\approx \bar{g}_{a_1\pi} \sin(\alpha+54.7^\circ) \sqrt{\frac{1}{2}},
\end{equation}
which gives the decay widths
\begin{eqnarray}
    \Gamma_{(K_1(1270)\bar{K})_0^+}&\approx& (23(19)~\mathrm{MeV}) \sin^2(\alpha+54.7^\circ)\nonumber\\
    \Gamma_{(K_1(1400)\bar{K})_0^+}&\approx& (18(15)~\mathrm{MeV}) \sin^2(\alpha+54.7^\circ).
\end{eqnarray}

The decay $\eta_1(2200)\to a_1\pi$ takes place also from the $\eta_1^{(l)}$ component of $\eta_1(2200)$. Thus using the value of the coupling constant $\bar{g}_{a_1\pi}=1.42(53)$, we estimate
\begin{equation}
    \Gamma_{a_1\pi}=(57(48)~\mathrm{MeV})\sin^2\alpha.
\end{equation}

$\eta_1(2200)$ also decays to $\rho\rho$ through its $\eta_1^{(l)}$ component, decays to $\phi\phi$ through its $\eta_1^{(s)}$ component, and also decays to $(K^*\bar{K}^*)_0^+$. The effective couplings are 
\begin{eqnarray}
    g_{\eta_1^{(H)}\rho\rho}&=&\bar{g}_{\rho\rho} \sin\alpha\nonumber\\
    g_{\eta_1^{(H)}\phi\phi}&=&\bar{g}_{\rho\rho} \cos\alpha\sqrt{\frac{2}{3}}\nonumber\\
    g_{\eta_1^{(H)}K^*\bar{K}^*}&\approx& \bar{g}_{\rho\rho} \sin(\alpha+54.7^\circ).
\end{eqnarray}
Then using $\bar{g}_{\rho\rho}=2.93(64)$ we have,
\begin{eqnarray}
    \Gamma_{\rho\rho}&=& (176(62)~\mathrm{MeV})\sin^2\alpha\nonumber\\
    \Gamma_{\phi\phi}&=&(10(4)~\mathrm{MeV})\cos^2\alpha\nonumber\\
    \Gamma_{K^*\bar{K}^*}&=& (76(27)~\mathrm{MeV})\sin^2(\alpha+54.7^\circ)
\end{eqnarray}
where the phase space factor $1/2$ has been considered for the two (generalized) identical particles in the final state $\rho\rho$, $\omega\omega$, $\phi\phi$ and also $K^*\bar{K}*$ given the definition of $(K^*\bar{K}^*)_0^+$ in Eq.~(\ref{eq:two-meson-comb}).

Similar to the $\eta_1(1855)\to K^*\bar{K}$, the effective coupling is 
\begin{equation}
     g_{\eta_1^{(H)}K^*\bar{K}}=\bar{g}_{\rho\pi}\sin(\alpha-54.7^\circ)\sqrt{\frac{3}{2}}.
\end{equation}
the partial decay width of $\eta_1(2200)\to K^*\bar{K}$ is estimated to be 
\begin{equation}
    \Gamma_{K^*\bar{K}}=(100(14)~\mathrm{MeV})\sin^2(54.7^\circ-\alpha).
\end{equation}

The decays $f_1(1285)\eta$ and $f_1(1420)\eta$ are now open for $\eta_1(2200)$, so we consider their partial decay widths. Similar to the discussion on $\eta_1(1855)\to f_1(1285)\eta$, we take $g\approx \bar{g}_{a_1\pi}=0.98(26)$ and ignore temporarily the contribution of $g_A, g_B, g_H$ in Eqs.~(\ref{eq:pi1AB}), (\ref{eq:lAB}), and (\ref{eq:sAB}), we have the estimate of the effective coupling 
\begin{eqnarray}
    g_{\eta_1^{(H)}f_1(1285)\eta}\approx\bar{g}_{f_1\pi}\big(&&\frac{1}{\sqrt{2}}\sin\alpha \cos\alpha_A \cos\alpha_P \nonumber\\
    &+&\cos\alpha \sin\alpha_A \sin\alpha_P \big)\nonumber\\
    g_{\eta_1^{(H)}f_1(1420)\eta}\approx\bar{g}_{f_1\pi}\big(&&\frac{1}{\sqrt{2}}\sin\alpha \sin\alpha_A \cos\alpha_P \nonumber\\
    &-&\cos\alpha \cos\alpha_A \sin\alpha_P \big).
\end{eqnarray}
If we take the values $\alpha_A\approx 30^\circ$, $\alpha_P\approx 45^\circ$, then the partial widths are 
\begin{eqnarray}
    \Gamma_{f_1(1285)\eta}&\approx (23(13)~\mathrm{MeV})(0.43\sin\alpha+0.36\cos\alpha)^2\nonumber\\
    \Gamma_{f_1(1420)\eta}&\approx (17(10)~\mathrm{MeV})(0.25\sin\alpha-0.61\cos\alpha)^2.
\end{eqnarray}

Both $\eta_1(1855)$ and $\eta_1(2200)$ decay into $\eta'\eta$ through their octet component since $\eta'\eta$ only appears in the flavor octet in the flavor SU(3) symmetry. So it is expected that 
\begin{equation}
    \frac{\Gamma(\eta_1(2200)\to\eta\eta')}{\Gamma(\eta_1(1855)\to\eta\eta')}\approx \frac{m_L^2 k_H^3}{m_H^2 k_L^3} \tan^2\theta\approx 1.3,
\end{equation}
where $m_{H/L}$ is the mass of $\eta_1^{(H/L)}$, $k_{H/L}$ is the decay momentum for $\eta_1^{(H/L)}\to \eta\eta'$ and $\theta=\alpha-54.7^\circ\approx -32^\circ$. Thus we estimate 
\begin{equation}
    \Gamma(\eta_1(2200\to\eta\eta')\approx 26~\mathrm{MeV}
\end{equation}
using $\Gamma(\eta_1(1855)\to\eta\eta')\approx 20~\mathrm{MeV}$. 
To this end, we can see that the dominant decay modes of $\eta_1(2200)$ are $K_1\bar{K}$, $K^*\bar{K}^*$, $\rho\rho$, $K^*\bar{K}$. $\eta_1(2200)$ also has sizeable decay fractions $\eta\eta'$ and $\phi\phi$. Given a major $s\bar{s}$ component of $\eta_1(2200)$, its decay pattern is very similar to its charmonium-like counterpart $\eta_{c1}$, which decays predominantly to $D_1\bar{D}$, $D^*\bar{D}^*$ and $D^*\bar{D}$~\cite{Shi:2023sdy}. 

%%%%%%%%%%%%%%%%%%%%%%%%%%%%%%%%%%%%%%%%%%%%%%%%%%%%%%%%%%%%%%%%%%%%%%%%%%%%%%%%%%
\begin{figure}[t]
    \centering
    \includegraphics[width=\linewidth]{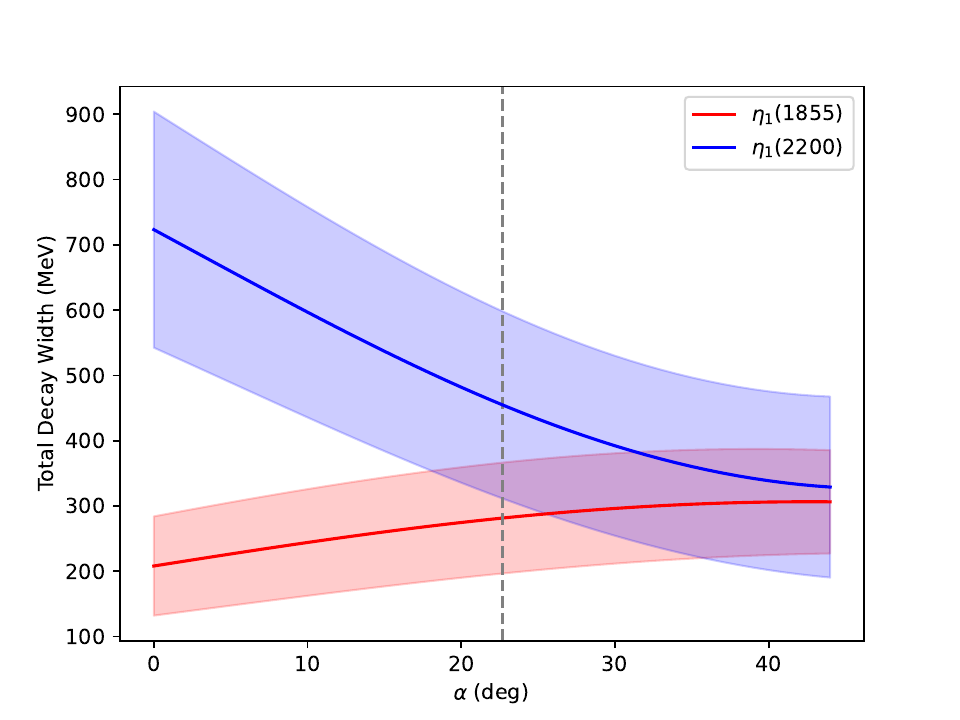}
    \vspace{-0.1in}
    \caption{The total decay widths of $\eta_1(1855)$ (red) and $\eta_1(2200)$ (blue) versus the mixing angle $\alpha$. The vertical dashed line indicates the value $\alpha \approx 22.7^\circ$.}
    \label{fig:widthvsalpha}
\end{figure}
%%%%%%%%%%%%%%%%%%%%%%%%%%%%%%%%%%%%%%%%%%%%%%%%%%%%%%%%%%%%%%%%%%%%%%%%%%%%%%%%%%

The major results of the $\eta_1(2200)$ decay are collected in Table~\ref{tab:partial-width-eta1}. All the partial decay widths, and therefore the total decay width, depend on the mixing angle $\alpha$, as shown in Fig.~\ref{fig:widthvsalpha}. Taking the lattice QCD value $\alpha\approx 22.7^\circ$ we estimate the total width of $\eta_1(2200)$ to be 
\begin{equation}
    \Gamma(\eta_1(2200))=\sum\limits_i \Gamma_i\approx 455(143)~\mathrm{MeV},
\end{equation}
which is roughly 1.6 times as large as $\Gamma(\eta_1(1855))$ and explains to some extent that the statistical significance of $\eta_1(2200)$ is lower than $\eta_1(1855)$ in the partial wave analysis of $J/\psi\to\gamma \eta\eta'$ by BESIII according to the expectation~\cite{Chen:2022isv}
\begin{equation}\label{eq:ratio}
    \frac{\mathrm{Br}(J/\psi\to \gamma\eta_1(1855)\to\gamma\eta\eta')}{\mathrm{Br}(J/\psi\to \gamma\eta_1(2200)\to\gamma\eta\eta')}
    \sim\frac{\Gamma(\eta_1(1855))}{\Gamma(\eta_1(2200))}.
\end{equation}
Our results indicate that $\eta_1(2200)$ can be searched in $K_1 \bar{K}$  $K^*\bar{K}^*$ systems. The processes $J/\psi\to\gamma (K_1\bar{K}, K^*\bar{K}^*)$ and $\psi(3686)\to \phi (K_1\bar{K}, K^*\bar{K}^*)$ might be good places for the $\eta_1(2200)$ hunting. 

%%%%%%%%%%%%%%%%%%%%%%%%%%%%%%%%%%%%%%%%%%%%%%%%%%%%%%%%%%%%%%%%%%%%%%%%%%%%%%%%%%%%%%%%%%%%%%%%%%%%%%%%%%%%%%%%%%%%%%%%%
\section{Summary}\label{sec:summary}
We study the decay properties of the isovector $1^{-+}$ hybrid meson $\pi_1$ and the isocalar $1^{-+}$ hybrid $\eta_1$ in the formalism of $N_f=2$ lattice QCD at a pion mass $m_\pi\approx 417~\mathrm{MeV}$. We adopt the Michael and McNeile method to extract the transition matrix elements, from which the effective couplings for the two-body decays are determined.

By using the PDG values of meson masses involved, the partial decay widths of $\pi_1(1600)$ (we use $m_{\pi_1}=1661_{-11}^{+15}~\mathrm{MeV}$ in PDG 2022~\cite{Workman:2022ynf}) are predicted to be $(\Gamma_{b_1\pi}, \Gamma_{f_1\pi}, \Gamma_{\rho\pi},\Gamma_{K^*\bar{K}}=(325\pm 75, \mathcal{O}(10), 52\pm 7, 8.6\pm 1.3)~\mathrm{MeV}$, and its total width is estimated to be around $396(90)~\mathrm{MeV}$. These results are compatible with the previous lattice QCD calculations using the L\"{u}scher method but with smaller uncertainties. This total width is larger than the PDG value $\Gamma(\pi_1(1600))=240\pm 50~\mathrm{MeV}$~\cite{Workman:2022ynf}, but consistent with the COMPASS result~\cite{COMPASS:2009xrl} and most of E852 ressults~\cite{E852:2004gpn,E852:2001ikk}. The dominant $b_1\pi$ decay mode of $\pi_1(1600)$ is also in line with the phenomenological expectation. We observe that $\Gamma_{\rho\pi}$ is large also. It is interesting to see that the effective coupling of the $1^{+(-)}0^{-+}$ is much larger than that of the $1^{+(+)}0^{-+}$ mode. This is intriguing and needs to be investigated in depth in future studies.

We obtain the effective couplings $g_{a_1\pi}$, $g_{f_1\eta}$ and $g_{\rho\rho}$ for the two-body decays of $\eta_1$ in $N_f=2$ QCD. There should be two $\eta_1$ mass eigenstates, $\eta_1^{(L)}$ and $\eta_1^{(H)}$ in the physical $N_f=2+1$ case. Based on the SU(3) flavor symmetry, the decay properties of $\pi_1$ and $\eta_1$ in $N_f=2$ QCD can be used to estimate the partial decay widths of $\eta_1^{(L)}$ and $\eta_1^{(H)}$. If $\eta_1(1855)$ and the $4.4\sigma$ signal (labeled as $\eta_1(2200)$) can be assigned to $\eta_1^{(L)}$ and $\eta_1^{(H)}$, respectively, using the mixing angle $\alpha=22.7^\circ$, their partial decay widths to $K_1(1270)\bar{K}$, $a_1\pi$, $\rho\rho$, $K^*\bar{K}$, $\omega\omega$, $\phi\phi$, $K^*\bar{K}^*$ are predicted and the values are listed in Table~\ref{tab:partial-width-eta1}. The major observation is that, for both states, the dominant decay channels are $K_1(1270)\bar{K}$ (for $\eta_1(1855)$ and $\eta_1(2200)$) and $K_1(1400)\bar{K}$ (for $\eta_1(2200)$) through the $1^{+(-)}0^{-+}$ mode. On the other hand, both states have large decay fractions to $VP$ and $VV$ mode ($K^*\bar{K},\rho\rho$ and $\omega\omega$ for $\eta_1(1855)$, and $K^*\bar{K}, K^*\bar{K}^*, \phi\phi$ for $\eta_1(2200)$).  It is surprising that the $VV$ decays has large decay fractions and is in sharp contrast to the phenomenological expectation that these decay channels are strictly prohibited. The partial decay widths of $a_1\pi$ is also sizable. Finally, the total widths of both states are estimated to be 
\begin{eqnarray}
    \Gamma_{\eta_1(1855)}=282(85)~\mathrm{MeV}\nonumber\\
    \Gamma_{\eta_1(2200)}=455(143)~\mathrm{MeV}
\end{eqnarray}
with $\alpha\approx 22.7^\circ$. The predicted $\Gamma_{\eta_1(1855)}$ at this mixing angle is compatible the experimental value $188\pm 18_{-8}^{+3}$. The dependence of the total widths on $\alpha$ is also illustrated in Fig.~\ref{fig:widthvsalpha}, where one can see that a smaller $\alpha$ would give a smaller $\Gamma_{\eta_1(1855)}$ and a larger value of $\Gamma_{\eta_1(2200)}/\Gamma_{\eta_1(1855)}$. Although many systematical uncertainties are not well under control, results in this study are qualitatively informative for the experimental search of light hybrid states. 

Our results suggest to search $\eta_1(1855)$ and $\eta_1(2200)$ in the $K_1\bar{K}$ systems. Actually, the discovery of the mass partner is crucial for $\eta_1(1855)$ to be assigned soundly as a hybrid state. If $\eta_1(1855)$ is surely the lighter state, then the heavier one, such as $\eta_1(2200)$, can be searched in the $K_1(1270)\bar{K}$ and $K_1(1400)\bar{K}$ systems in the radiative $J/\psi$ decays and also in the $\psi(3686)$ strong decays by recoiling against a $\phi$ meson, since the heavier state is expected to have a dominant $s\bar{s}g$ component.

\begin{acknowledgments}
    This work is supported by the National Natural Science Foundation of China (NNSFC) under Grants No. 11935017, No. 12293060, No. 12293065, No. 12293061, No. 12205311, No. 12070131001 (CRC 110 by DFG and NNSFC)), and the National Key Research and Development Program of China (No. 2020YFA0406400) and the Strategic Priority Research Program of Chinese Academy of Sciences (No. XDB34030302). The Chroma software system~\cite{Edwards:2004sx} and QUDA library~\cite{Clark:2009wm,Babich:2011np} are acknowledged. The computations were performed on the HPC clusters at Institute of High Energy Physics (Beijing) and China Spallation Neutron Source (Dongguan), and the ORISE computing environment.
\end{acknowledgments}

\section*{Appendix}
\setcounter{equation}{0}
\setcounter{figure}{0}
\setcounter{table}{0}
\renewcommand{\theequation}{A\arabic{equation}}
\renewcommand{\thefigure}{A\arabic{figure}}
\renewcommand{\thetable}{A\arabic{table}}
\renewcommand{\thesection}{}

We use the partial-wave method to construct the interpolating meson-meson (labeled as $A$ and $B$) operators for specific quantum numbers $J^P$ ~\cite{Feng:2010es, Wallace:2015pxa, Prelovsek:2016iyo}, assuming that all the irreducible representations (irreps) do not mix with lighter states. In general, let $\mathcal{O}_{X}^{M_{X}}(\vec{k})$ be the operator for the particle $X=A$ or $B$ with spin $S_X$ and spin projection $M_{X}$ in the $z$-direction. For the total angular momentum $J$ and the $z$-axis projection $M$, the relative orbital angular momentum $L$, and the total spin $S$, the explicit construction of the $AB$ operator is expressed as:
\begin{widetext}
\begin{equation}\label{eq:two-meson-sm}
\begin{aligned}
    \mathcal{O}_{AB;JLSP}^{M}(\hat{k}) = & \sum\limits_{M_L, M_S, M_{A}, M_{B}} \langle L, M_L; S,M_S| JM \rangle
                                    \langle S_AM_{A}; S_BM_{B}| S, M_S \rangle\\
                            & \times \sum\limits_{R\in O_h} Y^*_{LM_L}(R\circ\vec{k}) \mathcal{O}_A^{M_{A}}(R\circ \vec{k}) \mathcal{O}_B^{M_{B}}(-R\circ \vec{k}),
\end{aligned}
\end{equation}
\end{widetext}
where $\hat{k}=(n_1,n_2,n_3)$ is the momentum mode of $\vec{k}=\frac{2\pi}{La_s}\hat{k}$ with $n_1\ge n_2\ge n_3\ge 0$ by convention, $R\circ\vec{k}$ is the spatial momentum rotated from $\vec{k}$ by $R\in O_h$ with $O_h$ being the lattice symmetry group, $|S,M_S\rangle$ is the total spin state of the two particles involved, $|LM_L\rangle$ is the relative orbital angular momentum state, $|JM\rangle$ is the total angular momentum state, and $Y^*_{LM_L}(R\circ\vec{k})$ is the spherical harmonic function of the direction of $R\circ \vec{k}$.

For the case of this study, $\pi_1^0$ and $\eta_1$ have quantum numbers $I^G J^{PC}=1^- 1^{-+}$ and $0^+ 1^{-+}$, respectively. As addressed in the main context, the flavor wave function of the decay mode $AB$ that reflects the correct flavor quantum numbers $I^{G}$ and $C$ is properly normalized and applied implicitly in the practical calculation. Therefore, in this Appendix, we focus on the two-meson operators that have the desired quantum number $J^P=1^-$, which can be deduced from the $T_1^{P}$ representation of $O_h$. The two-body decays include the $S$-wave decay $AP$ (one axial vector meson and one pseudoscalar meson), the $P$-wave $VP$ mode (one vector and one pseudoscalar), and the $P$-wave $VV$ mode (two vector mesons).

Since the quantum numbers $J, L, S, P$ are perfectly known for each decay mode, we denote the two-meson operator by $\mathcal{O}_{AB}^M$ and omit the $JLSP$ subscripts in the following discussions and expressions. On the other hand, it is known that the $1^-$ ($T_{1u}$) operator has three components labeled by $i=1,2,3$ (corresponding to the $x, y, z$ components, respectively). In practice, we use the third component ($i=3$), which corresponds to the $M=0$ case in Eq.~(\ref{eq:two-meson-sm}).

The operators for the $AP$ mode are very simple. We choose the momentum modes $\hat{k}=(0,0,0)$ and $\hat{k}=(0,0,1)$, which result in the energy of $AP$ being close to the mass of $\pi_1$ and $\eta_1$ in this study. For simplicity, we abbreviate the single meson operators as $A$ and $P$, respectively, in the explicit expressions of two-meson operators. This convention also applies to other decay modes. For the quantum numbers $(T_{1u}, J=1, L=0, S=1, \hat{k}=(0,0,0))$, the operator is:
\begin{equation}\label{eq:AP-sm}
    \begin{aligned}
    \mathcal{O}_{AP}^{3}(\hat{k}) = & A^3(\vec{0}) P(\vec{0}),
    \end{aligned}
\end{equation}

For $\hat{k}=(0,0,1)$, an axial vector can be mixed with a pseudoscalar and a vector meson. The $f_1$ and $a_1$ mesons should be projected to the $A_1$ representation. The operator is written as:
\begin{equation}\label{eq:AP-sm-A1}
    \begin{aligned}
    \mathcal{O}_{AP}^{3}(\hat{k}) = A^3_{00+} P_{00-} + A^3_{00+} P_{00-}.
    \end{aligned}
\end{equation}
The $b_1$ meson should be projected to the $E$ representation. The operator is written as:
\begin{equation}\label{eq:AP-sm-E}
    \begin{aligned}
    \mathcal{O}_{AP}^{3}(\hat{k}) = A^3_{0+0} P_{0-0} + A^3_{0+0} P_{0-0} + A^3_{+00} P_{-00} + A^3_{+00} P_{-00}.
    \end{aligned}
\end{equation}

Here we omit the constant factor $1/\sqrt{4\pi}$ that comes from $Y_{00}$. For the momentum modes and spin configurations involved in this work, the Clebsch-Gordan coefficients and the spherical harmonic functions result in relative signs between different terms of a two-meson operator $\mathcal{O}_{AB}^3(\hat{k})$, apart from an overall constant factor. Since this constant factor can be canceled out by taking a proper ratio of correlation functions and is therefore irrelevant to the physical results, we omit it throughout the construction of two-meson operators.

The $VP$ mode is in the $P$-wave, and the two mesons have nonzero relative momentum. Since the momentum modes involved in this study are of the $\hat{k}=(0,0,n)$ type, the different orientations of the relative momentum are reflected by the signs of its nonzero components. Therefore, we introduce three subscripts, which are different combinations of $+, -, 0$, to the single meson operators. For example, $V^3_{+-0}$ denotes the third component of the operator for a vector meson with momentum $\vec{k}=\frac{2\pi}{La_s}(n,-n,0)$. Thus, for the momentum mode $\hat{k}=(0,0,1)$, the $\mathcal{O}_{VP}$ operator with quantum numbers $(T_{1u}, J=1, L=1, S=1)$ has four terms:
\begin{equation}\label{eq:PV100-sm}
    \mathcal{O}_{VP}^{3} = + V^1_{0+0} P_{0-0} - V^1_{0-0} P_{0+0} - V^2_{+00} P_{-00} + V^2_{-00} P_{+00}.
\end{equation}

\begin{widetext}
For $VV$ mode operators with quantum numbers $(T_{1u}, J=1, L=1, S=1)$, let $V^i$ and $V'^i$ be the operators for the two vector mesons, respectively. For the momentum mode $\hat{k}=(0,0,1)$, we have:
\begin{equation}\label{eq:VV100-sm}
    \begin{aligned}
      \mathcal{O}_{VV'}^3 =& - V^1_{+00}V'^3_{-00} + V^1_{-00}V'^3_{+00} 
                            - V^2_{0+0}V'^3_{0-0} + V^2_{0-0}V'^3_{0+0}\\ 
                          & + V^3_{+00}V'^1_{-00} - V^3_{-00}V'^1_{+00} 
                            + V^3_{0+0}V'^2_{0-0} - V^3_{0-0}V'^2_{0+0}.
    \end{aligned}
\end{equation}

For the momentum mode $\hat{k}=(0,1,1)$, the $VV$ operator reads,
\begin{equation}\label{eq:VVnn0-sm}
    \begin{aligned}
    \mathcal{O}_{VV'}^3 =
    &   - V^{1}_{++0}     V'^{3}_{--0}
        - V^{1}_{+0+}     V'^{3}_{-0-}
        - V^{1}_{+0-}     V'^{3}_{-0+}
        - V^{1}_{+-0}     V'^{3}_{-+0} \\
    &   + V^{1}_{-+0}     V'^{3}_{+-0}
        + V^{1}_{-0+}     V'^{3}_{+0-}
        + V^{1}_{-0-}     V'^{3}_{+0+}
        + V^{1}_{--0}     V'^{3}_{++0} \\
    &   - V^{2}_{++0}     V'^{3}_{--0}
        + V^{2}_{+-0}     V'^{3}_{-+0}
        - V^{2}_{0++}     V'^{3}_{0--}
        - V^{2}_{0+-}     V'^{3}_{0-+} \\
    &   + V^{2}_{0-+}     V'^{3}_{0+-}
        + V^{2}_{0--}     V'^{3}_{0++}
        - V^{2}_{-+0}     V'^{3}_{+-0}
        + V^{2}_{--0}     V'^{3}_{++0} \\
    &   + V^{3}_{++0}     V'^{2}_{--0}
        + V^{3}_{++0}     V'^{1}_{--0}
        + V^{3}_{+0+}     V'^{1}_{-0-}
        + V^{3}_{+0-}     V'^{1}_{-0+} \\
    &   - V^{3}_{+-0}     V'^{2}_{-+0}
        + V^{3}_{+-0}     V'^{1}_{-+0}
        + V^{3}_{0++}     V'^{2}_{0--}
        + V^{3}_{0+-}     V'^{2}_{0-+} \\
    &   - V^{3}_{0-+}     V'^{2}_{0+-}
        - V^{3}_{0--}     V'^{2}_{0++}
        + V^{3}_{-+0}     V'^{2}_{+-0}
        - V^{3}_{-+0}     V'^{1}_{+-0} \\
    &   - V^{3}_{-0+}     V'^{1}_{+0-}
        - V^{3}_{-0-}     V'^{1}_{+0+}
        - V^{3}_{--0}     V'^{2}_{++0}
        - V^{3}_{--0}     V'^{1}_{++0}.
    \end{aligned}
\end{equation}

\end{widetext}
\bibliography{ref}

\begin{thebibliography}{105}%
\makeatletter
\providecommand \@ifxundefined [1]{%
 \@ifx{#1\undefined}
}%
\providecommand \@ifnum [1]{%
 \ifnum #1\expandafter \@firstoftwo
 \else \expandafter \@secondoftwo
 \fi
}%
\providecommand \@ifx [1]{%
 \ifx #1\expandafter \@firstoftwo
 \else \expandafter \@secondoftwo
 \fi
}%
\providecommand \natexlab [1]{#1}%
\providecommand \enquote  [1]{``#1''}%
\providecommand \bibnamefont  [1]{#1}%
\providecommand \bibfnamefont [1]{#1}%
\providecommand \citenamefont [1]{#1}%
\providecommand \href@noop [0]{\@secondoftwo}%
\providecommand \href [0]{\begingroup \@sanitize@url \@href}%
\providecommand \@href[1]{\@@startlink{#1}\@@href}%
\providecommand \@@href[1]{\endgroup#1\@@endlink}%
\providecommand \@sanitize@url [0]{\catcode `\\12\catcode `\$12\catcode `\&12\catcode `\#12\catcode `\^12\catcode `\_12\catcode `\%12\relax}%
\providecommand \@@startlink[1]{}%
\providecommand \@@endlink[0]{}%
\providecommand \url  [0]{\begingroup\@sanitize@url \@url }%
\providecommand \@url [1]{\endgroup\@href {#1}{\urlprefix }}%
\providecommand \urlprefix  [0]{URL }%
\providecommand \Eprint [0]{\href }%
\providecommand \doibase [0]{http://dx.doi.org/}%
\providecommand \selectlanguage [0]{\@gobble}%
\providecommand \bibinfo  [0]{\@secondoftwo}%
\providecommand \bibfield  [0]{\@secondoftwo}%
\providecommand \translation [1]{[#1]}%
\providecommand \BibitemOpen [0]{}%
\providecommand \bibitemStop [0]{}%
\providecommand \bibitemNoStop [0]{.\EOS\space}%
\providecommand \EOS [0]{\spacefactor3000\relax}%
\providecommand \BibitemShut  [1]{\csname bibitem#1\endcsname}%
\let\auto@bib@innerbib\@empty
%</preamble>
\bibitem [{\citenamefont {Alde}\ \emph {et~al.}(1988)\citenamefont {Alde} \emph {et~al.}}]{IHEP-Brussels-LosAlamos-AnnecyLAPP:1988iqi}%
  \BibitemOpen
  \bibfield  {author} {\bibinfo {author} {\bibfnamefont {D.}~\bibnamefont {Alde}} \emph {et~al.} (\bibinfo {collaboration} {IHEP-Brussels-Los Alamos-Annecy (LAPP)}),\ }\bibfield  {title} {\enquote {\bibinfo {title} {{Evidence for a $1^{-+}$ exotic meson}},}\ }\href {\doibase 10.1016/0370-2693(88)91686-3} {\bibfield  {journal} {\bibinfo  {journal} {Phys. Lett. B}\ }\textbf {\bibinfo {volume} {205}},\ \bibinfo {pages} {397} (\bibinfo {year} {1988})}\BibitemShut {NoStop}%
\bibitem [{\citenamefont {Gouz}\ \emph {et~al.}(2008)\citenamefont {Gouz} \emph {et~al.}}]{VES:1992zkx}%
  \BibitemOpen
  \bibfield  {author} {\bibinfo {author} {\bibfnamefont {Yu.~P.}\ \bibnamefont {Gouz}} \emph {et~al.} (\bibinfo {collaboration} {VES}),\ }\bibfield  {title} {\enquote {\bibinfo {title} {{Study of the wave with $J^{PC} = 1^{-+}$ in the partial wave analysis of $\eta' \pi^-, \eta \pi^-, f_1 \pi^-$ and $\rho^0 \pi^-$ systems produced in $\pi^- N$ interactions at $p_{\pi^-} = 37$ GeV/c}},}\ }\href {\doibase 10.1063/1.43520} {\bibfield  {journal} {\bibinfo  {journal} {AIP Conf. Proc.}\ }\textbf {\bibinfo {volume} {272}},\ \bibinfo {pages} {572--576} (\bibinfo {year} {2008})}\BibitemShut {NoStop}%
\bibitem [{\citenamefont {Beladidze}\ \emph {et~al.}(1993)\citenamefont {Beladidze} \emph {et~al.}}]{VES:1993scg}%
  \BibitemOpen
  \bibfield  {author} {\bibinfo {author} {\bibfnamefont {G.~M.}\ \bibnamefont {Beladidze}} \emph {et~al.} (\bibinfo {collaboration} {VES}),\ }\bibfield  {title} {\enquote {\bibinfo {title} {{Study of $\pi^- N \to \eta \pi^- N$ and $\pi^- N \to \eta' \pi^- N$ reactions at $37~\text{GeV}/c$}},}\ }\href {\doibase 10.1016/0370-2693(93)91224-B} {\bibfield  {journal} {\bibinfo  {journal} {Phys. Lett. B}\ }\textbf {\bibinfo {volume} {313}},\ \bibinfo {pages} {276--282} (\bibinfo {year} {1993})}\BibitemShut {NoStop}%
\bibitem [{\citenamefont {Dorofeev}\ \emph {et~al.}(2002)\citenamefont {Dorofeev} \emph {et~al.}}]{VES:2001rwn}%
  \BibitemOpen
  \bibfield  {author} {\bibinfo {author} {\bibfnamefont {Valery}\ \bibnamefont {Dorofeev}} \emph {et~al.} (\bibinfo {collaboration} {VES}),\ }\bibfield  {title} {\enquote {\bibinfo {title} {{The $J^{PC} = 1^{-+}$ hunting season at VES}},}\ }\href {\doibase 10.1063/1.1482444} {\bibfield  {journal} {\bibinfo  {journal} {AIP Conf. Proc.}\ }\textbf {\bibinfo {volume} {619}},\ \bibinfo {pages} {143--154} (\bibinfo {year} {2002})},\ \Eprint {http://arxiv.org/abs/hep-ex/0110075} {arXiv:hep-ex/0110075} \BibitemShut {NoStop}%
\bibitem [{\citenamefont {Amelin}\ \emph {et~al.}(2005)\citenamefont {Amelin} \emph {et~al.}}]{Amelin:2005ry}%
  \BibitemOpen
  \bibfield  {author} {\bibinfo {author} {\bibfnamefont {D.~V.}\ \bibnamefont {Amelin}} \emph {et~al.},\ }\bibfield  {title} {\enquote {\bibinfo {title} {{Investigation of hybrid states in the VES experiment at the Institute for High Energy Physics (Protvino)}},}\ }\href {\doibase 10.1134/1.1891185} {\bibfield  {journal} {\bibinfo  {journal} {Phys. Atom. Nucl.}\ }\textbf {\bibinfo {volume} {68}},\ \bibinfo {pages} {359--371} (\bibinfo {year} {2005})}\BibitemShut {NoStop}%
\bibitem [{\citenamefont {Aoyagi}\ \emph {et~al.}(1993)\citenamefont {Aoyagi} \emph {et~al.}}]{Aoyagi:1993kn}%
  \BibitemOpen
  \bibfield  {author} {\bibinfo {author} {\bibfnamefont {H.}~\bibnamefont {Aoyagi}} \emph {et~al.},\ }\bibfield  {title} {\enquote {\bibinfo {title} {{Study of the $\eta \pi^-$ system in the $\pi^- p$ reaction at $6.3~\text{GeV}/c$}},}\ }\href {\doibase 10.1016/0370-2693(93)90456-R} {\bibfield  {journal} {\bibinfo  {journal} {Phys. Lett. B}\ }\textbf {\bibinfo {volume} {314}},\ \bibinfo {pages} {246--254} (\bibinfo {year} {1993})}\BibitemShut {NoStop}%
\bibitem [{\citenamefont {Thompson}\ \emph {et~al.}(1997)\citenamefont {Thompson} \emph {et~al.}}]{E852:1997gvf}%
  \BibitemOpen
  \bibfield  {author} {\bibinfo {author} {\bibfnamefont {D.~R.}\ \bibnamefont {Thompson}} \emph {et~al.} (\bibinfo {collaboration} {E852}),\ }\bibfield  {title} {\enquote {\bibinfo {title} {{Evidence for exotic meson production in the reaction $pi^- p \to \eta \pi^- p$ at $18~\text{GeV}/c$}},}\ }\href {\doibase 10.1103/PhysRevLett.79.1630} {\bibfield  {journal} {\bibinfo  {journal} {Phys. Rev. Lett.}\ }\textbf {\bibinfo {volume} {79}},\ \bibinfo {pages} {1630--1633} (\bibinfo {year} {1997})},\ \Eprint {http://arxiv.org/abs/hep-ex/9705011} {arXiv:hep-ex/9705011} \BibitemShut {NoStop}%
\bibitem [{\citenamefont {Chung}\ \emph {et~al.}(1999)\citenamefont {Chung} \emph {et~al.}}]{E852:1999xev}%
  \BibitemOpen
  \bibfield  {author} {\bibinfo {author} {\bibfnamefont {S.~U.}\ \bibnamefont {Chung}} \emph {et~al.} (\bibinfo {collaboration} {E852}),\ }\bibfield  {title} {\enquote {\bibinfo {title} {{Evidence for exotic $J^{PC} = 1^{-+}$ meson production in the reaction $\pi^- p \to \eta \pi^- p$ at $18~\text{GeV}/c$}},}\ }\href {\doibase 10.1103/PhysRevD.60.092001} {\bibfield  {journal} {\bibinfo  {journal} {Phys. Rev. D}\ }\textbf {\bibinfo {volume} {60}},\ \bibinfo {pages} {092001} (\bibinfo {year} {1999})},\ \Eprint {http://arxiv.org/abs/hep-ex/9902003} {arXiv:hep-ex/9902003} \BibitemShut {NoStop}%
\bibitem [{\citenamefont {Adams}\ \emph {et~al.}(2007)\citenamefont {Adams} \emph {et~al.}}]{E862:2006cfp}%
  \BibitemOpen
  \bibfield  {author} {\bibinfo {author} {\bibfnamefont {G.~S.}\ \bibnamefont {Adams}} \emph {et~al.} (\bibinfo {collaboration} {E862}),\ }\bibfield  {title} {\enquote {\bibinfo {title} {{Confirmation of a $pi_1^0$ exotic meson in the $\eta \pi^0$ system}},}\ }\href {\doibase 10.1016/j.physletb.2007.07.068} {\bibfield  {journal} {\bibinfo  {journal} {Phys. Lett. B}\ }\textbf {\bibinfo {volume} {657}},\ \bibinfo {pages} {27--31} (\bibinfo {year} {2007})},\ \Eprint {http://arxiv.org/abs/hep-ex/0612062} {arXiv:hep-ex/0612062} \BibitemShut {NoStop}%
\bibitem [{\citenamefont {Abele}\ \emph {et~al.}(1998)\citenamefont {Abele} \emph {et~al.}}]{CrystalBarrel:1998cfz}%
  \BibitemOpen
  \bibfield  {author} {\bibinfo {author} {\bibfnamefont {A.}~\bibnamefont {Abele}} \emph {et~al.} (\bibinfo {collaboration} {Crystal Barrel}),\ }\bibfield  {title} {\enquote {\bibinfo {title} {{Exotic $\eta \pi$ state in $\bar{p} d$ annihilation at rest into $\pi^- \pi^0 \eta p$(spectator)}},}\ }\href {\doibase 10.1016/S0370-2693(98)00123-3} {\bibfield  {journal} {\bibinfo  {journal} {Phys. Lett. B}\ }\textbf {\bibinfo {volume} {423}},\ \bibinfo {pages} {175--184} (\bibinfo {year} {1998})}\BibitemShut {NoStop}%
\bibitem [{\citenamefont {Abele}\ \emph {et~al.}(1999)\citenamefont {Abele} \emph {et~al.}}]{CrystalBarrel:1999reg}%
  \BibitemOpen
  \bibfield  {author} {\bibinfo {author} {\bibfnamefont {A.}~\bibnamefont {Abele}} \emph {et~al.} (\bibinfo {collaboration} {Crystal Barrel}),\ }\bibfield  {title} {\enquote {\bibinfo {title} {{Evidence for a $\pi \eta$-$P$-wave in $\bar{p} p$ annihilations at rest into $\pi^0 \pi^0 \eta$}},}\ }\href {\doibase 10.1016/S0370-2693(98)01544-5} {\bibfield  {journal} {\bibinfo  {journal} {Phys. Lett. B}\ }\textbf {\bibinfo {volume} {446}},\ \bibinfo {pages} {349--355} (\bibinfo {year} {1999})}\BibitemShut {NoStop}%
\bibitem [{\citenamefont {Salvini}\ \emph {et~al.}(2004)\citenamefont {Salvini} \emph {et~al.}}]{OBELIX:2004oio}%
  \BibitemOpen
  \bibfield  {author} {\bibinfo {author} {\bibfnamefont {P.}~\bibnamefont {Salvini}} \emph {et~al.} (\bibinfo {collaboration} {OBELIX}),\ }\bibfield  {title} {\enquote {\bibinfo {title} {{$\bar{p} p$ annihilation into four charged pions at rest and in flight}},}\ }\href {\doibase 10.1140/epjc/s2004-01811-8} {\bibfield  {journal} {\bibinfo  {journal} {Eur. Phys. J. C}\ }\textbf {\bibinfo {volume} {35}},\ \bibinfo {pages} {21--33} (\bibinfo {year} {2004})}\BibitemShut {NoStop}%
\bibitem [{\citenamefont {Zaitsev}(2000)}]{Zaitsev:2000rc}%
  \BibitemOpen
  \bibfield  {author} {\bibinfo {author} {\bibfnamefont {A.}~\bibnamefont {Zaitsev}} (\bibinfo {collaboration} {VES}),\ }\bibfield  {title} {\enquote {\bibinfo {title} {{Study of exotic resonances in diffractive reactions}},}\ }\href {\doibase 10.1016/S0375-9474(00)00238-4} {\bibfield  {journal} {\bibinfo  {journal} {Nucl. Phys. A}\ }\textbf {\bibinfo {volume} {675}},\ \bibinfo {pages} {155C--160C} (\bibinfo {year} {2000})}\BibitemShut {NoStop}%
\bibitem [{\citenamefont {Khokhlov}(2000)}]{Khokhlov:2000tk}%
  \BibitemOpen
  \bibfield  {author} {\bibinfo {author} {\bibfnamefont {Yu.~A.}\ \bibnamefont {Khokhlov}} (\bibinfo {collaboration} {VES}),\ }\bibfield  {title} {\enquote {\bibinfo {title} {{Study of $X(1600)$ $1^{-+}$ hybrid}},}\ }\href {\doibase 10.1016/S0375-9474(99)00663-6} {\bibfield  {journal} {\bibinfo  {journal} {Nucl. Phys. A}\ }\textbf {\bibinfo {volume} {663}},\ \bibinfo {pages} {596--599} (\bibinfo {year} {2000})}\BibitemShut {NoStop}%
\bibitem [{\citenamefont {Kuhn}\ \emph {et~al.}(2004)\citenamefont {Kuhn} \emph {et~al.}}]{E852:2004gpn}%
  \BibitemOpen
  \bibfield  {author} {\bibinfo {author} {\bibfnamefont {Joachim}\ \bibnamefont {Kuhn}} \emph {et~al.} (\bibinfo {collaboration} {E852}),\ }\bibfield  {title} {\enquote {\bibinfo {title} {{Exotic meson production in the $f_1(1285) \pi^-$ system observed in the reaction $\pi^- p \to \eta \pi^+ \pi^- \pi^- p$ at $18~\text{GeV}/c$}},}\ }\href {\doibase 10.1016/j.physletb.2004.05.032} {\bibfield  {journal} {\bibinfo  {journal} {Phys. Lett. B}\ }\textbf {\bibinfo {volume} {595}},\ \bibinfo {pages} {109--117} (\bibinfo {year} {2004})},\ \Eprint {http://arxiv.org/abs/hep-ex/0401004} {arXiv:hep-ex/0401004} \BibitemShut {NoStop}%
\bibitem [{\citenamefont {Adams}\ \emph {et~al.}(2011)\citenamefont {Adams} \emph {et~al.}}]{CLEO:2011upl}%
  \BibitemOpen
  \bibfield  {author} {\bibinfo {author} {\bibfnamefont {G.~S.}\ \bibnamefont {Adams}} \emph {et~al.} (\bibinfo {collaboration} {CLEO}),\ }\bibfield  {title} {\enquote {\bibinfo {title} {{Amplitude analyses of the decays $\chi_{c1} \to \eta \pi^+ \pi^-$ and $\chi_{c1} \to \eta' \pi^+ \pi^-$}},}\ }\href {\doibase 10.1103/PhysRevD.84.112009} {\bibfield  {journal} {\bibinfo  {journal} {Phys. Rev. D}\ }\textbf {\bibinfo {volume} {84}},\ \bibinfo {pages} {112009} (\bibinfo {year} {2011})},\ \Eprint {http://arxiv.org/abs/1109.5843} {arXiv:1109.5843 [hep-ex]} \BibitemShut {NoStop}%
\bibitem [{\citenamefont {Adolph}\ \emph {et~al.}(2015)\citenamefont {Adolph} \emph {et~al.}}]{COMPASS:2014vkj}%
  \BibitemOpen
  \bibfield  {author} {\bibinfo {author} {\bibfnamefont {C.}~\bibnamefont {Adolph}} \emph {et~al.} (\bibinfo {collaboration} {COMPASS}),\ }\bibfield  {title} {\enquote {\bibinfo {title} {{Odd and even partial waves of $\eta\pi^-$ and $\eta'\pi^-$ in $\pi^-p\to\eta^{(\prime)}\pi^-p$ at $191\,\textrm{GeV}/c$}},}\ }\href {\doibase 10.1016/j.physletb.2014.11.058} {\bibfield  {journal} {\bibinfo  {journal} {Phys. Lett. B}\ }\textbf {\bibinfo {volume} {740}},\ \bibinfo {pages} {303--311} (\bibinfo {year} {2015})},\ \bibinfo {note} {[Erratum: Phys.Lett.B 811, 135913 (2020)]},\ \Eprint {http://arxiv.org/abs/1408.4286} {arXiv:1408.4286 [hep-ex]} \BibitemShut {NoStop}%
\bibitem [{\citenamefont {Lu}\ \emph {et~al.}(2005)\citenamefont {Lu} \emph {et~al.}}]{E852:2004rfa}%
  \BibitemOpen
  \bibfield  {author} {\bibinfo {author} {\bibfnamefont {M.}~\bibnamefont {Lu}} \emph {et~al.} (\bibinfo {collaboration} {E852}),\ }\bibfield  {title} {\enquote {\bibinfo {title} {{Exotic Meson Decay to $\omega \pi^0 \pi^-$}},}\ }\href {\doibase 10.1103/PhysRevLett.94.032002} {\bibfield  {journal} {\bibinfo  {journal} {Phys. Rev. Lett.}\ }\textbf {\bibinfo {volume} {94}},\ \bibinfo {pages} {032002} (\bibinfo {year} {2005})},\ \Eprint {http://arxiv.org/abs/hep-ex/0405044} {arXiv:hep-ex/0405044} \BibitemShut {NoStop}%
\bibitem [{\citenamefont {Baker}\ \emph {et~al.}(2003)\citenamefont {Baker} \emph {et~al.}}]{Baker:2003jh}%
  \BibitemOpen
  \bibfield  {author} {\bibinfo {author} {\bibfnamefont {C.~A.}\ \bibnamefont {Baker}} \emph {et~al.},\ }\bibfield  {title} {\enquote {\bibinfo {title} {{Confirmation of $a_0(1450)$ and $\pi_1(1600)$ in $\bar{p} p \to \omega \pi^+ \pi^- \pi^0$ at rest}},}\ }\href {\doibase 10.1016/S0370-2693(03)00643-9} {\bibfield  {journal} {\bibinfo  {journal} {Phys. Lett. B}\ }\textbf {\bibinfo {volume} {563}},\ \bibinfo {pages} {140--149} (\bibinfo {year} {2003})}\BibitemShut {NoStop}%
\bibitem [{\citenamefont {Adams}\ \emph {et~al.}(1998)\citenamefont {Adams} \emph {et~al.}}]{E852:1998mbq}%
  \BibitemOpen
  \bibfield  {author} {\bibinfo {author} {\bibfnamefont {G.~S.}\ \bibnamefont {Adams}} \emph {et~al.} (\bibinfo {collaboration} {E852}),\ }\bibfield  {title} {\enquote {\bibinfo {title} {{Observation of a New $J^{PC} = 1^{-+}$ Exotic State in the Reaction $\pi^- p\to \pi^+ \pi^- \pi^- p$ at $18~\text{GeV}/c$}},}\ }\href {\doibase 10.1103/PhysRevLett.81.5760} {\bibfield  {journal} {\bibinfo  {journal} {Phys. Rev. Lett.}\ }\textbf {\bibinfo {volume} {81}},\ \bibinfo {pages} {5760--5763} (\bibinfo {year} {1998})}\BibitemShut {NoStop}%
\bibitem [{\citenamefont {Chung}\ \emph {et~al.}(2002)\citenamefont {Chung} \emph {et~al.}}]{Chung:2002pu}%
  \BibitemOpen
  \bibfield  {author} {\bibinfo {author} {\bibfnamefont {S.~U.}\ \bibnamefont {Chung}} \emph {et~al.},\ }\bibfield  {title} {\enquote {\bibinfo {title} {{Exotic and $q\bar{q}$ resonances in the $\pi^+ \pi^- \pi^-$ system produced in $\pi^- p$ collisions at $18~\text{GeV}/c$}},}\ }\href {\doibase 10.1103/PhysRevD.65.072001} {\bibfield  {journal} {\bibinfo  {journal} {Phys. Rev. D}\ }\textbf {\bibinfo {volume} {65}},\ \bibinfo {pages} {072001} (\bibinfo {year} {2002})}\BibitemShut {NoStop}%
\bibitem [{\citenamefont {Alekseev}\ \emph {et~al.}(2010)\citenamefont {Alekseev} \emph {et~al.}}]{COMPASS:2009xrl}%
  \BibitemOpen
  \bibfield  {author} {\bibinfo {author} {\bibfnamefont {M.}~\bibnamefont {Alekseev}} \emph {et~al.} (\bibinfo {collaboration} {COMPASS}),\ }\bibfield  {title} {\enquote {\bibinfo {title} {{Observation of a $J^{PC} = 1^{-+}$ exotic resonance in diffractive dissociation of $190~\text{GeV}/c$ $\pi^-$ into $\pi^- \pi^- \pi^+$}},}\ }\href {\doibase 10.1103/PhysRevLett.104.241803} {\bibfield  {journal} {\bibinfo  {journal} {Phys. Rev. Lett.}\ }\textbf {\bibinfo {volume} {104}},\ \bibinfo {pages} {241803} (\bibinfo {year} {2010})},\ \Eprint {http://arxiv.org/abs/0910.5842} {arXiv:0910.5842 [hep-ex]} \BibitemShut {NoStop}%
\bibitem [{\citenamefont {Aghasyan}\ \emph {et~al.}(2018)\citenamefont {Aghasyan} \emph {et~al.}}]{COMPASS:2018uzl}%
  \BibitemOpen
  \bibfield  {author} {\bibinfo {author} {\bibfnamefont {M.}~\bibnamefont {Aghasyan}} \emph {et~al.} (\bibinfo {collaboration} {COMPASS}),\ }\bibfield  {title} {\enquote {\bibinfo {title} {{Light isovector resonances in $\pi^- p \to \pi^-\pi^-\pi^+ p$ at 190 GeV/${\it c}$}},}\ }\href {\doibase 10.1103/PhysRevD.98.092003} {\bibfield  {journal} {\bibinfo  {journal} {Phys. Rev. D}\ }\textbf {\bibinfo {volume} {98}},\ \bibinfo {pages} {092003} (\bibinfo {year} {2018})},\ \Eprint {http://arxiv.org/abs/1802.05913} {arXiv:1802.05913 [hep-ex]} \BibitemShut {NoStop}%
\bibitem [{\citenamefont {Alexeev}\ \emph {et~al.}(2022)\citenamefont {Alexeev} \emph {et~al.}}]{COMPASS:2021ogp}%
  \BibitemOpen
  \bibfield  {author} {\bibinfo {author} {\bibfnamefont {M.~G.}\ \bibnamefont {Alexeev}} \emph {et~al.} (\bibinfo {collaboration} {COMPASS}),\ }\bibfield  {title} {\enquote {\bibinfo {title} {{Exotic meson $\pi_1(1600)$ with $J^{PC} = 1^{-+}$ and its decay into $\rho(770)\pi$}},}\ }\href {\doibase 10.1103/PhysRevD.105.012005} {\bibfield  {journal} {\bibinfo  {journal} {Phys. Rev. D}\ }\textbf {\bibinfo {volume} {105}},\ \bibinfo {pages} {012005} (\bibinfo {year} {2022})},\ \Eprint {http://arxiv.org/abs/2108.01744} {arXiv:2108.01744 [hep-ex]} \BibitemShut {NoStop}%
\bibitem [{\citenamefont {Levinson}\ \emph {et~al.}(1964)\citenamefont {Levinson}, \citenamefont {Lipkin},\ and\ \citenamefont {Meshkov}}]{Levinson:1964xx}%
  \BibitemOpen
  \bibfield  {author} {\bibinfo {author} {\bibfnamefont {C.~A.}\ \bibnamefont {Levinson}}, \bibinfo {author} {\bibfnamefont {H.~J.}\ \bibnamefont {Lipkin}}, \ and\ \bibinfo {author} {\bibfnamefont {S.}~\bibnamefont {Meshkov}},\ }\bibfield  {title} {\enquote {\bibinfo {title} {{A Unitary Symmetry Selection Rule and its Application to New Resonances}},}\ }\href {\doibase 10.1007/BF02726076} {\bibfield  {journal} {\bibinfo  {journal} {Nuovo Cim.}\ }\textbf {\bibinfo {volume} {32}},\ \bibinfo {pages} {1376} (\bibinfo {year} {1964})}\BibitemShut {NoStop}%
\bibitem [{\citenamefont {Close}\ and\ \citenamefont {Lipkin}(1987)}]{Close:1987aw}%
  \BibitemOpen
  \bibfield  {author} {\bibinfo {author} {\bibfnamefont {F.~E.}\ \bibnamefont {Close}}\ and\ \bibinfo {author} {\bibfnamefont {Harry~J.}\ \bibnamefont {Lipkin}},\ }\bibfield  {title} {\enquote {\bibinfo {title} {{New experimental evidence for four quark exotics. the serpukhov $\phi \pi$ resonance and the GAMS $\eta \pi$ enhancement}},}\ }\href {\doibase 10.1016/0370-2693(87)90613-7} {\bibfield  {journal} {\bibinfo  {journal} {Phys. Lett. B}\ }\textbf {\bibinfo {volume} {196}},\ \bibinfo {pages} {245} (\bibinfo {year} {1987})}\BibitemShut {NoStop}%
\bibitem [{\citenamefont {Rodas}\ \emph {et~al.}(2019)\citenamefont {Rodas} \emph {et~al.}}]{JPAC:2018zyd}%
  \BibitemOpen
  \bibfield  {author} {\bibinfo {author} {\bibfnamefont {A.}~\bibnamefont {Rodas}} \emph {et~al.} (\bibinfo {collaboration} {JPAC}),\ }\bibfield  {title} {\enquote {\bibinfo {title} {{Determination of the Pole Position of the Lightest Hybrid Meson Candidate}},}\ }\href {\doibase 10.1103/PhysRevLett.122.042002} {\bibfield  {journal} {\bibinfo  {journal} {Phys. Rev. Lett.}\ }\textbf {\bibinfo {volume} {122}},\ \bibinfo {pages} {042002} (\bibinfo {year} {2019})},\ \Eprint {http://arxiv.org/abs/1810.04171} {arXiv:1810.04171 [hep-ph]} \BibitemShut {NoStop}%
\bibitem [{\citenamefont {Kopf}\ \emph {et~al.}(2021)\citenamefont {Kopf}, \citenamefont {Albrecht}, \citenamefont {Koch}, \citenamefont {K\"u\ss{}ner}, \citenamefont {Pychy}, \citenamefont {Qin},\ and\ \citenamefont {Wiedner}}]{Kopf:2020yoa}%
  \BibitemOpen
  \bibfield  {author} {\bibinfo {author} {\bibfnamefont {B.}~\bibnamefont {Kopf}}, \bibinfo {author} {\bibfnamefont {M.}~\bibnamefont {Albrecht}}, \bibinfo {author} {\bibfnamefont {H.}~\bibnamefont {Koch}}, \bibinfo {author} {\bibfnamefont {M.}~\bibnamefont {K\"u\ss{}ner}}, \bibinfo {author} {\bibfnamefont {J.}~\bibnamefont {Pychy}}, \bibinfo {author} {\bibfnamefont {X.}~\bibnamefont {Qin}}, \ and\ \bibinfo {author} {\bibfnamefont {U.}~\bibnamefont {Wiedner}},\ }\bibfield  {title} {\enquote {\bibinfo {title} {{Investigation of the lightest hybrid meson candidate with a coupled-channel analysis of ${{\bar{p}}p}$-, $\pi ^- p$- and ${\pi \pi }$-Data}},}\ }\href {\doibase 10.1140/epjc/s10052-021-09821-2} {\bibfield  {journal} {\bibinfo  {journal} {Eur. Phys. J. C}\ }\textbf {\bibinfo {volume} {81}},\ \bibinfo {pages} {1056} (\bibinfo {year} {2021})},\ \Eprint {http://arxiv.org/abs/2008.11566} {arXiv:2008.11566 [hep-ph]} \BibitemShut {NoStop}%
\bibitem [{\citenamefont {Workman}\ and\ \citenamefont {Others}(2022)}]{Workman:2022ynf}%
  \BibitemOpen
  \bibfield  {author} {\bibinfo {author} {\bibfnamefont {R.~L.}\ \bibnamefont {Workman}}\ and\ \bibinfo {author} {\bibnamefont {Others}} (\bibinfo {collaboration} {Particle Data Group}),\ }\bibfield  {title} {\enquote {\bibinfo {title} {{Review of Particle Physics}},}\ }\href {\doibase 10.1093/ptep/ptac097} {\bibfield  {journal} {\bibinfo  {journal} {PTEP}\ }\textbf {\bibinfo {volume} {2022}},\ \bibinfo {pages} {083C01} (\bibinfo {year} {2022})}\BibitemShut {NoStop}%
\bibitem [{\citenamefont {Navas}\ \emph {et~al.}(2024)\citenamefont {Navas} \emph {et~al.}}]{ParticleDataGroup:2024cfk}%
  \BibitemOpen
  \bibfield  {author} {\bibinfo {author} {\bibfnamefont {S.}~\bibnamefont {Navas}} \emph {et~al.} (\bibinfo {collaboration} {Particle Data Group}),\ }\bibfield  {title} {\enquote {\bibinfo {title} {{Review of particle physics}},}\ }\href {\doibase 10.1103/PhysRevD.110.030001} {\bibfield  {journal} {\bibinfo  {journal} {Phys. Rev. D}\ }\textbf {\bibinfo {volume} {110}},\ \bibinfo {pages} {030001} (\bibinfo {year} {2024})}\BibitemShut {NoStop}%
\bibitem [{\citenamefont {Ablikim}\ \emph {et~al.}(2022{\natexlab{a}})\citenamefont {Ablikim} \emph {et~al.}}]{BESIII:2022riz}%
  \BibitemOpen
  \bibfield  {author} {\bibinfo {author} {\bibfnamefont {M.}~\bibnamefont {Ablikim}} \emph {et~al.} (\bibinfo {collaboration} {BESIII}),\ }\bibfield  {title} {\enquote {\bibinfo {title} {{Observation of an Isoscalar Resonance with Exotic $J^{PC}=1^{-+}$ Quantum Numbers in $J/\psi\to \gamma\eta\eta'$}},}\ }\href {\doibase 10.1103/PhysRevLett.129.192002} {\bibfield  {journal} {\bibinfo  {journal} {Phys. Rev. Lett.}\ }\textbf {\bibinfo {volume} {129}},\ \bibinfo {pages} {192002} (\bibinfo {year} {2022}{\natexlab{a}})},\ \bibinfo {note} {[Erratum: Phys.Rev.Lett. 130, 159901 (2023)]},\ \Eprint {http://arxiv.org/abs/2202.00621} {arXiv:2202.00621 [hep-ex]} \BibitemShut {NoStop}%
\bibitem [{\citenamefont {Ablikim}\ \emph {et~al.}(2022{\natexlab{b}})\citenamefont {Ablikim} \emph {et~al.}}]{BESIII:2022iwi}%
  \BibitemOpen
  \bibfield  {author} {\bibinfo {author} {\bibfnamefont {M.}~\bibnamefont {Ablikim}} \emph {et~al.} (\bibinfo {collaboration} {BESIII}),\ }\bibfield  {title} {\enquote {\bibinfo {title} {{Partial wave analysis of $J/\psi\to\gamma\eta\eta'$}},}\ }\href {\doibase 10.1103/PhysRevD.106.072012} {\bibfield  {journal} {\bibinfo  {journal} {Phys. Rev. D}\ }\textbf {\bibinfo {volume} {106}},\ \bibinfo {pages} {072012} (\bibinfo {year} {2022}{\natexlab{b}})},\ \bibinfo {note} {[Erratum: Phys.Rev.D 107, 079901 (2023)]},\ \Eprint {http://arxiv.org/abs/2202.00623} {arXiv:2202.00623 [hep-ex]} \BibitemShut {NoStop}%
\bibitem [{\citenamefont {Barnes}\ \emph {et~al.}(1983)\citenamefont {Barnes}, \citenamefont {Close},\ and\ \citenamefont {de~Viron}}]{Barnes:1982tx}%
  \BibitemOpen
  \bibfield  {author} {\bibinfo {author} {\bibfnamefont {Ted}\ \bibnamefont {Barnes}}, \bibinfo {author} {\bibfnamefont {F.~E.}\ \bibnamefont {Close}}, \ and\ \bibinfo {author} {\bibfnamefont {F.}~\bibnamefont {de~Viron}},\ }\bibfield  {title} {\enquote {\bibinfo {title} {{Q anti-Q G Hermaphrodite Mesons in the MIT Bag Model}},}\ }\href {\doibase 10.1016/0550-3213(83)90004-4} {\bibfield  {journal} {\bibinfo  {journal} {Nucl. Phys. B}\ }\textbf {\bibinfo {volume} {224}},\ \bibinfo {pages} {241} (\bibinfo {year} {1983})}\BibitemShut {NoStop}%
\bibitem [{\citenamefont {Chanowitz}\ and\ \citenamefont {Sharpe}(1983)}]{Chanowitz:1982qj}%
  \BibitemOpen
  \bibfield  {author} {\bibinfo {author} {\bibfnamefont {Michael~S.}\ \bibnamefont {Chanowitz}}\ and\ \bibinfo {author} {\bibfnamefont {Stephen~R.}\ \bibnamefont {Sharpe}},\ }\bibfield  {title} {\enquote {\bibinfo {title} {{Hybrids: Mixed states of quarks and gluons}},}\ }\href {\doibase 10.1016/0550-3213(83)90635-1} {\bibfield  {journal} {\bibinfo  {journal} {Nucl. Phys. B}\ }\textbf {\bibinfo {volume} {222}},\ \bibinfo {pages} {211--244} (\bibinfo {year} {1983})},\ \bibinfo {note} {[Erratum: Nucl.Phys.B 228, 588--588 (1983)]}\BibitemShut {NoStop}%
\bibitem [{\citenamefont {Horn}\ and\ \citenamefont {Mandula}(1978)}]{Horn:1977rq}%
  \BibitemOpen
  \bibfield  {author} {\bibinfo {author} {\bibfnamefont {D.}~\bibnamefont {Horn}}\ and\ \bibinfo {author} {\bibfnamefont {J.}~\bibnamefont {Mandula}},\ }\bibfield  {title} {\enquote {\bibinfo {title} {{A model of mesons with constituent gluons}},}\ }\href {\doibase 10.1103/PhysRevD.17.898} {\bibfield  {journal} {\bibinfo  {journal} {Phys. Rev. D}\ }\textbf {\bibinfo {volume} {17}},\ \bibinfo {pages} {898} (\bibinfo {year} {1978})}\BibitemShut {NoStop}%
\bibitem [{\citenamefont {Ishida}\ \emph {et~al.}(1993)\citenamefont {Ishida}, \citenamefont {Sawazaki}, \citenamefont {Oda},\ and\ \citenamefont {Yamada}}]{Ishida:1991mx}%
  \BibitemOpen
  \bibfield  {author} {\bibinfo {author} {\bibfnamefont {Shin}\ \bibnamefont {Ishida}}, \bibinfo {author} {\bibfnamefont {Haruhiko}\ \bibnamefont {Sawazaki}}, \bibinfo {author} {\bibfnamefont {Masuho}\ \bibnamefont {Oda}}, \ and\ \bibinfo {author} {\bibfnamefont {Kenji}\ \bibnamefont {Yamada}},\ }\bibfield  {title} {\enquote {\bibinfo {title} {{Decay properties of hybrid mesons with a massive constituent gluon and search for their candidates}},}\ }\href {\doibase 10.1103/PhysRevD.47.179} {\bibfield  {journal} {\bibinfo  {journal} {Phys. Rev. D}\ }\textbf {\bibinfo {volume} {47}},\ \bibinfo {pages} {179--198} (\bibinfo {year} {1993})}\BibitemShut {NoStop}%
\bibitem [{\citenamefont {Balitsky}\ \emph {et~al.}(1986)\citenamefont {Balitsky}, \citenamefont {Diakonov},\ and\ \citenamefont {Yung}}]{Balitsky:1986hf}%
  \BibitemOpen
  \bibfield  {author} {\bibinfo {author} {\bibfnamefont {I.~I.}\ \bibnamefont {Balitsky}}, \bibinfo {author} {\bibfnamefont {Dmitri}\ \bibnamefont {Diakonov}}, \ and\ \bibinfo {author} {\bibfnamefont {A.~V.}\ \bibnamefont {Yung}},\ }\bibfield  {title} {\enquote {\bibinfo {title} {{Exotic mesons with $J^{PC}=1^{-+}$, strange and nonstrange}},}\ }\href {\doibase 10.1007/BF01411145} {\bibfield  {journal} {\bibinfo  {journal} {Z. Phys. C}\ }\textbf {\bibinfo {volume} {33}},\ \bibinfo {pages} {265--273} (\bibinfo {year} {1986})}\BibitemShut {NoStop}%
\bibitem [{\citenamefont {Latorre}\ \emph {et~al.}(1987)\citenamefont {Latorre}, \citenamefont {Pascual},\ and\ \citenamefont {Narison}}]{Latorre:1985tg}%
  \BibitemOpen
  \bibfield  {author} {\bibinfo {author} {\bibfnamefont {J.~I.}\ \bibnamefont {Latorre}}, \bibinfo {author} {\bibfnamefont {P.}~\bibnamefont {Pascual}}, \ and\ \bibinfo {author} {\bibfnamefont {Stephan}\ \bibnamefont {Narison}},\ }\bibfield  {title} {\enquote {\bibinfo {title} {{Spectra and hadronic couplings of light hermaphrodite mesons}},}\ }\href {\doibase 10.1007/BF01548817} {\bibfield  {journal} {\bibinfo  {journal} {Z. Phys. C}\ }\textbf {\bibinfo {volume} {34}},\ \bibinfo {pages} {347} (\bibinfo {year} {1987})}\BibitemShut {NoStop}%
\bibitem [{\citenamefont {Govaerts}\ \emph {et~al.}(1987)\citenamefont {Govaerts}, \citenamefont {Reinders}, \citenamefont {Francken}, \citenamefont {Gonze},\ and\ \citenamefont {Weyers}}]{Govaerts:1986pp}%
  \BibitemOpen
  \bibfield  {author} {\bibinfo {author} {\bibfnamefont {J.}~\bibnamefont {Govaerts}}, \bibinfo {author} {\bibfnamefont {L.~J.}\ \bibnamefont {Reinders}}, \bibinfo {author} {\bibfnamefont {P.}~\bibnamefont {Francken}}, \bibinfo {author} {\bibfnamefont {X.}~\bibnamefont {Gonze}}, \ and\ \bibinfo {author} {\bibfnamefont {J.}~\bibnamefont {Weyers}},\ }\bibfield  {title} {\enquote {\bibinfo {title} {{Coupled {QCD} sum rules for hybrid mesons}},}\ }\href {\doibase 10.1016/0550-3213(87)90056-3} {\bibfield  {journal} {\bibinfo  {journal} {Nucl. Phys. B}\ }\textbf {\bibinfo {volume} {284}},\ \bibinfo {pages} {674} (\bibinfo {year} {1987})}\BibitemShut {NoStop}%
\bibitem [{\citenamefont {Isgur}\ \emph {et~al.}(1985)\citenamefont {Isgur}, \citenamefont {Kokoski},\ and\ \citenamefont {Paton}}]{Isgur:1985vy}%
  \BibitemOpen
  \bibfield  {author} {\bibinfo {author} {\bibfnamefont {Nathan}\ \bibnamefont {Isgur}}, \bibinfo {author} {\bibfnamefont {Richard}\ \bibnamefont {Kokoski}}, \ and\ \bibinfo {author} {\bibfnamefont {Jack}\ \bibnamefont {Paton}},\ }\bibfield  {title} {\enquote {\bibinfo {title} {{Gluonic Excitations of Mesons: Why They Are Missing and Where to Find Them}},}\ }\href {\doibase 10.1103/PhysRevLett.54.869} {\bibfield  {journal} {\bibinfo  {journal} {Phys. Rev. Lett.}\ }\textbf {\bibinfo {volume} {54}},\ \bibinfo {pages} {869} (\bibinfo {year} {1985})}\BibitemShut {NoStop}%
\bibitem [{\citenamefont {Close}\ and\ \citenamefont {Page}(1995)}]{Close:1994hc}%
  \BibitemOpen
  \bibfield  {author} {\bibinfo {author} {\bibfnamefont {Frank~E.}\ \bibnamefont {Close}}\ and\ \bibinfo {author} {\bibfnamefont {Philip~R.}\ \bibnamefont {Page}},\ }\bibfield  {title} {\enquote {\bibinfo {title} {{The Production and decay of hybrid mesons by flux tube breaking}},}\ }\href {\doibase 10.1016/0550-3213(95)00085-7} {\bibfield  {journal} {\bibinfo  {journal} {Nucl. Phys. B}\ }\textbf {\bibinfo {volume} {443}},\ \bibinfo {pages} {233--254} (\bibinfo {year} {1995})},\ \Eprint {http://arxiv.org/abs/hep-ph/9411301} {arXiv:hep-ph/9411301} \BibitemShut {NoStop}%
\bibitem [{\citenamefont {Lacock}\ \emph {et~al.}(1997)\citenamefont {Lacock}, \citenamefont {Michael}, \citenamefont {Boyle},\ and\ \citenamefont {Rowland}}]{Lacock:1996ny}%
  \BibitemOpen
  \bibfield  {author} {\bibinfo {author} {\bibfnamefont {P.}~\bibnamefont {Lacock}}, \bibinfo {author} {\bibfnamefont {Christopher}\ \bibnamefont {Michael}}, \bibinfo {author} {\bibfnamefont {P.}~\bibnamefont {Boyle}}, \ and\ \bibinfo {author} {\bibfnamefont {P.}~\bibnamefont {Rowland}} (\bibinfo {collaboration} {UKQCD}),\ }\bibfield  {title} {\enquote {\bibinfo {title} {{Hybrid mesons from quenched QCD}},}\ }\href {\doibase 10.1016/S0370-2693(97)00384-5} {\bibfield  {journal} {\bibinfo  {journal} {Phys. Lett. B}\ }\textbf {\bibinfo {volume} {401}},\ \bibinfo {pages} {308--312} (\bibinfo {year} {1997})},\ \Eprint {http://arxiv.org/abs/hep-lat/9611011} {arXiv:hep-lat/9611011} \BibitemShut {NoStop}%
\bibitem [{\citenamefont {Bernard}\ \emph {et~al.}(1997)\citenamefont {Bernard} \emph {et~al.}}]{MILC:1997usn}%
  \BibitemOpen
  \bibfield  {author} {\bibinfo {author} {\bibfnamefont {Claude~W.}\ \bibnamefont {Bernard}} \emph {et~al.} (\bibinfo {collaboration} {MILC}),\ }\bibfield  {title} {\enquote {\bibinfo {title} {{Exotic mesons in quenched lattice QCD}},}\ }\href {\doibase 10.1103/PhysRevD.56.7039} {\bibfield  {journal} {\bibinfo  {journal} {Phys. Rev. D}\ }\textbf {\bibinfo {volume} {56}},\ \bibinfo {pages} {7039--7051} (\bibinfo {year} {1997})},\ \Eprint {http://arxiv.org/abs/hep-lat/9707008} {arXiv:hep-lat/9707008} \BibitemShut {NoStop}%
\bibitem [{\citenamefont {Mei}\ and\ \citenamefont {Luo}(2003)}]{Mei:2002ip}%
  \BibitemOpen
  \bibfield  {author} {\bibinfo {author} {\bibfnamefont {Zhong-Hao}\ \bibnamefont {Mei}}\ and\ \bibinfo {author} {\bibfnamefont {Xiang-Qian}\ \bibnamefont {Luo}},\ }\bibfield  {title} {\enquote {\bibinfo {title} {{Exotic mesons from quantum chromodynamics with improved gluon and quark actions on the anisotropic lattice}},}\ }\href {\doibase 10.1142/S0217751X03017038} {\bibfield  {journal} {\bibinfo  {journal} {Int. J. Mod. Phys. A}\ }\textbf {\bibinfo {volume} {18}},\ \bibinfo {pages} {5713} (\bibinfo {year} {2003})},\ \Eprint {http://arxiv.org/abs/hep-lat/0206012} {arXiv:hep-lat/0206012} \BibitemShut {NoStop}%
\bibitem [{\citenamefont {Bernard}\ \emph {et~al.}(2003)\citenamefont {Bernard}, \citenamefont {Burch}, \citenamefont {Gregory}, \citenamefont {Toussaint}, \citenamefont {DeTar}, \citenamefont {Osborn}, \citenamefont {Gottlieb}, \citenamefont {Heller},\ and\ \citenamefont {Sugar}}]{Bernard:2003jd}%
  \BibitemOpen
  \bibfield  {author} {\bibinfo {author} {\bibfnamefont {C.}~\bibnamefont {Bernard}}, \bibinfo {author} {\bibfnamefont {T.}~\bibnamefont {Burch}}, \bibinfo {author} {\bibfnamefont {E.~B.}\ \bibnamefont {Gregory}}, \bibinfo {author} {\bibfnamefont {D.}~\bibnamefont {Toussaint}}, \bibinfo {author} {\bibfnamefont {Carleton~E.}\ \bibnamefont {DeTar}}, \bibinfo {author} {\bibfnamefont {J.}~\bibnamefont {Osborn}}, \bibinfo {author} {\bibfnamefont {Steven~A.}\ \bibnamefont {Gottlieb}}, \bibinfo {author} {\bibfnamefont {U.~M.}\ \bibnamefont {Heller}}, \ and\ \bibinfo {author} {\bibfnamefont {R.}~\bibnamefont {Sugar}},\ }\bibfield  {title} {\enquote {\bibinfo {title} {{Lattice calculation of $1^{-+}$ hybrid mesons with improved Kogut-Susskind fermions}},}\ }\href {\doibase 10.1103/PhysRevD.68.074505} {\bibfield  {journal} {\bibinfo  {journal} {Phys. Rev. D}\ }\textbf {\bibinfo {volume} {68}},\ \bibinfo {pages} {074505} (\bibinfo {year} {2003})},\ \Eprint {http://arxiv.org/abs/hep-lat/0301024} {arXiv:hep-lat/0301024} \BibitemShut {NoStop}%
\bibitem [{\citenamefont {Hedditch}\ \emph {et~al.}(2005)\citenamefont {Hedditch}, \citenamefont {Kamleh}, \citenamefont {Lasscock}, \citenamefont {Leinweber}, \citenamefont {Williams},\ and\ \citenamefont {Zanotti}}]{Hedditch:2005zf}%
  \BibitemOpen
  \bibfield  {author} {\bibinfo {author} {\bibfnamefont {J.~N.}\ \bibnamefont {Hedditch}}, \bibinfo {author} {\bibfnamefont {W.}~\bibnamefont {Kamleh}}, \bibinfo {author} {\bibfnamefont {B.~G.}\ \bibnamefont {Lasscock}}, \bibinfo {author} {\bibfnamefont {D.~B.}\ \bibnamefont {Leinweber}}, \bibinfo {author} {\bibfnamefont {A.~G.}\ \bibnamefont {Williams}}, \ and\ \bibinfo {author} {\bibfnamefont {J.~M.}\ \bibnamefont {Zanotti}},\ }\bibfield  {title} {\enquote {\bibinfo {title} {{$1^{-+}$ exotic meson at light quark masses}},}\ }\href {\doibase 10.1103/PhysRevD.72.114507} {\bibfield  {journal} {\bibinfo  {journal} {Phys. Rev. D}\ }\textbf {\bibinfo {volume} {72}},\ \bibinfo {pages} {114507} (\bibinfo {year} {2005})},\ \Eprint {http://arxiv.org/abs/hep-lat/0509106} {arXiv:hep-lat/0509106} \BibitemShut {NoStop}%
\bibitem [{\citenamefont {McNeile}\ and\ \citenamefont {Michael}(2006{\natexlab{a}})}]{McNeile:2006bz}%
  \BibitemOpen
  \bibfield  {author} {\bibinfo {author} {\bibfnamefont {C.}~\bibnamefont {McNeile}}\ and\ \bibinfo {author} {\bibfnamefont {Christopher}\ \bibnamefont {Michael}} (\bibinfo {collaboration} {UKQCD}),\ }\bibfield  {title} {\enquote {\bibinfo {title} {{Decay width of light quark hybrid meson from the lattice}},}\ }\href {\doibase 10.1103/PhysRevD.73.074506} {\bibfield  {journal} {\bibinfo  {journal} {Phys. Rev. D}\ }\textbf {\bibinfo {volume} {73}},\ \bibinfo {pages} {074506} (\bibinfo {year} {2006}{\natexlab{a}})},\ \Eprint {http://arxiv.org/abs/hep-lat/0603007} {arXiv:hep-lat/0603007} \BibitemShut {NoStop}%
\bibitem [{\citenamefont {Dudek}\ \emph {et~al.}(2013)\citenamefont {Dudek}, \citenamefont {Edwards}, \citenamefont {Guo},\ and\ \citenamefont {Thomas}}]{Dudek:2013yja}%
  \BibitemOpen
  \bibfield  {author} {\bibinfo {author} {\bibfnamefont {Jozef~J.}\ \bibnamefont {Dudek}}, \bibinfo {author} {\bibfnamefont {Robert~G.}\ \bibnamefont {Edwards}}, \bibinfo {author} {\bibfnamefont {Peng}\ \bibnamefont {Guo}}, \ and\ \bibinfo {author} {\bibfnamefont {Christopher~E.}\ \bibnamefont {Thomas}} (\bibinfo {collaboration} {Hadron Spectrum}),\ }\bibfield  {title} {\enquote {\bibinfo {title} {{Toward the excited isoscalar meson spectrum from lattice QCD}},}\ }\href {\doibase 10.1103/PhysRevD.88.094505} {\bibfield  {journal} {\bibinfo  {journal} {Phys. Rev. D}\ }\textbf {\bibinfo {volume} {88}},\ \bibinfo {pages} {094505} (\bibinfo {year} {2013})},\ \Eprint {http://arxiv.org/abs/1309.2608} {arXiv:1309.2608 [hep-lat]} \BibitemShut {NoStop}%
\bibitem [{\citenamefont {Woss}\ \emph {et~al.}(2021)\citenamefont {Woss}, \citenamefont {Dudek}, \citenamefont {Edwards}, \citenamefont {Thomas},\ and\ \citenamefont {Wilson}}]{Woss:2020ayi}%
  \BibitemOpen
  \bibfield  {author} {\bibinfo {author} {\bibfnamefont {Antoni~J.}\ \bibnamefont {Woss}}, \bibinfo {author} {\bibfnamefont {Jozef~J.}\ \bibnamefont {Dudek}}, \bibinfo {author} {\bibfnamefont {Robert~G.}\ \bibnamefont {Edwards}}, \bibinfo {author} {\bibfnamefont {Christopher~E.}\ \bibnamefont {Thomas}}, \ and\ \bibinfo {author} {\bibfnamefont {David~J.}\ \bibnamefont {Wilson}} (\bibinfo {collaboration} {Hadron Spectrum}),\ }\bibfield  {title} {\enquote {\bibinfo {title} {{Decays of an exotic $1^{-+}$ hybrid meson resonance in QCD}},}\ }\href {\doibase 10.1103/PhysRevD.103.054502} {\bibfield  {journal} {\bibinfo  {journal} {Phys. Rev. D}\ }\textbf {\bibinfo {volume} {103}},\ \bibinfo {pages} {054502} (\bibinfo {year} {2021})},\ \Eprint {http://arxiv.org/abs/2009.10034} {arXiv:2009.10034 [hep-lat]} \BibitemShut {NoStop}%
\bibitem [{\citenamefont {Chen}\ \emph {et~al.}(2023{\natexlab{a}})\citenamefont {Chen}, \citenamefont {Jiang}, \citenamefont {Chen}, \citenamefont {Gong}, \citenamefont {Liu}, \citenamefont {Shi},\ and\ \citenamefont {Sun}}]{Chen:2022isv}%
  \BibitemOpen
  \bibfield  {author} {\bibinfo {author} {\bibfnamefont {Feiyu}\ \bibnamefont {Chen}}, \bibinfo {author} {\bibfnamefont {Xiangyu}\ \bibnamefont {Jiang}}, \bibinfo {author} {\bibfnamefont {Ying}\ \bibnamefont {Chen}}, \bibinfo {author} {\bibfnamefont {Ming}\ \bibnamefont {Gong}}, \bibinfo {author} {\bibfnamefont {Zhaofeng}\ \bibnamefont {Liu}}, \bibinfo {author} {\bibfnamefont {Chunjiang}\ \bibnamefont {Shi}}, \ and\ \bibinfo {author} {\bibfnamefont {Wei}\ \bibnamefont {Sun}},\ }\bibfield  {title} {\enquote {\bibinfo {title} {{$1^{-+}$ hybrid meson in $J/\psi$ radiative decays from lattice QCD}},}\ }\href {\doibase 10.1103/PhysRevD.107.054511} {\bibfield  {journal} {\bibinfo  {journal} {Phys. Rev. D}\ }\textbf {\bibinfo {volume} {107}},\ \bibinfo {pages} {054511} (\bibinfo {year} {2023}{\natexlab{a}})},\ \Eprint {http://arxiv.org/abs/2207.04694} {arXiv:2207.04694 [hep-lat]} \BibitemShut {NoStop}%
\bibitem [{\citenamefont {Ackleh}\ \emph {et~al.}(1996)\citenamefont {Ackleh}, \citenamefont {Barnes},\ and\ \citenamefont {Swanson}}]{Ackleh:1996yt}%
  \BibitemOpen
  \bibfield  {author} {\bibinfo {author} {\bibfnamefont {E.~S.}\ \bibnamefont {Ackleh}}, \bibinfo {author} {\bibfnamefont {Ted}\ \bibnamefont {Barnes}}, \ and\ \bibinfo {author} {\bibfnamefont {E.~S.}\ \bibnamefont {Swanson}},\ }\bibfield  {title} {\enquote {\bibinfo {title} {{On the mechanism of open flavor strong decays}},}\ }\href {\doibase 10.1103/PhysRevD.54.6811} {\bibfield  {journal} {\bibinfo  {journal} {Phys. Rev. D}\ }\textbf {\bibinfo {volume} {54}},\ \bibinfo {pages} {6811--6829} (\bibinfo {year} {1996})},\ \Eprint {http://arxiv.org/abs/hep-ph/9604355} {arXiv:hep-ph/9604355} \BibitemShut {NoStop}%
\bibitem [{\citenamefont {Barnes}\ \emph {et~al.}(1997)\citenamefont {Barnes}, \citenamefont {Close}, \citenamefont {Page},\ and\ \citenamefont {Swanson}}]{Barnes:1996ff}%
  \BibitemOpen
  \bibfield  {author} {\bibinfo {author} {\bibfnamefont {Ted}\ \bibnamefont {Barnes}}, \bibinfo {author} {\bibfnamefont {F.~E.}\ \bibnamefont {Close}}, \bibinfo {author} {\bibfnamefont {P.~R.}\ \bibnamefont {Page}}, \ and\ \bibinfo {author} {\bibfnamefont {E.~S.}\ \bibnamefont {Swanson}},\ }\bibfield  {title} {\enquote {\bibinfo {title} {{Higher quarkonia}},}\ }\href {\doibase 10.1103/PhysRevD.55.4157} {\bibfield  {journal} {\bibinfo  {journal} {Phys. Rev. D}\ }\textbf {\bibinfo {volume} {55}},\ \bibinfo {pages} {4157--4188} (\bibinfo {year} {1997})},\ \Eprint {http://arxiv.org/abs/hep-ph/9609339} {arXiv:hep-ph/9609339} \BibitemShut {NoStop}%
\bibitem [{\citenamefont {Ivanov}\ \emph {et~al.}(2001)\citenamefont {Ivanov} \emph {et~al.}}]{E852:2001ikk}%
  \BibitemOpen
  \bibfield  {author} {\bibinfo {author} {\bibfnamefont {E.~I.}\ \bibnamefont {Ivanov}} \emph {et~al.} (\bibinfo {collaboration} {E852}),\ }\bibfield  {title} {\enquote {\bibinfo {title} {{Observation of exotic meson production in the reaction $\pi^- p \to \eta' \pi^- p$ at $18~\text{GeV} / c$}},}\ }\href {\doibase 10.1103/PhysRevLett.86.3977} {\bibfield  {journal} {\bibinfo  {journal} {Phys. Rev. Lett.}\ }\textbf {\bibinfo {volume} {86}},\ \bibinfo {pages} {3977--3980} (\bibinfo {year} {2001})},\ \Eprint {http://arxiv.org/abs/hep-ex/0101058} {arXiv:hep-ex/0101058} \BibitemShut {NoStop}%
\bibitem [{\citenamefont {Chen}\ \emph {et~al.}(2022)\citenamefont {Chen}, \citenamefont {Su},\ and\ \citenamefont {Zhu}}]{Chen:2022qpd}%
  \BibitemOpen
  \bibfield  {author} {\bibinfo {author} {\bibfnamefont {Hua-Xing}\ \bibnamefont {Chen}}, \bibinfo {author} {\bibfnamefont {Niu}\ \bibnamefont {Su}}, \ and\ \bibinfo {author} {\bibfnamefont {Shi-Lin}\ \bibnamefont {Zhu}},\ }\bibfield  {title} {\enquote {\bibinfo {title} {{QCD Axial Anomaly Enhances the $\eta\eta'$ Decay of the Hybrid Candidate $\eta_{1}(1855)$}},}\ }\href {\doibase 10.1088/0256-307X/39/5/051201} {\bibfield  {journal} {\bibinfo  {journal} {Chin. Phys. Lett.}\ }\textbf {\bibinfo {volume} {39}},\ \bibinfo {pages} {051201} (\bibinfo {year} {2022})},\ \Eprint {http://arxiv.org/abs/2202.04918} {arXiv:2202.04918 [hep-ph]} \BibitemShut {NoStop}%
\bibitem [{\citenamefont {Qiu}\ and\ \citenamefont {Zhao}(2022)}]{Qiu:2022ktc}%
  \BibitemOpen
  \bibfield  {author} {\bibinfo {author} {\bibfnamefont {Lin}\ \bibnamefont {Qiu}}\ and\ \bibinfo {author} {\bibfnamefont {Qiang}\ \bibnamefont {Zhao}},\ }\bibfield  {title} {\enquote {\bibinfo {title} {{Towards the establishment of the light $J^{P(C)}=1^{–(+)}$ hybrid nonet}},}\ }\href {\doibase 10.1088/1674-1137/ac567e} {\bibfield  {journal} {\bibinfo  {journal} {Chin. Phys. C}\ }\textbf {\bibinfo {volume} {46}},\ \bibinfo {pages} {051001} (\bibinfo {year} {2022})},\ \Eprint {http://arxiv.org/abs/2202.00904} {arXiv:2202.00904 [hep-ph]} \BibitemShut {NoStop}%
\bibitem [{\citenamefont {Shastry}\ \emph {et~al.}(2022)\citenamefont {Shastry}, \citenamefont {Fischer},\ and\ \citenamefont {Giacosa}}]{Shastry:2022mhk}%
  \BibitemOpen
  \bibfield  {author} {\bibinfo {author} {\bibfnamefont {Vanamali}\ \bibnamefont {Shastry}}, \bibinfo {author} {\bibfnamefont {Christian~S.}\ \bibnamefont {Fischer}}, \ and\ \bibinfo {author} {\bibfnamefont {Francesco}\ \bibnamefont {Giacosa}},\ }\bibfield  {title} {\enquote {\bibinfo {title} {{The phenomenology of the exotic hybrid nonet with $\pi_1(1600)$ and $\eta_1(1855)$}},}\ }\href {\doibase 10.1016/j.physletb.2022.137478} {\  (\bibinfo {year} {2022}),\ 10.1016/j.physletb.2022.137478},\ \Eprint {http://arxiv.org/abs/2203.04327} {arXiv:2203.04327 [hep-ph]} \BibitemShut {NoStop}%
\bibitem [{\citenamefont {Wang}\ \emph {et~al.}(2022)\citenamefont {Wang}, \citenamefont {Zeng},\ and\ \citenamefont {Liu}}]{Wang:2022sib}%
  \BibitemOpen
  \bibfield  {author} {\bibinfo {author} {\bibfnamefont {Xiao-Yun}\ \bibnamefont {Wang}}, \bibinfo {author} {\bibfnamefont {Fan-Cong}\ \bibnamefont {Zeng}}, \ and\ \bibinfo {author} {\bibfnamefont {Xiang}\ \bibnamefont {Liu}},\ }\bibfield  {title} {\enquote {\bibinfo {title} {{Production of the $\eta_1(1855)$ through kaon induced reactions under the assumptions that it is a molecular or a hybrid state}},}\ }\href {\doibase 10.1103/PhysRevD.106.036005} {\bibfield  {journal} {\bibinfo  {journal} {Phys. Rev. D}\ }\textbf {\bibinfo {volume} {106}},\ \bibinfo {pages} {036005} (\bibinfo {year} {2022})},\ \Eprint {http://arxiv.org/abs/2205.09283} {arXiv:2205.09283 [hep-ph]} \BibitemShut {NoStop}%
\bibitem [{\citenamefont {Swanson}(2023)}]{Swanson:2023zlm}%
  \BibitemOpen
  \bibfield  {author} {\bibinfo {author} {\bibfnamefont {E.~S.}\ \bibnamefont {Swanson}},\ }\bibfield  {title} {\enquote {\bibinfo {title} {{Light hybrid meson mixing and phenomenology}},}\ }\href {\doibase 10.1103/PhysRevD.107.074028} {\bibfield  {journal} {\bibinfo  {journal} {Phys. Rev. D}\ }\textbf {\bibinfo {volume} {107}},\ \bibinfo {pages} {074028} (\bibinfo {year} {2023})},\ \Eprint {http://arxiv.org/abs/2302.01372} {arXiv:2302.01372 [hep-ph]} \BibitemShut {NoStop}%
\bibitem [{\citenamefont {Chen}\ \emph {et~al.}(2023{\natexlab{b}})\citenamefont {Chen}, \citenamefont {Luo},\ and\ \citenamefont {Liu}}]{Chen:2023ukh}%
  \BibitemOpen
  \bibfield  {author} {\bibinfo {author} {\bibfnamefont {Bing}\ \bibnamefont {Chen}}, \bibinfo {author} {\bibfnamefont {Si-Qiang}\ \bibnamefont {Luo}}, \ and\ \bibinfo {author} {\bibfnamefont {Xiang}\ \bibnamefont {Liu}},\ }\bibfield  {title} {\enquote {\bibinfo {title} {{Constructing the $J^{P(C)}=1^{-(+)}$ light flavor hybrid nonet with the newly observed $\eta_1(1855)$}},}\ }\href {\doibase 10.1103/PhysRevD.108.054034} {\bibfield  {journal} {\bibinfo  {journal} {Phys. Rev. D}\ }\textbf {\bibinfo {volume} {108}},\ \bibinfo {pages} {054034} (\bibinfo {year} {2023}{\natexlab{b}})},\ \Eprint {http://arxiv.org/abs/2302.06785} {arXiv:2302.06785 [hep-ph]} \BibitemShut {NoStop}%
\bibitem [{\citenamefont {Shastry}\ and\ \citenamefont {Giacosa}(2023)}]{Shastry:2023ths}%
  \BibitemOpen
  \bibfield  {author} {\bibinfo {author} {\bibfnamefont {Vanamali}\ \bibnamefont {Shastry}}\ and\ \bibinfo {author} {\bibfnamefont {Francesco}\ \bibnamefont {Giacosa}},\ }\bibfield  {title} {\enquote {\bibinfo {title} {{Radiative production and decays of the exotic $\eta_1'(1855)$ and its siblings}},}\ }\href {\doibase 10.1016/j.nuclphysa.2023.122683} {\bibfield  {journal} {\bibinfo  {journal} {Nucl. Phys. A}\ }\textbf {\bibinfo {volume} {1037}},\ \bibinfo {pages} {122683} (\bibinfo {year} {2023})},\ \Eprint {http://arxiv.org/abs/2302.07687} {arXiv:2302.07687 [hep-ph]} \BibitemShut {NoStop}%
\bibitem [{\citenamefont {Farina}\ and\ \citenamefont {Swanson}(2024)}]{Farina:2023oqk}%
  \BibitemOpen
  \bibfield  {author} {\bibinfo {author} {\bibfnamefont {Christian}\ \bibnamefont {Farina}}\ and\ \bibinfo {author} {\bibfnamefont {Eric~S.}\ \bibnamefont {Swanson}},\ }\bibfield  {title} {\enquote {\bibinfo {title} {{Constituent model of light hybrid meson decays}},}\ }\href {\doibase 10.1103/PhysRevD.109.094015} {\bibfield  {journal} {\bibinfo  {journal} {Phys. Rev. D}\ }\textbf {\bibinfo {volume} {109}},\ \bibinfo {pages} {094015} (\bibinfo {year} {2024})},\ \Eprint {http://arxiv.org/abs/2312.05370} {arXiv:2312.05370 [hep-ph]} \BibitemShut {NoStop}%
\bibitem [{\citenamefont {Barsbay}\ \emph {et~al.}(2024)\citenamefont {Barsbay}, \citenamefont {Azizi},\ and\ \citenamefont {Sundu}}]{Barsbay:2024vjt}%
  \BibitemOpen
  \bibfield  {author} {\bibinfo {author} {\bibfnamefont {B.}~\bibnamefont {Barsbay}}, \bibinfo {author} {\bibfnamefont {K.}~\bibnamefont {Azizi}}, \ and\ \bibinfo {author} {\bibfnamefont {H.}~\bibnamefont {Sundu}},\ }\bibfield  {title} {\enquote {\bibinfo {title} {{Light quarkonium hybrid mesons}},}\ }\href {\doibase 10.1103/PhysRevD.109.094034} {\bibfield  {journal} {\bibinfo  {journal} {Phys. Rev. D}\ }\textbf {\bibinfo {volume} {109}},\ \bibinfo {pages} {094034} (\bibinfo {year} {2024})},\ \Eprint {http://arxiv.org/abs/2402.19006} {arXiv:2402.19006 [hep-ph]} \BibitemShut {NoStop}%
\bibitem [{\citenamefont {Tan}\ \emph {et~al.}(2024)\citenamefont {Tan}, \citenamefont {Su},\ and\ \citenamefont {Chen}}]{Tan:2024grd}%
  \BibitemOpen
  \bibfield  {author} {\bibinfo {author} {\bibfnamefont {Wei-Han}\ \bibnamefont {Tan}}, \bibinfo {author} {\bibfnamefont {Niu}\ \bibnamefont {Su}}, \ and\ \bibinfo {author} {\bibfnamefont {Hua-Xing}\ \bibnamefont {Chen}},\ }\bibfield  {title} {\enquote {\bibinfo {title} {{Light single-gluon hybrid states with various exotic quantum numbers}},}\ }\href {\doibase 10.1103/PhysRevD.110.034031} {\bibfield  {journal} {\bibinfo  {journal} {Phys. Rev. D}\ }\textbf {\bibinfo {volume} {110}},\ \bibinfo {pages} {034031} (\bibinfo {year} {2024})},\ \Eprint {http://arxiv.org/abs/2404.09538} {arXiv:2404.09538 [hep-ph]} \BibitemShut {NoStop}%
\bibitem [{\citenamefont {Giacosa}\ \emph {et~al.}(2024)\citenamefont {Giacosa}, \citenamefont {Kov\'acs},\ and\ \citenamefont {Jafarzade}}]{Giacosa:2024epf}%
  \BibitemOpen
  \bibfield  {author} {\bibinfo {author} {\bibfnamefont {Francesco}\ \bibnamefont {Giacosa}}, \bibinfo {author} {\bibfnamefont {P\'eter}\ \bibnamefont {Kov\'acs}}, \ and\ \bibinfo {author} {\bibfnamefont {Shahriyar}\ \bibnamefont {Jafarzade}},\ }\bibfield  {title} {\enquote {\bibinfo {title} {{Ordinary and exotic mesons in the extended Linear Sigma Model}},}\ }\href@noop {} {\  (\bibinfo {year} {2024})},\ \Eprint {http://arxiv.org/abs/2407.18348} {arXiv:2407.18348 [hep-ph]} \BibitemShut {NoStop}%
\bibitem [{\citenamefont {Dong}\ \emph {et~al.}(2022)\citenamefont {Dong}, \citenamefont {Lin},\ and\ \citenamefont {Zou}}]{Dong:2022cuw}%
  \BibitemOpen
  \bibfield  {author} {\bibinfo {author} {\bibfnamefont {Xiang-Kun}\ \bibnamefont {Dong}}, \bibinfo {author} {\bibfnamefont {Yong-Hui}\ \bibnamefont {Lin}}, \ and\ \bibinfo {author} {\bibfnamefont {Bing-Song}\ \bibnamefont {Zou}},\ }\bibfield  {title} {\enquote {\bibinfo {title} {{Interpretation of the $\eta_1(1855)$ as a $K\bar{K}_{1}(1400) + c.c.$ molecule}},}\ }\href {\doibase 10.1007/s11433-022-1887-5} {\bibfield  {journal} {\bibinfo  {journal} {Sci. China Phys. Mech. Astron.}\ }\textbf {\bibinfo {volume} {65}},\ \bibinfo {pages} {261011} (\bibinfo {year} {2022})},\ \Eprint {http://arxiv.org/abs/2202.00863} {arXiv:2202.00863 [hep-ph]} \BibitemShut {NoStop}%
\bibitem [{\citenamefont {L\"uscher}(1986)}]{Luscher:1986pf}%
  \BibitemOpen
  \bibfield  {author} {\bibinfo {author} {\bibfnamefont {M.}~\bibnamefont {L\"uscher}},\ }\bibfield  {title} {\enquote {\bibinfo {title} {{Volume dependence of the energy spectrum in massive quantum field theories. II. Scattering States}},}\ }\href {\doibase 10.1007/BF01211097} {\bibfield  {journal} {\bibinfo  {journal} {Commun. Math. Phys.}\ }\textbf {\bibinfo {volume} {105}},\ \bibinfo {pages} {153--188} (\bibinfo {year} {1986})}\BibitemShut {NoStop}%
\bibitem [{\citenamefont {L\"uscher}(1991{\natexlab{a}})}]{Luscher:1990ux}%
  \BibitemOpen
  \bibfield  {author} {\bibinfo {author} {\bibfnamefont {M.}~\bibnamefont {L\"uscher}},\ }\bibfield  {title} {\enquote {\bibinfo {title} {{Two particle states on a torus and their relation to the scattering matrix}},}\ }\href {\doibase 10.1016/0550-3213(91)90366-6} {\bibfield  {journal} {\bibinfo  {journal} {Nucl. Phys. B}\ }\textbf {\bibinfo {volume} {354}},\ \bibinfo {pages} {531--578} (\bibinfo {year} {1991}{\natexlab{a}})}\BibitemShut {NoStop}%
\bibitem [{\citenamefont {L\"uscher}(1991{\natexlab{b}})}]{Luscher:1991cf}%
  \BibitemOpen
  \bibfield  {author} {\bibinfo {author} {\bibfnamefont {Martin}\ \bibnamefont {L\"uscher}},\ }\bibfield  {title} {\enquote {\bibinfo {title} {{Signatures of unstable particles in finite volume}},}\ }\href {\doibase 10.1016/0550-3213(91)90584-K} {\bibfield  {journal} {\bibinfo  {journal} {Nucl. Phys. B}\ }\textbf {\bibinfo {volume} {364}},\ \bibinfo {pages} {237--251} (\bibinfo {year} {1991}{\natexlab{b}})}\BibitemShut {NoStop}%
\bibitem [{\citenamefont {Brice\~no}\ \emph {et~al.}(2018)\citenamefont {Brice\~no}, \citenamefont {Dudek},\ and\ \citenamefont {Young}}]{Briceno:2017max}%
  \BibitemOpen
  \bibfield  {author} {\bibinfo {author} {\bibfnamefont {Ra\'ul~A.}\ \bibnamefont {Brice\~no}}, \bibinfo {author} {\bibfnamefont {Jozef~J.}\ \bibnamefont {Dudek}}, \ and\ \bibinfo {author} {\bibfnamefont {Ross~D.}\ \bibnamefont {Young}},\ }\bibfield  {title} {\enquote {\bibinfo {title} {{Scattering processes and resonances from lattice QCD}},}\ }\href {\doibase 10.1103/RevModPhys.90.025001} {\bibfield  {journal} {\bibinfo  {journal} {Rev. Mod. Phys.}\ }\textbf {\bibinfo {volume} {90}},\ \bibinfo {pages} {025001} (\bibinfo {year} {2018})},\ \Eprint {http://arxiv.org/abs/1706.06223} {arXiv:1706.06223 [hep-lat]} \BibitemShut {NoStop}%
\bibitem [{\citenamefont {Mai}\ \emph {et~al.}(2023)\citenamefont {Mai}, \citenamefont {Mei\ss{}ner},\ and\ \citenamefont {Urbach}}]{Mai:2022eur}%
  \BibitemOpen
  \bibfield  {author} {\bibinfo {author} {\bibfnamefont {Maxim}\ \bibnamefont {Mai}}, \bibinfo {author} {\bibfnamefont {Ulf-G.}\ \bibnamefont {Mei\ss{}ner}}, \ and\ \bibinfo {author} {\bibfnamefont {Carsten}\ \bibnamefont {Urbach}},\ }\bibfield  {title} {\enquote {\bibinfo {title} {{Towards a theory of hadron resonances}},}\ }\href {\doibase 10.1016/j.physrep.2022.11.005} {\bibfield  {journal} {\bibinfo  {journal} {Phys. Rept.}\ }\textbf {\bibinfo {volume} {1001}},\ \bibinfo {pages} {1--66} (\bibinfo {year} {2023})},\ \Eprint {http://arxiv.org/abs/2206.01477} {arXiv:2206.01477 [hep-ph]} \BibitemShut {NoStop}%
\bibitem [{\citenamefont {McNeile}\ \emph {et~al.}(2002)\citenamefont {McNeile}, \citenamefont {Michael},\ and\ \citenamefont {Pennanen}}]{McNeile:2002az}%
  \BibitemOpen
  \bibfield  {author} {\bibinfo {author} {\bibfnamefont {C.}~\bibnamefont {McNeile}}, \bibinfo {author} {\bibfnamefont {Christopher}\ \bibnamefont {Michael}}, \ and\ \bibinfo {author} {\bibfnamefont {P.}~\bibnamefont {Pennanen}} (\bibinfo {collaboration} {UKQCD}),\ }\bibfield  {title} {\enquote {\bibinfo {title} {{Hybrid meson decay from the lattice}},}\ }\href {\doibase 10.1103/PhysRevD.65.094505} {\bibfield  {journal} {\bibinfo  {journal} {Phys. Rev. D}\ }\textbf {\bibinfo {volume} {65}},\ \bibinfo {pages} {094505} (\bibinfo {year} {2002})},\ \Eprint {http://arxiv.org/abs/hep-lat/0201006} {arXiv:hep-lat/0201006} \BibitemShut {NoStop}%
\bibitem [{\citenamefont {McNeile}\ and\ \citenamefont {Michael}(2003)}]{McNeile:2002fh}%
  \BibitemOpen
  \bibfield  {author} {\bibinfo {author} {\bibfnamefont {C.}~\bibnamefont {McNeile}}\ and\ \bibinfo {author} {\bibfnamefont {Christopher}\ \bibnamefont {Michael}} (\bibinfo {collaboration} {UKQCD}),\ }\bibfield  {title} {\enquote {\bibinfo {title} {{Hadronic decay of a vector meson from the lattice}},}\ }\href {\doibase 10.1016/S0370-2693(03)00130-8} {\bibfield  {journal} {\bibinfo  {journal} {Phys. Lett. B}\ }\textbf {\bibinfo {volume} {556}},\ \bibinfo {pages} {177--184} (\bibinfo {year} {2003})},\ \Eprint {http://arxiv.org/abs/hep-lat/0212020} {arXiv:hep-lat/0212020} \BibitemShut {NoStop}%
\bibitem [{\citenamefont {McNeile}\ \emph {et~al.}(2004)\citenamefont {McNeile}, \citenamefont {Michael},\ and\ \citenamefont {Thompson}}]{McNeile:2004rf}%
  \BibitemOpen
  \bibfield  {author} {\bibinfo {author} {\bibfnamefont {C.}~\bibnamefont {McNeile}}, \bibinfo {author} {\bibfnamefont {Christopher}\ \bibnamefont {Michael}}, \ and\ \bibinfo {author} {\bibfnamefont {G.}~\bibnamefont {Thompson}} (\bibinfo {collaboration} {UKQCD}),\ }\bibfield  {title} {\enquote {\bibinfo {title} {{Hadronic decay of a scalar B meson from the lattice}},}\ }\href {\doibase 10.1103/PhysRevD.70.054501} {\bibfield  {journal} {\bibinfo  {journal} {Phys. Rev. D}\ }\textbf {\bibinfo {volume} {70}},\ \bibinfo {pages} {054501} (\bibinfo {year} {2004})},\ \Eprint {http://arxiv.org/abs/hep-lat/0404010} {arXiv:hep-lat/0404010} \BibitemShut {NoStop}%
\bibitem [{\citenamefont {Michael}(2006)}]{Michael:2005kw}%
  \BibitemOpen
  \bibfield  {author} {\bibinfo {author} {\bibfnamefont {Christopher}\ \bibnamefont {Michael}},\ }\bibfield  {title} {\enquote {\bibinfo {title} {{Hadronic decays}},}\ }\href {\doibase 10.22323/1.020.0008} {\bibfield  {journal} {\bibinfo  {journal} {PoS}\ }\textbf {\bibinfo {volume} {LAT2005}},\ \bibinfo {pages} {008} (\bibinfo {year} {2006})},\ \Eprint {http://arxiv.org/abs/hep-lat/0509023} {arXiv:hep-lat/0509023} \BibitemShut {NoStop}%
\bibitem [{\citenamefont {McNeile}\ and\ \citenamefont {Michael}(2006{\natexlab{b}})}]{McNeile:2006nv}%
  \BibitemOpen
  \bibfield  {author} {\bibinfo {author} {\bibfnamefont {C.}~\bibnamefont {McNeile}}\ and\ \bibinfo {author} {\bibfnamefont {Christopher}\ \bibnamefont {Michael}} (\bibinfo {collaboration} {UKQCD}),\ }\bibfield  {title} {\enquote {\bibinfo {title} {{Properties of light scalar mesons from lattice QCD}},}\ }\href {\doibase 10.1103/PhysRevD.74.014508} {\bibfield  {journal} {\bibinfo  {journal} {Phys. Rev. D}\ }\textbf {\bibinfo {volume} {74}},\ \bibinfo {pages} {014508} (\bibinfo {year} {2006}{\natexlab{b}})},\ \Eprint {http://arxiv.org/abs/hep-lat/0604009} {arXiv:hep-lat/0604009} \BibitemShut {NoStop}%
\bibitem [{\citenamefont {Hart}\ \emph {et~al.}(2006)\citenamefont {Hart}, \citenamefont {McNeile}, \citenamefont {Michael},\ and\ \citenamefont {Pickavance}}]{Hart:2006ps}%
  \BibitemOpen
  \bibfield  {author} {\bibinfo {author} {\bibfnamefont {A.}~\bibnamefont {Hart}}, \bibinfo {author} {\bibfnamefont {C.}~\bibnamefont {McNeile}}, \bibinfo {author} {\bibfnamefont {Christopher}\ \bibnamefont {Michael}}, \ and\ \bibinfo {author} {\bibfnamefont {J.}~\bibnamefont {Pickavance}} (\bibinfo {collaboration} {UKQCD}),\ }\bibfield  {title} {\enquote {\bibinfo {title} {{A lattice study of the masses of singlet $0^++$ mesons}},}\ }\href {\doibase 10.1103/PhysRevD.74.114504} {\bibfield  {journal} {\bibinfo  {journal} {Phys. Rev. D}\ }\textbf {\bibinfo {volume} {74}},\ \bibinfo {pages} {114504} (\bibinfo {year} {2006})},\ \Eprint {http://arxiv.org/abs/hep-lat/0608026} {arXiv:hep-lat/0608026} \BibitemShut {NoStop}%
\bibitem [{\citenamefont {Michael}(2007)}]{Michael:2006hf}%
  \BibitemOpen
  \bibfield  {author} {\bibinfo {author} {\bibfnamefont {Christopher}\ \bibnamefont {Michael}},\ }\bibfield  {title} {\enquote {\bibinfo {title} {{Hadronic decays from the lattice}},}\ }\href {\doibase 10.1140/epja/i2006-10177-6} {\bibfield  {journal} {\bibinfo  {journal} {Eur. Phys. J. A}\ }\textbf {\bibinfo {volume} {31}},\ \bibinfo {pages} {793--798} (\bibinfo {year} {2007})},\ \Eprint {http://arxiv.org/abs/hep-lat/0609008} {arXiv:hep-lat/0609008} \BibitemShut {NoStop}%
\bibitem [{\citenamefont {Bali}\ \emph {et~al.}(2016)\citenamefont {Bali}, \citenamefont {Collins}, \citenamefont {Cox}, \citenamefont {Donald}, \citenamefont {G\"ockeler}, \citenamefont {Lang},\ and\ \citenamefont {Sch\"afer}}]{Bali:2015gji}%
  \BibitemOpen
  \bibfield  {author} {\bibinfo {author} {\bibfnamefont {Gunnar~S.}\ \bibnamefont {Bali}}, \bibinfo {author} {\bibfnamefont {Sara}\ \bibnamefont {Collins}}, \bibinfo {author} {\bibfnamefont {Antonio}\ \bibnamefont {Cox}}, \bibinfo {author} {\bibfnamefont {Gordon}\ \bibnamefont {Donald}}, \bibinfo {author} {\bibfnamefont {Meinulf}\ \bibnamefont {G\"ockeler}}, \bibinfo {author} {\bibfnamefont {C.~B.}\ \bibnamefont {Lang}}, \ and\ \bibinfo {author} {\bibfnamefont {Andreas}\ \bibnamefont {Sch\"afer}} (\bibinfo {collaboration} {RQCD}),\ }\bibfield  {title} {\enquote {\bibinfo {title} {{$\rho$ and $K^*$ resonances on the lattice at nearly physical quark masses and $N_f=2$}},}\ }\href {\doibase 10.1103/PhysRevD.93.054509} {\bibfield  {journal} {\bibinfo  {journal} {Phys. Rev. D}\ }\textbf {\bibinfo {volume} {93}},\ \bibinfo {pages} {054509} (\bibinfo {year} {2016})},\ \Eprint {http://arxiv.org/abs/1512.08678} {arXiv:1512.08678 [hep-lat]} \BibitemShut {NoStop}%
\bibitem [{\citenamefont {Zhang}\ \emph {et~al.}(2022)\citenamefont {Zhang}, \citenamefont {Sun}, \citenamefont {Chen}, \citenamefont {Gong}, \citenamefont {Gui},\ and\ \citenamefont {Liu}}]{Zhang:2021xvl}%
  \BibitemOpen
  \bibfield  {author} {\bibinfo {author} {\bibfnamefont {Renqiang}\ \bibnamefont {Zhang}}, \bibinfo {author} {\bibfnamefont {Wei}\ \bibnamefont {Sun}}, \bibinfo {author} {\bibfnamefont {Ying}\ \bibnamefont {Chen}}, \bibinfo {author} {\bibfnamefont {Ming}\ \bibnamefont {Gong}}, \bibinfo {author} {\bibfnamefont {Long-Cheng}\ \bibnamefont {Gui}}, \ and\ \bibinfo {author} {\bibfnamefont {Zhaofeng}\ \bibnamefont {Liu}},\ }\bibfield  {title} {\enquote {\bibinfo {title} {{The glueball content of $\eta_c$}},}\ }\href {\doibase 10.1016/j.physletb.2022.136960} {\bibfield  {journal} {\bibinfo  {journal} {Phys. Lett. B}\ }\textbf {\bibinfo {volume} {827}},\ \bibinfo {pages} {136960} (\bibinfo {year} {2022})},\ \Eprint {http://arxiv.org/abs/2107.12749} {arXiv:2107.12749 [hep-lat]} \BibitemShut {NoStop}%
\bibitem [{\citenamefont {Jiang}\ \emph {et~al.}(2023{\natexlab{a}})\citenamefont {Jiang}, \citenamefont {Sun}, \citenamefont {Chen}, \citenamefont {Chen}, \citenamefont {Gong}, \citenamefont {Liu},\ and\ \citenamefont {Zhang}}]{Jiang:2022ffl}%
  \BibitemOpen
  \bibfield  {author} {\bibinfo {author} {\bibfnamefont {Xiangyu}\ \bibnamefont {Jiang}}, \bibinfo {author} {\bibfnamefont {Wei}\ \bibnamefont {Sun}}, \bibinfo {author} {\bibfnamefont {Feiyu}\ \bibnamefont {Chen}}, \bibinfo {author} {\bibfnamefont {Ying}\ \bibnamefont {Chen}}, \bibinfo {author} {\bibfnamefont {Ming}\ \bibnamefont {Gong}}, \bibinfo {author} {\bibfnamefont {Zhaofeng}\ \bibnamefont {Liu}}, \ and\ \bibinfo {author} {\bibfnamefont {Renqiang}\ \bibnamefont {Zhang}},\ }\bibfield  {title} {\enquote {\bibinfo {title} {{$\eta$-glueball mixing from $N_f=2$ lattice QCD}},}\ }\href {\doibase 10.1103/PhysRevD.107.094510} {\bibfield  {journal} {\bibinfo  {journal} {Phys. Rev. D}\ }\textbf {\bibinfo {volume} {107}},\ \bibinfo {pages} {094510} (\bibinfo {year} {2023}{\natexlab{a}})},\ \Eprint {http://arxiv.org/abs/2205.12541} {arXiv:2205.12541 [hep-lat]} \BibitemShut {NoStop}%
\bibitem [{\citenamefont {Shi}\ \emph {et~al.}(2024{\natexlab{a}})\citenamefont {Shi}, \citenamefont {Chen}, \citenamefont {Gong}, \citenamefont {Jiang}, \citenamefont {Liu},\ and\ \citenamefont {Sun}}]{Shi:2023sdy}%
  \BibitemOpen
  \bibfield  {author} {\bibinfo {author} {\bibfnamefont {Chunjiang}\ \bibnamefont {Shi}}, \bibinfo {author} {\bibfnamefont {Ying}\ \bibnamefont {Chen}}, \bibinfo {author} {\bibfnamefont {Ming}\ \bibnamefont {Gong}}, \bibinfo {author} {\bibfnamefont {Xiangyu}\ \bibnamefont {Jiang}}, \bibinfo {author} {\bibfnamefont {Zhaofeng}\ \bibnamefont {Liu}}, \ and\ \bibinfo {author} {\bibfnamefont {Wei}\ \bibnamefont {Sun}},\ }\bibfield  {title} {\enquote {\bibinfo {title} {{Decays of $1^{-+}$ charmoniumlike hybrid using lattice QCD}},}\ }\href {\doibase 10.1103/PhysRevD.109.094513} {\bibfield  {journal} {\bibinfo  {journal} {Phys. Rev. D}\ }\textbf {\bibinfo {volume} {109}},\ \bibinfo {pages} {094513} (\bibinfo {year} {2024}{\natexlab{a}})},\ \Eprint {http://arxiv.org/abs/2306.12884} {arXiv:2306.12884 [hep-lat]} \BibitemShut {NoStop}%
\bibitem [{\citenamefont {Alexandrou}\ \emph {et~al.}(2013)\citenamefont {Alexandrou}, \citenamefont {Negele}, \citenamefont {Petschlies}, \citenamefont {Strelchenko},\ and\ \citenamefont {Tsapalis}}]{Alexandrou:2013ata}%
  \BibitemOpen
  \bibfield  {author} {\bibinfo {author} {\bibfnamefont {C.}~\bibnamefont {Alexandrou}}, \bibinfo {author} {\bibfnamefont {J.~W.}\ \bibnamefont {Negele}}, \bibinfo {author} {\bibfnamefont {M.}~\bibnamefont {Petschlies}}, \bibinfo {author} {\bibfnamefont {A.}~\bibnamefont {Strelchenko}}, \ and\ \bibinfo {author} {\bibfnamefont {A.}~\bibnamefont {Tsapalis}},\ }\bibfield  {title} {\enquote {\bibinfo {title} {{Determination of $\Delta$ Resonance Parameters from Lattice QCD}},}\ }\href {\doibase 10.1103/PhysRevD.88.031501} {\bibfield  {journal} {\bibinfo  {journal} {Phys. Rev. D}\ }\textbf {\bibinfo {volume} {88}},\ \bibinfo {pages} {031501} (\bibinfo {year} {2013})},\ \Eprint {http://arxiv.org/abs/1305.6081} {arXiv:1305.6081 [hep-lat]} \BibitemShut {NoStop}%
\bibitem [{\citenamefont {Alexandrou}\ \emph {et~al.}(2016)\citenamefont {Alexandrou}, \citenamefont {Negele}, \citenamefont {Petschlies}, \citenamefont {Pochinsky},\ and\ \citenamefont {Syritsyn}}]{Alexandrou:2015hxa}%
  \BibitemOpen
  \bibfield  {author} {\bibinfo {author} {\bibfnamefont {Constantia}\ \bibnamefont {Alexandrou}}, \bibinfo {author} {\bibfnamefont {John~W.}\ \bibnamefont {Negele}}, \bibinfo {author} {\bibfnamefont {Marcus}\ \bibnamefont {Petschlies}}, \bibinfo {author} {\bibfnamefont {Andrew~V.}\ \bibnamefont {Pochinsky}}, \ and\ \bibinfo {author} {\bibfnamefont {Sergey~N.}\ \bibnamefont {Syritsyn}},\ }\bibfield  {title} {\enquote {\bibinfo {title} {{Study of decuplet baryon resonances from lattice QCD}},}\ }\href {\doibase 10.1103/PhysRevD.93.114515} {\bibfield  {journal} {\bibinfo  {journal} {Phys. Rev. D}\ }\textbf {\bibinfo {volume} {93}},\ \bibinfo {pages} {114515} (\bibinfo {year} {2016})},\ \Eprint {http://arxiv.org/abs/1507.02724} {arXiv:1507.02724 [hep-lat]} \BibitemShut {NoStop}%
\bibitem [{\citenamefont {Andersen}\ \emph {et~al.}(2018)\citenamefont {Andersen}, \citenamefont {Bulava}, \citenamefont {H\"orz},\ and\ \citenamefont {Morningstar}}]{Andersen:2017una}%
  \BibitemOpen
  \bibfield  {author} {\bibinfo {author} {\bibfnamefont {Christian~Walther}\ \bibnamefont {Andersen}}, \bibinfo {author} {\bibfnamefont {John}\ \bibnamefont {Bulava}}, \bibinfo {author} {\bibfnamefont {Ben}\ \bibnamefont {H\"orz}}, \ and\ \bibinfo {author} {\bibfnamefont {Colin}\ \bibnamefont {Morningstar}},\ }\bibfield  {title} {\enquote {\bibinfo {title} {{Elastic $I=3/2$ $p$-wave nucleon-pion scattering amplitude and the $\Delta$(1232) resonance from $N_f=2+1$ lattice QCD}},}\ }\href {\doibase 10.1103/PhysRevD.97.014506} {\bibfield  {journal} {\bibinfo  {journal} {Phys. Rev. D}\ }\textbf {\bibinfo {volume} {97}},\ \bibinfo {pages} {014506} (\bibinfo {year} {2018})},\ \Eprint {http://arxiv.org/abs/1710.01557} {arXiv:1710.01557 [hep-lat]} \BibitemShut {NoStop}%
\bibitem [{\citenamefont {Silvi}\ \emph {et~al.}(2021)\citenamefont {Silvi} \emph {et~al.}}]{Silvi:2021uya}%
  \BibitemOpen
  \bibfield  {author} {\bibinfo {author} {\bibfnamefont {Giorgio}\ \bibnamefont {Silvi}} \emph {et~al.},\ }\bibfield  {title} {\enquote {\bibinfo {title} {{$P$-wave nucleon-pion scattering amplitude in the $\Delta$(1232) channel from lattice QCD}},}\ }\href {\doibase 10.1103/PhysRevD.103.094508} {\bibfield  {journal} {\bibinfo  {journal} {Phys. Rev. D}\ }\textbf {\bibinfo {volume} {103}},\ \bibinfo {pages} {094508} (\bibinfo {year} {2021})},\ \Eprint {http://arxiv.org/abs/2101.00689} {arXiv:2101.00689 [hep-lat]} \BibitemShut {NoStop}%
\bibitem [{\citenamefont {Morningstar}\ \emph {et~al.}(2022)\citenamefont {Morningstar}, \citenamefont {Bulava}, \citenamefont {Hanlon}, \citenamefont {H\"orz}, \citenamefont {Mohler}, \citenamefont {Nicholson}, \citenamefont {Skinner},\ and\ \citenamefont {Walker-Loud}}]{Morningstar:2021ewk}%
  \BibitemOpen
  \bibfield  {author} {\bibinfo {author} {\bibfnamefont {Colin}\ \bibnamefont {Morningstar}}, \bibinfo {author} {\bibfnamefont {John}\ \bibnamefont {Bulava}}, \bibinfo {author} {\bibfnamefont {Andrew~D.}\ \bibnamefont {Hanlon}}, \bibinfo {author} {\bibfnamefont {Ben}\ \bibnamefont {H\"orz}}, \bibinfo {author} {\bibfnamefont {Daniel}\ \bibnamefont {Mohler}}, \bibinfo {author} {\bibfnamefont {Amy}\ \bibnamefont {Nicholson}}, \bibinfo {author} {\bibfnamefont {Sarah}\ \bibnamefont {Skinner}}, \ and\ \bibinfo {author} {\bibfnamefont {Andr\'e}\ \bibnamefont {Walker-Loud}},\ }\bibfield  {title} {\enquote {\bibinfo {title} {{Progress on Meson-Baryon Scattering}},}\ }\href {\doibase 10.22323/1.396.0170} {\bibfield  {journal} {\bibinfo  {journal} {PoS}\ }\textbf {\bibinfo {volume} {LATTICE2021}},\ \bibinfo {pages} {170} (\bibinfo {year} {2022})},\ \Eprint {http://arxiv.org/abs/2111.07755} {arXiv:2111.07755 [hep-lat]} \BibitemShut {NoStop}%
\bibitem [{\citenamefont {Pascalutsa}\ and\ \citenamefont {Vanderhaeghen}(2006)}]{Pascalutsa:2005vq}%
  \BibitemOpen
  \bibfield  {author} {\bibinfo {author} {\bibfnamefont {Vladimir}\ \bibnamefont {Pascalutsa}}\ and\ \bibinfo {author} {\bibfnamefont {Marc}\ \bibnamefont {Vanderhaeghen}},\ }\bibfield  {title} {\enquote {\bibinfo {title} {{Chiral effective-field theory in the $\Delta (1232)$ region: I. Pion electroproduction on the nucleon}},}\ }\href {\doibase 10.1103/PhysRevD.73.034003} {\bibfield  {journal} {\bibinfo  {journal} {Phys. Rev. D}\ }\textbf {\bibinfo {volume} {73}},\ \bibinfo {pages} {034003} (\bibinfo {year} {2006})},\ \Eprint {http://arxiv.org/abs/hep-ph/0512244} {arXiv:hep-ph/0512244} \BibitemShut {NoStop}%
\bibitem [{\citenamefont {Hemmert}\ \emph {et~al.}(1995)\citenamefont {Hemmert}, \citenamefont {Holstein},\ and\ \citenamefont {Mukhopadhyay}}]{Hemmert:1994ky}%
  \BibitemOpen
  \bibfield  {author} {\bibinfo {author} {\bibfnamefont {Thomas~R.}\ \bibnamefont {Hemmert}}, \bibinfo {author} {\bibfnamefont {Barry~R.}\ \bibnamefont {Holstein}}, \ and\ \bibinfo {author} {\bibfnamefont {Nimai~C.}\ \bibnamefont {Mukhopadhyay}},\ }\bibfield  {title} {\enquote {\bibinfo {title} {{$N N$, $N \Delta$ couplings and the quark model}},}\ }\href {\doibase 10.1103/PhysRevD.51.158} {\bibfield  {journal} {\bibinfo  {journal} {Phys. Rev. D}\ }\textbf {\bibinfo {volume} {51}},\ \bibinfo {pages} {158--167} (\bibinfo {year} {1995})},\ \Eprint {http://arxiv.org/abs/hep-ph/9409323} {arXiv:hep-ph/9409323} \BibitemShut {NoStop}%
\bibitem [{\citenamefont {Peardon}\ \emph {et~al.}(2009)\citenamefont {Peardon}, \citenamefont {Bulava}, \citenamefont {Foley}, \citenamefont {Morningstar}, \citenamefont {Dudek}, \citenamefont {Edwards}, \citenamefont {Joo}, \citenamefont {Lin}, \citenamefont {Richards},\ and\ \citenamefont {Juge}}]{Peardon:2009gh}%
  \BibitemOpen
  \bibfield  {author} {\bibinfo {author} {\bibfnamefont {Michael}\ \bibnamefont {Peardon}}, \bibinfo {author} {\bibfnamefont {John}\ \bibnamefont {Bulava}}, \bibinfo {author} {\bibfnamefont {Justin}\ \bibnamefont {Foley}}, \bibinfo {author} {\bibfnamefont {Colin}\ \bibnamefont {Morningstar}}, \bibinfo {author} {\bibfnamefont {Jozef}\ \bibnamefont {Dudek}}, \bibinfo {author} {\bibfnamefont {Robert~G.}\ \bibnamefont {Edwards}}, \bibinfo {author} {\bibfnamefont {Balint}\ \bibnamefont {Joo}}, \bibinfo {author} {\bibfnamefont {Huey-Wen}\ \bibnamefont {Lin}}, \bibinfo {author} {\bibfnamefont {David~G.}\ \bibnamefont {Richards}}, \ and\ \bibinfo {author} {\bibfnamefont {Keisuke~Jimmy}\ \bibnamefont {Juge}} (\bibinfo {collaboration} {Hadron Spectrum}),\ }\bibfield  {title} {\enquote {\bibinfo {title} {{A Novel quark-field creation operator construction for hadronic physics in lattice QCD}},}\ }\href {\doibase 10.1103/PhysRevD.80.054506} {\bibfield  {journal} {\bibinfo  {journal} {Phys. Rev. D}\ }\textbf {\bibinfo {volume} {80}},\ \bibinfo {pages} {054506} (\bibinfo {year} {2009})},\ \Eprint {http://arxiv.org/abs/0905.2160} {arXiv:0905.2160 [hep-lat]} \BibitemShut {NoStop}%
\bibitem [{\citenamefont {Li}\ \emph {et~al.}(2024)\citenamefont {Li}, \citenamefont {Shi}, \citenamefont {Chen}, \citenamefont {Gong}, \citenamefont {Liang}, \citenamefont {Liu},\ and\ \citenamefont {Sun}}]{Li:2024pfg}%
  \BibitemOpen
  \bibfield  {author} {\bibinfo {author} {\bibfnamefont {Haozheng}\ \bibnamefont {Li}}, \bibinfo {author} {\bibfnamefont {Chunjiang}\ \bibnamefont {Shi}}, \bibinfo {author} {\bibfnamefont {Ying}\ \bibnamefont {Chen}}, \bibinfo {author} {\bibfnamefont {Ming}\ \bibnamefont {Gong}}, \bibinfo {author} {\bibfnamefont {Juzheng}\ \bibnamefont {Liang}}, \bibinfo {author} {\bibfnamefont {Zhaofeng}\ \bibnamefont {Liu}}, \ and\ \bibinfo {author} {\bibfnamefont {Wei}\ \bibnamefont {Sun}},\ }\bibfield  {title} {\enquote {\bibinfo {title} {{$X(3872)$ Relevant $D\bar{D}^*$ Scattering in $N_f=2$ Lattice QCD}},}\ }\href@noop {} {\  (\bibinfo {year} {2024})},\ \Eprint {http://arxiv.org/abs/2402.14541} {arXiv:2402.14541 [hep-lat]} \BibitemShut {NoStop}%
\bibitem [{\citenamefont {Dudek}\ \emph {et~al.}(2008)\citenamefont {Dudek}, \citenamefont {Edwards}, \citenamefont {Mathur},\ and\ \citenamefont {Richards}}]{Dudek:2007wv}%
  \BibitemOpen
  \bibfield  {author} {\bibinfo {author} {\bibfnamefont {Jozef~J.}\ \bibnamefont {Dudek}}, \bibinfo {author} {\bibfnamefont {Robert~G.}\ \bibnamefont {Edwards}}, \bibinfo {author} {\bibfnamefont {Nilmani}\ \bibnamefont {Mathur}}, \ and\ \bibinfo {author} {\bibfnamefont {David~G.}\ \bibnamefont {Richards}},\ }\bibfield  {title} {\enquote {\bibinfo {title} {{Charmonium excited state spectrum in lattice QCD}},}\ }\href {\doibase 10.1103/PhysRevD.77.034501} {\bibfield  {journal} {\bibinfo  {journal} {Phys. Rev. D}\ }\textbf {\bibinfo {volume} {77}},\ \bibinfo {pages} {034501} (\bibinfo {year} {2008})},\ \Eprint {http://arxiv.org/abs/0707.4162} {arXiv:0707.4162 [hep-lat]} \BibitemShut {NoStop}%
\bibitem [{\citenamefont {Shi}\ \emph {et~al.}(2024{\natexlab{b}})\citenamefont {Shi}, \citenamefont {Chen}, \citenamefont {Jiang}, \citenamefont {Gong}, \citenamefont {Liu},\ and\ \citenamefont {Sun}}]{Shi:2024fyv}%
  \BibitemOpen
  \bibfield  {author} {\bibinfo {author} {\bibfnamefont {Chunjiang}\ \bibnamefont {Shi}}, \bibinfo {author} {\bibfnamefont {Ying}\ \bibnamefont {Chen}}, \bibinfo {author} {\bibfnamefont {Xiangyu}\ \bibnamefont {Jiang}}, \bibinfo {author} {\bibfnamefont {Ming}\ \bibnamefont {Gong}}, \bibinfo {author} {\bibfnamefont {Zhaofeng}\ \bibnamefont {Liu}}, \ and\ \bibinfo {author} {\bibfnamefont {Wei}\ \bibnamefont {Sun}},\ }\bibfield  {title} {\enquote {\bibinfo {title} {{Form factor for Dalitz decays from $J/\psi$ to light pseudoscalars}},}\ }\href {\doibase 10.1088/1674-1137/ad641b} {\bibfield  {journal} {\bibinfo  {journal} {Chin. Phys. C}\ }\textbf {\bibinfo {volume} {48}},\ \bibinfo {pages} {113105} (\bibinfo {year} {2024}{\natexlab{b}})},\ \Eprint {http://arxiv.org/abs/2403.11842} {arXiv:2403.11842 [hep-lat]} \BibitemShut {NoStop}%
\bibitem [{\citenamefont {Jiang}\ \emph {et~al.}(2023{\natexlab{b}})\citenamefont {Jiang}, \citenamefont {Chen}, \citenamefont {Chen}, \citenamefont {Gong}, \citenamefont {Li}, \citenamefont {Liu}, \citenamefont {Sun},\ and\ \citenamefont {Zhang}}]{Jiang:2022gnd}%
  \BibitemOpen
  \bibfield  {author} {\bibinfo {author} {\bibfnamefont {Xiangyu}\ \bibnamefont {Jiang}}, \bibinfo {author} {\bibfnamefont {Feiyu}\ \bibnamefont {Chen}}, \bibinfo {author} {\bibfnamefont {Ying}\ \bibnamefont {Chen}}, \bibinfo {author} {\bibfnamefont {Ming}\ \bibnamefont {Gong}}, \bibinfo {author} {\bibfnamefont {Ning}\ \bibnamefont {Li}}, \bibinfo {author} {\bibfnamefont {Zhaofeng}\ \bibnamefont {Liu}}, \bibinfo {author} {\bibfnamefont {Wei}\ \bibnamefont {Sun}}, \ and\ \bibinfo {author} {\bibfnamefont {Renqiang}\ \bibnamefont {Zhang}},\ }\bibfield  {title} {\enquote {\bibinfo {title} {{Radiative Decay Width of $J/\psi\to\gamma\eta_{(2)}$ from $N_f=2$ Lattice QCD}},}\ }\href {\doibase 10.1103/PhysRevLett.130.061901} {\bibfield  {journal} {\bibinfo  {journal} {Phys. Rev. Lett.}\ }\textbf {\bibinfo {volume} {130}},\ \bibinfo {pages} {061901} (\bibinfo {year} {2023}{\natexlab{b}})},\ \Eprint {http://arxiv.org/abs/2206.02724} {arXiv:2206.02724 [hep-lat]} \BibitemShut {NoStop}%
\bibitem [{\citenamefont {Feng}\ \emph {et~al.}(2011)\citenamefont {Feng}, \citenamefont {Jansen},\ and\ \citenamefont {Renner}}]{Feng:2010es}%
  \BibitemOpen
  \bibfield  {author} {\bibinfo {author} {\bibfnamefont {Xu}~\bibnamefont {Feng}}, \bibinfo {author} {\bibfnamefont {Karl}\ \bibnamefont {Jansen}}, \ and\ \bibinfo {author} {\bibfnamefont {Dru~B.}\ \bibnamefont {Renner}},\ }\bibfield  {title} {\enquote {\bibinfo {title} {{Resonance parameters of the $\rho$-meson from lattice QCD}},}\ }\href {\doibase 10.1103/PhysRevD.83.094505} {\bibfield  {journal} {\bibinfo  {journal} {Phys. Rev. D}\ }\textbf {\bibinfo {volume} {83}},\ \bibinfo {pages} {094505} (\bibinfo {year} {2011})},\ \Eprint {http://arxiv.org/abs/1011.5288} {arXiv:1011.5288 [hep-lat]} \BibitemShut {NoStop}%
\bibitem [{\citenamefont {Wallace}(2015)}]{Wallace:2015pxa}%
  \BibitemOpen
  \bibfield  {author} {\bibinfo {author} {\bibfnamefont {Stephen~J.}\ \bibnamefont {Wallace}},\ }\bibfield  {title} {\enquote {\bibinfo {title} {{Partial-wave and helicity operators for the scattering of two hadrons in lattice QCD}},}\ }\href {\doibase 10.1103/PhysRevD.92.034520} {\bibfield  {journal} {\bibinfo  {journal} {Phys. Rev. D}\ }\textbf {\bibinfo {volume} {92}},\ \bibinfo {pages} {034520} (\bibinfo {year} {2015})},\ \Eprint {http://arxiv.org/abs/1506.05492} {arXiv:1506.05492 [hep-lat]} \BibitemShut {NoStop}%
\bibitem [{\citenamefont {Prelovsek}\ \emph {et~al.}(2017)\citenamefont {Prelovsek}, \citenamefont {Skerbis},\ and\ \citenamefont {Lang}}]{Prelovsek:2016iyo}%
  \BibitemOpen
  \bibfield  {author} {\bibinfo {author} {\bibfnamefont {S.}~\bibnamefont {Prelovsek}}, \bibinfo {author} {\bibfnamefont {U.}~\bibnamefont {Skerbis}}, \ and\ \bibinfo {author} {\bibfnamefont {C.~B.}\ \bibnamefont {Lang}},\ }\bibfield  {title} {\enquote {\bibinfo {title} {{Lattice operators for scattering of particles with spin}},}\ }\href {\doibase 10.1007/JHEP01(2017)129} {\bibfield  {journal} {\bibinfo  {journal} {JHEP}\ }\textbf {\bibinfo {volume} {01}},\ \bibinfo {pages} {129} (\bibinfo {year} {2017})},\ \Eprint {http://arxiv.org/abs/1607.06738} {arXiv:1607.06738 [hep-lat]} \BibitemShut {NoStop}%
\bibitem [{\citenamefont {Suzuki}(1993)}]{Suzuki:1993yc}%
  \BibitemOpen
  \bibfield  {author} {\bibinfo {author} {\bibfnamefont {M.}~\bibnamefont {Suzuki}},\ }\bibfield  {title} {\enquote {\bibinfo {title} {{Strange axial - vector mesons}},}\ }\href {\doibase 10.1103/PhysRevD.47.1252} {\bibfield  {journal} {\bibinfo  {journal} {Phys. Rev. D}\ }\textbf {\bibinfo {volume} {47}},\ \bibinfo {pages} {1252--1255} (\bibinfo {year} {1993})}\BibitemShut {NoStop}%
\bibitem [{\citenamefont {Burakovsky}\ and\ \citenamefont {Goldman}(1997)}]{Burakovsky:1997dd}%
  \BibitemOpen
  \bibfield  {author} {\bibinfo {author} {\bibfnamefont {L.}~\bibnamefont {Burakovsky}}\ and\ \bibinfo {author} {\bibfnamefont {J.~Terrance}\ \bibnamefont {Goldman}},\ }\bibfield  {title} {\enquote {\bibinfo {title} {{Constraint on axial - vector meson mixing angle from nonrelativistic constituent quark model}},}\ }\href {\doibase 10.1103/PhysRevD.56.R1368} {\bibfield  {journal} {\bibinfo  {journal} {Phys. Rev. D}\ }\textbf {\bibinfo {volume} {56}},\ \bibinfo {pages} {R1368--R1372} (\bibinfo {year} {1997})},\ \Eprint {http://arxiv.org/abs/hep-ph/9703274} {arXiv:hep-ph/9703274} \BibitemShut {NoStop}%
\bibitem [{\citenamefont {Cheng}(2003)}]{Cheng:2003bn}%
  \BibitemOpen
  \bibfield  {author} {\bibinfo {author} {\bibfnamefont {Hai-Yang}\ \bibnamefont {Cheng}},\ }\bibfield  {title} {\enquote {\bibinfo {title} {{Hadronic charmed meson decays involving axial vector mesons}},}\ }\href {\doibase 10.1103/PhysRevD.67.094007} {\bibfield  {journal} {\bibinfo  {journal} {Phys. Rev. D}\ }\textbf {\bibinfo {volume} {67}},\ \bibinfo {pages} {094007} (\bibinfo {year} {2003})},\ \Eprint {http://arxiv.org/abs/hep-ph/0301198} {arXiv:hep-ph/0301198} \BibitemShut {NoStop}%
\bibitem [{\citenamefont {Isgur}\ and\ \citenamefont {Paton}(1985)}]{Isgur:1984bm}%
  \BibitemOpen
  \bibfield  {author} {\bibinfo {author} {\bibfnamefont {Nathan}\ \bibnamefont {Isgur}}\ and\ \bibinfo {author} {\bibfnamefont {Jack~E.}\ \bibnamefont {Paton}},\ }\bibfield  {title} {\enquote {\bibinfo {title} {{Flux tube model for hadrons in QCD}},}\ }\href {\doibase 10.1103/PhysRevD.31.2910} {\bibfield  {journal} {\bibinfo  {journal} {Phys. Rev. D}\ }\textbf {\bibinfo {volume} {31}},\ \bibinfo {pages} {2910} (\bibinfo {year} {1985})}\BibitemShut {NoStop}%
\bibitem [{\citenamefont {Page}(1997)}]{Page:1996rj}%
  \BibitemOpen
  \bibfield  {author} {\bibinfo {author} {\bibfnamefont {Philip~R.}\ \bibnamefont {Page}},\ }\bibfield  {title} {\enquote {\bibinfo {title} {{Why hybrid meson coupling to two $S$-wave mesons is suppressed}},}\ }\href {\doibase 10.1016/S0370-2693(97)00438-3} {\bibfield  {journal} {\bibinfo  {journal} {Phys. Lett. B}\ }\textbf {\bibinfo {volume} {402}},\ \bibinfo {pages} {183--188} (\bibinfo {year} {1997})},\ \Eprint {http://arxiv.org/abs/hep-ph/9611375} {arXiv:hep-ph/9611375} \BibitemShut {NoStop}%
\bibitem [{\citenamefont {Page}\ \emph {et~al.}(1999)\citenamefont {Page}, \citenamefont {Swanson},\ and\ \citenamefont {Szczepaniak}}]{Page:1998gz}%
  \BibitemOpen
  \bibfield  {author} {\bibinfo {author} {\bibfnamefont {Philip~R.}\ \bibnamefont {Page}}, \bibinfo {author} {\bibfnamefont {Eric~S.}\ \bibnamefont {Swanson}}, \ and\ \bibinfo {author} {\bibfnamefont {Adam~P.}\ \bibnamefont {Szczepaniak}},\ }\bibfield  {title} {\enquote {\bibinfo {title} {{Hybrid meson decay phenomenology}},}\ }\href {\doibase 10.1103/PhysRevD.59.034016} {\bibfield  {journal} {\bibinfo  {journal} {Phys. Rev. D}\ }\textbf {\bibinfo {volume} {59}},\ \bibinfo {pages} {034016} (\bibinfo {year} {1999})},\ \Eprint {http://arxiv.org/abs/hep-ph/9808346} {arXiv:hep-ph/9808346} \BibitemShut {NoStop}%
\bibitem [{\citenamefont {Edwards}\ and\ \citenamefont {Joo}(2005)}]{Edwards:2004sx}%
  \BibitemOpen
  \bibfield  {author} {\bibinfo {author} {\bibfnamefont {Robert~G.}\ \bibnamefont {Edwards}}\ and\ \bibinfo {author} {\bibfnamefont {Balint}\ \bibnamefont {Joo}} (\bibinfo {collaboration} {SciDAC, LHPC, UKQCD}),\ }\bibfield  {title} {\enquote {\bibinfo {title} {{The Chroma Software System for Lattice QCD}},}\ }\href {\doibase 10.1016/j.nuclphysbps.2004.11.254} {\bibfield  {journal} {\bibinfo  {journal} {Nucl. Phys. B Proc. Suppl.}\ }\textbf {\bibinfo {volume} {140}},\ \bibinfo {pages} {832} (\bibinfo {year} {2005})},\ \Eprint {http://arxiv.org/abs/hep-lat/0409003} {arXiv:hep-lat/0409003} \BibitemShut {NoStop}%
\bibitem [{\citenamefont {Clark}\ \emph {et~al.}(2010)\citenamefont {Clark}, \citenamefont {Babich}, \citenamefont {Barros}, \citenamefont {Brower},\ and\ \citenamefont {Rebbi}}]{Clark:2009wm}%
  \BibitemOpen
  \bibfield  {author} {\bibinfo {author} {\bibfnamefont {M.~A.}\ \bibnamefont {Clark}}, \bibinfo {author} {\bibfnamefont {R.}~\bibnamefont {Babich}}, \bibinfo {author} {\bibfnamefont {K.}~\bibnamefont {Barros}}, \bibinfo {author} {\bibfnamefont {R.~C.}\ \bibnamefont {Brower}}, \ and\ \bibinfo {author} {\bibfnamefont {C.}~\bibnamefont {Rebbi}},\ }\bibfield  {title} {\enquote {\bibinfo {title} {{Solving Lattice QCD systems of equations using mixed precision solvers on GPUs}},}\ }\href {\doibase 10.1016/j.cpc.2010.05.002} {\bibfield  {journal} {\bibinfo  {journal} {Comput. Phys. Commun.}\ }\textbf {\bibinfo {volume} {181}},\ \bibinfo {pages} {1517--1528} (\bibinfo {year} {2010})},\ \Eprint {http://arxiv.org/abs/0911.3191} {arXiv:0911.3191 [hep-lat]} \BibitemShut {NoStop}%
\bibitem [{\citenamefont {Babich}\ \emph {et~al.}(2011)\citenamefont {Babich}, \citenamefont {Clark}, \citenamefont {Joo}, \citenamefont {Shi}, \citenamefont {Brower},\ and\ \citenamefont {Gottlieb}}]{Babich:2011np}%
  \BibitemOpen
  \bibfield  {author} {\bibinfo {author} {\bibfnamefont {R.}~\bibnamefont {Babich}}, \bibinfo {author} {\bibfnamefont {M.~A.}\ \bibnamefont {Clark}}, \bibinfo {author} {\bibfnamefont {B.}~\bibnamefont {Joo}}, \bibinfo {author} {\bibfnamefont {G.}~\bibnamefont {Shi}}, \bibinfo {author} {\bibfnamefont {R.~C.}\ \bibnamefont {Brower}}, \ and\ \bibinfo {author} {\bibfnamefont {S.}~\bibnamefont {Gottlieb}},\ }\bibfield  {title} {\enquote {\bibinfo {title} {{Scaling lattice QCD beyond 100 GPUs}},}\ }in\ \href {\doibase 10.1145/2063384.2063478} {\emph {\bibinfo {booktitle} {{SC11 International Conference for High Performance Computing, Networking, Storage and Analysis}}}}\ (\bibinfo {year} {2011})\ \Eprint {http://arxiv.org/abs/1109.2935} {arXiv:1109.2935 [hep-lat]} \BibitemShut {NoStop}%
\end{thebibliography}%

\end{document}